\newif\ifpdf
\ifx\pdfoutput\undefined
  \pdffalse    
\else
  \pdfoutput=1 
  \pdftrue
\fi

\documentclass[rmp,onecolumn,amsmath,amssymb]{revtex4}

\ifpdf
  \usepackage[pdftex]{graphicx}
  \usepackage[colorlinks=true, pdfstartview=FitV, linkcolor=blue, 
            citecolor=blue, urlcolor=blue]{hyperref}
\else
  \usepackage{graphicx}
\fi

\newcommand{\br}{{\bf r}}

\newcommand{\xor}{{\text{\sc XOR}}}
\newcommand{\cpf}{{\text{\sc CPF}}}
\newcommand{\p}{{\cal P}}
\newcommand{\boldphi}{{\mbox{\boldmath $\phi$}}}  

\newcommand{\bra}[1]{{\langle #1 |}}
\newcommand{\ket}[1]{{| #1 \rangle}}

\newcommand{\spupup}{\ket{\!\uparrow\uparrow}}
\newcommand{\spupdown}{\ket{\!\uparrow\downarrow}}
\newcommand{\spdownup}{\ket{\!\downarrow\uparrow}}
\newcommand{\spdowndown}{\ket{\!\downarrow\downarrow}}

\newcommand{\bomega}{{\mbox{\boldmath $\omega$}}}
\newcommand{\bdelta}{{\mbox{\boldmath $\delta$}}}
\newcommand{\bphi}{{\mbox{\boldmath $\varphi$}}}

\newcommand{\Tr}{\mathrm{Tr}}

\newcommand{\bb}{{\mbox{\boldmath $\beta$}}}

\begin{document}


\title{Theory of solid state quantum information processing}
\author{Guido Burkard}
\affiliation{IBM T.\ J.\ Watson Research Center,
         P.\ O.\ Box 218,
         Yorktown Heights, NY 10598, USA}

\begin{abstract}
Recent theoretical work on solid-state proposals for the implementation
of quantum computation and quantum information processing is reviewed.  
The differences and similarities between microscopic and macroscopic qubits are 
highlighted and exemplified by the spin qubit proposal on one side and 
the superconducting qubits on the other.  Before explaining the spin
and supercondcuting qubits in detail, some general concepts that are
relevant for both types of solid-state qubits are reviewed.
The controlled production of entanglement in solid-state devices, the transport
of carriers of entanglement, and entanglement detection will be discussed
in the final part of this review.
\end{abstract}

\maketitle

\tableofcontents


\section{\label{introduction}Introduction}

The capabilities of information processing devices are derived
from their physical properties;  in Landauer's words, 
``Information is physical'' \cite{landauer91}.
The pioneers of quantum information processing recognized 
that if a device was quantum mechanical, then it could have computational 
powers exceeding those of a classical machine.
A sign for the superiority of quantum hardware is that typical 
simulations of quantum systems on classical computers appear to be
computationally hard.

This article is intended to give an overview of the theory of 
solid-state quantum information processing.  
For a general introduction to quantum computation (QC) and 
quantum information, we refer the reader to \cite{nielsen00}.
Although the distinction
between different quantum devices is probably less fundamental than that 
between quantum and classical devices, Landauer's motto can also be applied here. 
In other words, the specific physical properties of the quantum 
hardware \textit{do} matter.  Two rather different categories of this hardware
are those involving atomic systems, e.g.,
\begin{itemize}
\item atoms in an ion trap,
\item atoms in an optical lattice,
\item ensemble of nuclear spins in a liquid,
\end{itemize}
and those involving solid-state systems, e.g.,
\begin{itemize}
\item spins of electrons in semiconductor quantum dots,
\item nuclear spins of donor atoms in a semiconductor,
\item superconducting microcircuits containing Josephson junctions.
\end{itemize}
This list is by no means complete; an informative collection
of various proposals can be found in \cite{braunstein00}.
While there have so far been more successful demonstrations
involving atomic systems in the laboratory, many solid-state systems
are scalable, i.e., one can fabricate systems with many quantum
bits (qubits) using essentially the same fabrication technique that is
proposed or used for a single qubit.

\subsection{\label{divincenzofive}What actually has to be achieved?  DiVincenzo's criteria}
For the following discussion of attempts to implement a quantum
computer (or parts of it) in solid-state systems, it may be
useful to review what actually has to be achieved.  An excellent
summary of the criteria for the physical implementation of quantum 
computation are DiVincenzo's following ``five requirements'' 
\cite{divincenzo97,divincenzo00b}.

\subsubsection{A scalable physical system with well characterized qubits}
A quantum bit, or qubit, is a suitable quantum-mechanical two-state
system (see item \ref{ss-decoherence} for more about what it means for the qubit to be quantum
mechanical).  A pure state of the two-state system then takes the form
\begin{equation}
  \label{qubitstate}
  |\psi\rangle = \alpha |0\rangle +\beta |1\rangle,
\end{equation}
where the amplitudes $\alpha$ and $\beta$ are complex numbers
such that $|\alpha|^2+|\beta|^2=1$. The states $|0\rangle$ and 
$|1\rangle$ form an orthonormal basis of the Hilbert space
${\cal H}_2 = {\rm span}\{|0\rangle, |1\rangle\}$ of the qubit.
A good example of a quantum two-state
system is the spin 1/2 of an electron, where
$|\!\uparrow\rangle \equiv|0\rangle$ and 
$|\!\downarrow\rangle \equiv|1\rangle$.
The Hilbert space of the 
entire system then needs to be a tensor product of 
a large number $n$ of such two-state systems,
\begin{equation}
  \label{hilbertspace}
  {\cal H} = {\cal H}_2^{\otimes n} 
           = \underbrace{{\cal H}_2\otimes{\cal H}_2\otimes\cdots\otimes{\cal H}_2}_{n\,{\rm factors}}.
\end{equation}
An excellent tutorial on the physical meaning of the tensor product
in Eq.~(\ref{hilbertspace}) and the difference between classical and
quantum bits can be found in \cite{mermin03}.
A system is \textit{scalable} if it can be realized
(in principle) for arbitrary $n$.  Some early atomic qubit realizations
are not (easily) scalable, and one of the biggest
motivations for studying solid-state qubits is the hope
that they will be scalable like conventional solid-state
integrated circuits.  A collection of identical particles,
e.g., the Fermi sea of electrons in a metal, typically
does not represent a set of \textit{well characterized}
qubits.  The qubits need to be ``labeled'' in order to
make them distinguishable, e.g., in an arrangement where
single electrons sit on localized sites (quantum dots,
donor levels of impurity atoms) and can be addressed, e.g., 
as ``spin of the $i$-th dot''.

\subsubsection{The ability to initialize the state of the qubits}
Before a quantum computation is started, a fresh register
of qubits, e.g., in the state
\begin{equation}
  \label{zeroregister}
  |\psi\rangle = |0\rangle^{\otimes n} 
               = \underbrace{|0\rangle\otimes|0\rangle\otimes\cdots\otimes|0\rangle}_{n\,{\rm factors}},
\end{equation}
is required.  This requirement looks more innocent than
it actually is, since it is not always easy to create such
low-entropy states, e.g., if the temperature is not 
sufficiently low.  Suppling a quantum computer
with fresh ``zeros'' is also essential for quantum
error correction, where the entropy that accumulates
due to decoherence is pumped out of the quantum memory \cite{nielsen00}.
For this purpose, it also matters \textit{how fast} the
fresh ``zeros'' can be supplied.

\subsubsection{\label{ss-decoherence}Long relevant decoherence times, much longer than the gate operation times}
A decoherence time characterizes how long it takes until the quantum phase coherence
of a system (e.g., a qubit) is lost due to its interaction with the environment.
Frequently used figures of merit are the so-called energy-relaxation time $T_1$
and the decoherence time $T_2$ of a single qubit (the notation originates from 
the NMR literature).  To illustrate the meaning of these
two quantities, let us assume for the moment that $T_1 \gg T_2$ (this need not
be the case;  $T_1$ and $T_2$ can also be of the same order of magnitude).
A pure state of a qubit, Eq.~(\ref{qubitstate}), has degraded to an incoherent
mixture after a time of the order of $T_2$, described by the density matrix
\begin{equation}
  \label{mixedstate}
  \rho = |\alpha|^2 |0\rangle\langle 0|+|\beta|^2 |1\rangle\langle 1|.
\end{equation}
An elementary introduction into the meaning of the density operator $\rho$
can be found in quantum mechanics textbooks or in \cite{nielsen00}.
Note that, in the case where $|0\rangle$ and $|1\rangle$ are
eigenstates with different energies, the processes involved in
the decoherence of the qubit so far did not involve any energy 
exchange with the environment.  Nevertheless, this loss of the phase
information is sufficient to disrupt a quantum computation.
After a time of the order of $T_1$, energy relaxation has taken
place and the system is in the thermal equilibrium state $\rho=Z^{-1}\exp(-H/k_B T)$,
with the partition sum $Z={\rm Tr}\,\exp(-H/k_B T)$, the qubit Hamiltonian $H$,
the temperature $T$, and Boltzmann's constant $k_B$.
The requirement for quantum computation is that $T_2 \gg T_{\rm op}$
where $T_{\rm op}$ denotes the time to perform a typical operation
from the universal set (see Eq.~(\ref{eq:universal2}) below).  In order to
achieve quantum computations of arbitrary length with the help of quantum
error correction, it is required that the
error probability per gate $\epsilon \approx T_{\rm op}/T_2$ is below
its threshold value $\epsilon_{\rm thres}$ for fault-tolerant quantum
computation \cite{nielsen00}.  The number $\epsilon_{\rm thres}$
depends on the error-correcting codes used and the type of errors
they have to protect against.

\subsubsection{A universal set of quantum gates}\label{universal-set}
It is required that there is a set ${\cal S}$ of unitary operators,
called \textit{gates} or \textit{quantum gates}, operating on a bounded
number of qubits at a time, from which all unitary operators $U$ on any
number of qubits can be composed by applying them in series,
\begin{equation}
  \label{eq:universal1}
  U = U_k U_{k-1} \cdots U_2 U_1,
\end{equation}
where $U_i\in{\cal S}$.
It has been shown that there are universal sets consisting of quantum gates
that operate only on one or two qubits \cite{divincenzo95b}, 
e.g., the union of one suitable two-qubit gate $U_{(2)}$ with
the set of all operations on a single qubit,
\begin{equation}
  \label{eq:universal2}
  {\cal S} = \{ U_{(2)}\} \cup SU(2).
\end{equation}
Examples of suitable two-quits gates $U_{(2)}$ are the 
CNOT gate, also known as quantum-XOR or simply XOR \cite{barenco95b},
with the following matrix representation in ${\cal H}_2 \otimes {\cal H}_2$,
\begin{equation}
  \label{eq:XOR}
  U_{\sc XOR} = \left(\begin{array}{c c c c}
      1 & 0 & 0 & 0\\
      0 & 1 & 0 & 0\\
      0 & 0 & 0 & 1\\
      0 & 0 & 1 & 0
\end{array}\right),
\end{equation}
or the square-root of SWAP \cite{loss98},
\begin{equation}
  \label{eq:SSWAP}
  S = \left(\begin{array}{c c c c}
      1 & 0 & 0 & 0\\
      0 & \frac{1-i}{2} & \frac{1+i}{2} & 0\\
      0 & \frac{1+i}{2} & \frac{1-i}{2} & 0\\
      0 & 0 & 0 & 1
\end{array}\right).
\end{equation}

It should be added here that there are ways of achieving unitary gates
by performing non-unitary operations on a larger Hilbert space.  There
have been several proposals for doing universal quantum computation by
performing von Neumann measurements on a subset of the qubits of a
large entangled state or cluster state \cite{raussendorf01,nielsen03,leung04}.
Another example where measurements are used to generate unitary gates
is that of free-electron quantum computation \cite{beenakker04}.

\subsubsection{The ability to measure specific single qubits}
At the end of a computation, the qubits (or, at least, a subset of them)
need to be measured individually in some fixed basis, e.g., the
computational basis given by the states $|0\rangle$ and $|1\rangle$.
The observable that is measured in this procedure is the Pauli matrix
\begin{equation}
  \label{Pauliz}
  \sigma_z = \left(\begin{array}{c c} 1 & 0 \\ 0 & -1 \end{array}\right).
\end{equation}

\subsection{\label{micromacro}Microscopic vs.\ macroscopic solid-state qubits}
The existing and proposed solid-state qubits can roughly be grouped
into two categories.  The qubits of the first category, which we will label 
\textit{microscopic}, are similar to the atomic qubits in the sense that 
they are based on quantum objects on the atomic scale whose states
$|0\rangle$ and $|1\rangle$ are 
distinguishable only by measuring a microscopic observable, such as an
angular momentum on the order of Planck's constant $\hbar$ or 
a magnetic dipole moment
of the order of one Bohr magneton, $\mu_B$. 
Electron and nuclear spin qubits, as well as the orbital state of an electron
in a semiconductor quantum dot, fall under this category.  
The second category of qubits we call \textit{macroscopic},
for their distinguishability under measurement of a macroscopic
observable, such as a current carried by a large number of electrons,
the magnetic field induced by such a current, or the position of an
electron charge in a system with two macroscopically distinguishable sites.
The typical examples in this category are the superconducting qubits
(with exceptions).

\subsection{\label{scope}Scope of this review article}
This is not intended to be a comprehensive review of all
theoretical work that has been done in the field of solid-state
quantum computation.  Besides the discussion of some general concepts that
apply for a broad range of possible implementations in Sec.~\ref{concepts}, 
we concentrate on qubits based on the electron spin (Sec.~\ref{spin})
in semiconductor structures (quantum dots) and on superconducting 
circuits (Sec.~\ref{sc}), representing an example of a microscopic and
a macroscopic qubit.

Other solid-state proposals for quantum computation include 
quantum Hall systems \cite{privman98,yang02}, anyons
in fractional quantum Hall systems \cite{kitaev03}, the
nuclear spin of donors in a semiconductor \cite{kane98},
electron charge degrees of freedom in quantum dots
\cite{barenco95a,landauer96,brum97,zanardi98,tanamoto00},
``flying'' electron spin qubits in surface acoustic waves \cite{barnes00}
or ballistic quantum wires \cite{popescu04},
ferroelectrically coupled quantum dots \cite{levy01},
excitons \cite{biolatti00,troiani00,chen01}, SiGe quantum
dots \cite{vrijen00}, paramagnetic impurities in semiconductor 
quantum wells \cite{bao03}, Si-based solid-state NMR \cite{ladd02},
and electrons on the surface of liquid He \cite{platzman99}.



\section{General concepts}
\label{concepts}

\subsection{The Loss-DiVincenzo proposal}\label{LD}
The underlying idea of this proposal is that the spins of single electrons confined 
in semiconductor quantum dots (e.g., in a two-dimensional semiconductor heterostructure)
are to be used as qubits \cite{loss98}.
The required coupling between the qubits in this case is provided by the tunneling 
between adjacent quantum dots, giving rise to a nearest-neighbor exchange coupling.  
The resulting spin Hamiltonian is that of the Heisenberg model,
\begin{equation}
  \label{spin-Hamiltonian}
  H(t) = \sum_{\langle i,j\rangle}   J_{ij}(t) {\bf S}_i  \cdot {\bf S}_j   
               +   \mu_B \sum_i  g_i {\bf B}({\bf r}_i)\cdot{\bf S}_i,
\end{equation}
where ${\bf S}_i$ denotes the spin operator of the electron in the $i$-th quantum dot and
$J_{ij}$ the exchange energy between spins $i$ and $j$.  

It has to be noted, however, that this proposal for exchange-based QC
extends far beyond electron spins in quantum dots.  Subsequent proposals for QC,
using the nuclear spins of donor atoms buried in a silicon substrate \cite{kane98},
or using electron spins in SiGe quantum dots \cite{vrijen00}, 
electrons trapped by surface acoustic waves \cite{barnes00},
and spins of paramagnetic impurities \cite{bao03},
rely on the same type of interaction.

In Eq.~(\ref{spin-Hamiltonian}), we have also taken into account the Zeeman coupling to an
external magnetic field ${\bf B}$ which may be spatially varying.  It may also be that
the Lande g-factor $g_i$ is also be site-dependent in some semiconductor heterostructures.  The Bohr magnetic moment is denoted by $\mu_B$.
\begin{figure}
\centerline{\includegraphics[width=14cm]{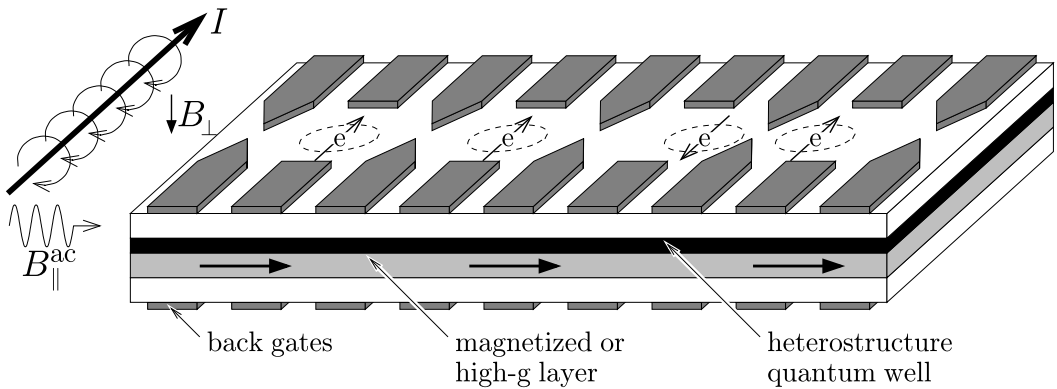}}
\caption{\label{qd-array}
Schematics of a quantum-dot array for quantum computing according to \cite{loss98}.
Quantum dots (dashed circles) are defined in a two-dimensional semiconductor 
heterostructure with metal gates (shown schematically in grey) and host one
(excess) electron (e) with a spin 1/2 each.  By controlling the gate voltages,
the coupling of adjacent quantum dots is switched on and off for quantum gate
operations.}
\end{figure}
Structures with two coupled quantum dots where the electron number could be controlled
one-by-one down to a single electron per dot have recently been demonstrated in
GaAs-AlGaAs heterostructures \cite{elzerman03}, see Fig.~\ref{delft-dots}.

In the ``idle'' phase, i.e., when no quantum gates are performed on the register,
the exchange coupling would be switched off $J_{ij}=0$ between all dots $i$ and $j$.
In order to perform an elementary two-qubit gate between dots $i$ and $j$, the exchange
coupling between dots $i$ and $j$ is temporarily switched on, while leaving the other
exchange couplings off. Several non-overlapping pairs of qubits can be coupled 
simultaneously in this way.  A pulse $J_{ij}(t)$ with the property
\begin{equation}
  \label{Jpluse}
  \frac{1}{\hbar}\int J_{ij}(t')dt' = \frac{\pi}{2}\quad \left({\rm mod}\, 2\pi\right)
\end{equation}
generates the above-mentioned square-root of SWAP gate (up to an unimportant global phase factor
$e^{-i\pi/8}$ which we omit below),
\begin{equation}
S \simeq \exp\left(\frac{i}{\hbar}\int dt' H(t')\right)  = \exp\left(i\frac{\pi}{2}{\bf S}_i\cdot{\bf S}_j\right).
\end{equation}
The quantum gate $S$ can be combined with single-spin rotations 
\begin{equation}
U_i(\boldphi) = \exp(i \boldphi \cdot{\bf S}_i),\label{single_qubit}
\end{equation}
to produce a controlled phase flip (CPF) \cite{loss98},
\begin{equation}\label{cpf_circuit}
U_{\sc CPF}=e^{-i\frac{\pi}{2}}e^{i\frac{\pi}{2}S_1^z}e^{-i\frac{\pi}{2}S_2^z} S e^{i\pi S_1^z} S,
\end{equation}
which, up to a basis change, equals the quantum XOR gate:
\begin{eqnarray}
U_{\sc XOR} &=& V  U_{\sc CPF}  V^{\dagger},\label{basis_change}\\
V       &=& \exp(-i\pi S_2^y/2).\label{basis_change_unitary}
\end{eqnarray}
\begin{figure}
\begin{minipage}{11cm}
\includegraphics[width=10.7cm]{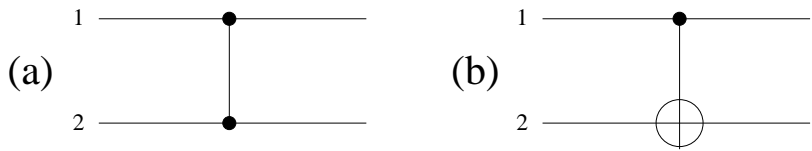}
\end{minipage}
\hfill
\begin{minipage}{6cm}
\caption{\label{ugates}\small
Circuit notation of two examples of two-qubit gates that are
universal for quantum computation when combined with single-qubit gates.
(a) The `square-root-of-swap' (S) gate, (b) the XOR gate.}
\end{minipage}
\end{figure}
\begin{figure}[b]
\begin{minipage}{9.3cm}
\includegraphics[width=9cm]{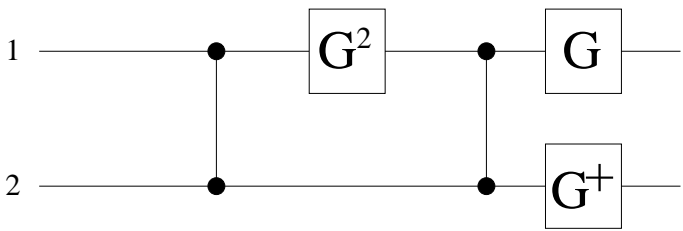}
\end{minipage}
\hfill
\begin{minipage}{5cm}
\caption{\label{cpf_fig}\small
A circuit representation for the conditional phase flip ($\cpf$), 
Eq.~(\ref{cpf_circuit}).
The single qubit rotations are 
$G=e^{i\frac{\pi}{2}S^z}$,
$G^\dagger=e^{-i\frac{\pi}{2}S^z}$,
and $G^2=e^{i\pi S^z}$. The $\cpf$ is related to the 
XOR gate Eq.~(\ref{eq:XOR})
by the basis transformation Eq.~(\ref{basis_change}).}
\end{minipage}
\end{figure}

The effect of an inhomogeneous external magnetic field on the 
exchange interaction and the robustness of the procedure described
here are discussed in \cite{sousa01}.

\subsection{QC with anisotropic couplings}\label{QC-anisotropic}

\subsubsection{Ising and transverse (XY) coupling}
Both for photon-mediated spin-spin coupling in a semiconductor microcavity \cite{imamoglu99}
and for inductively coupled superconducting qubits \cite{makhlin99,makhlin01},
which will both be further discussed in Secs.~\ref{qed} and \ref{sc} below, 
the coupling takes an anisotropic form instead of being described by the isotropic
Heisenberg Hamiltonian Eq.~(\ref{spin-Hamiltonian}).  
In both cases, the form of the coupling
turns out to be that of the XY (transverse) spin Hamiltonian,
\begin{equation}\label{XY}
H_{XY} = J \sum_{i,j}(S_i^x S_j^x + S_i^y S_j^y)
= \frac{J}{2} \left(\begin{array}{c c c c}
0 & 0 & 0 & 0\\
0 & 0 & 1 & 0\\
0 & 1 & 0 & 0\\
0 & 0 & 0 & 0
\end{array}\right),
\end{equation}
where we chose the $S^z$ basis of the two interacting qubits
for the matrix representation of $H_{XY}$.

It is known that any generic two-qubit Hamiltonian gives rise to a universal set of 
gates when combined with single-qubit operations.
In two notable cases of anisotropic spin couplings, the Ising and the XY interactions, it is known how the $\cpf$ and $\xor$ gates can be constructed.  In the case of a system described by the Ising Hamiltonian $H_I = J S_1^z S_2^z$ and a homogeneous magnetic field in $z$ direction, there is a particularly simple realization of the $\cpf$ gate with constant parameters, namely $U_{\cpf} = \exp(i\pi(1-2S_1^z-2S_2^z+4S_1^z S_2^z)/4)$ \cite{loss98}. 

For the transverse spin-spin coupling of Eq.~(\ref{XY}), 
we find that a useful two-qubit gate, such as the
conditional-phase-flip (CPF) operation, can be carried out by
combining $H_{XY}$ with one-bit rotations.
The unitary evolution operator generated by 
the Hamiltonian of Eq.~(\ref{XY}) is
\begin{eqnarray}
U_{XY}(\phi) & = &  T \exp\left[ i \int dt H_{XY} \right]
= \exp\left[ i \phi (S_i^x S_j^x + S_i^y S_j^y)\right]
\label{UXY}
\end{eqnarray}
where $\phi = \int dt J(t)$. The CPF gate ($U_{\cpf}$) can be
realized by the sequence of operators \cite{imamoglu99,burkard99b}
\begin{equation}
U_{\cpf}  =   e^{i \pi/4} e^{i \pi {\bf n}_i \cdot
{\mbox{\boldmath $\sigma$}}_i/3} e^{i \pi {\bf n}_j \cdot
{\mbox{\boldmath $\sigma$}}_j/3}
U_{XY}(\pi/2) 
e^{i \pi \sigma_z^i/2} U_{XY}(\pi/2)
e^{i \pi \sigma_y^i/4} e^{i \pi \sigma_y^j/4}
\label{CPF-XY}
\end{equation}
where ${\mbox{\boldmath $\sigma$}}$ denotes the vector Pauli operator,
where ${\bf S} = {\mbox{\boldmath $\sigma$}}/2$,
and ${\bf n}_i=(1,1,-1)/\sqrt{3}$ and ${\bf n}_j=(-1,1,1)/\sqrt{3}$.
The $\xor$
gate can be realized by combining the CPF operation with single-qubit
rotations as in Eqs.~(\ref{basis_change}) and (\ref{basis_change_unitary}).

While it is impossible to generate the CNOT gate with a single use of
the XY Hamiltonian \cite{burkard99b}, it is possible to generate a different
universal quantum gate with the XY interaction in a
single pulse;  the CNOT + SWAP (CNS) gate 
$U_{\text{\sc CNS}} = U_{\text{\sc SWAP}} U_{\xor}$,
is generated as \cite{schuch03}
\begin{equation}
  \label{CNS}
  U_{\text{\sc CNS}} = H_1 U_{XY}(\pi)e^{-i \pi \sigma_z^i/4} e^{-i \pi \sigma_z^j/4}   H_2,
\end{equation}
where $H_i$ is the Hadamard gate
\begin{equation}
  \label{eq:5}
  H = \frac{1}{\sqrt{2}}\left(\begin{array}{c c} 1 & 1 \\ 1 & -1 \end{array}\right),
\end{equation}
applied to qubit $i$.

Gate errors due to unwanted inhomogeneous magnetic fields during an
otherwise isotropic coupling, effectively creating an anisotropy,
have been studied and quantified in \cite{hu03}.

\subsubsection{Anisotropy due to the spin-orbit coupling}

The exchange interaction, Eq.~(\ref{spin-Hamiltonian}),
between electron spins in tunnel-coupled sites
(such as quantum dots)
can acquire anisotropic terms due to spin-orbit coupling
during tunneling between the sites \cite{kavokin01}.
Surprisingly, it turns out that the first-order effect of the 
spin-orbit coupling during quantum gate operations can be eliminated
by using time-symmetric pulse shapes for the
coupling between the spins \cite{bonesteel01}.
A related, but independent, result shows that 
the spin-orbit effects {\em exactly}
cancel in the gate sequence on the right hand side of 
Eq.~(\ref{cpf_circuit}) required to produce the
quantum {\sc XOR} gate, provided that
the pulse form for the spin-orbit and the exchange couplings
are identical \cite{burkard02a}.   The {\sc XOR} gate being
universal when complemented with single-qubit operations, this result
implies that the spin-orbit coupling can be dealt with in
any quantum computation.  In any real implementation, there
will be some (small) discrepancy between the pulse shapes for the
exchange and the spin-orbit coupling; however, one can
choose two pulse shapes which are very similar.  It 
was shown that the cancellation still holds to a 
very good approximation in such a case,
i.e.\ the effect of the spin-orbit coupling will still be
strongly suppressed \cite{burkard02a}.  There is also an effect
of dipolar interactions between adjacent spins, providing
another anisotropic coupling; this coupling can also be treated
as an anisotropic contribution to Eq.~(\ref{spin-Hamiltonian})
and therefore cancels out in the gate sequence Eq.~(\ref{cpf_circuit})
for the same reasons as the spin-orbit interaction.

The spin-orbit coupling for a conduction-band electron
is given by the following Hamiltonian \cite{gantmakher87}, 
being linear in the 2D
momentum operator $p_i$, $i=x,y$ ([100] orientation of the 2D plane),
\begin{equation}
\label{s-o}
H_{\rm so}= \sum_{i,j=x,y} \beta_{i j} \sigma_i p_j ,
\end{equation}
where the constants $\beta_{ij}$
depend on the strength of the confinement in z-direction  and
are of the order $(1\div 3) \cdot 10^5 \,{\rm cm/s}$ for GaAs heterostructures.
Combining the isotropic Heisenberg coupling (\ref{spin-Hamiltonian})
with the anisotropic exchange between two localized spins ${\bf S}_1$ and ${\bf S}_2$ 
one obtains the Hamiltonian \cite{burkard02a}
\begin{equation}
H(t) = J(t) \left( {\bf S}_1 \cdot {\bf S}_2 +   {\cal A}(t)\right),
\end{equation}
where the anisotropic part is given by the expression \cite{kavokin01},
\begin{equation}
   \label{a}
   {\cal A}(t) = \bb(t) \cdot ({\bf S}_1\times {\bf S}_2)
            +  \gamma(t) (\bb(t)\cdot{\bf S}_1) (\bb(t)\cdot{\bf S}_2),
\end{equation}
and $\beta_i = \sum_j\beta_{ij}\langle \psi_1 | i p_j|\psi_2\rangle$
is the spin-orbit field,
$|\psi_i\rangle$ the ground state in site (dot) $i=1,2$,
and $\gamma\approx O(\beta^0)$.
As was discussed in Sec.~\ref{LD}, for ${\cal A}=0$, the quantum {\sc XOR} 
gate can be obtained by applying $H(t)$ twice, together with single-spin rotations,
see Eqs.~(\ref{cpf_circuit}) and (\ref{basis_change_unitary}).
Moreover, if ${\cal A}=0$, then $H(t)$ commutes
with itself at different times and the time-ordered exponential
\begin{equation}
\label{t-exp}
U(\varphi) = T\exp\left(-i \int_{-\tau_s/2}^{\tau_s/2} \! H(t) \, dt \right)
\end{equation}
is a function of the integrated interaction strength only,
$\varphi = \int_{-\tau_s/2}^{\tau_s/2} J(t) dt$.
In particular, $U(\varphi=\pi/2)=U_{\rm sw}^{1/2}=S$ is the
``square-root of swap'' gate.

The interesting situation is of course ${\cal A} \neq 0$.
If in this case, $\bb$ and $\gamma$ (and thus ${\cal A}$) are 
time-independent, then $H(t)$ \textit{still} commutes with itself 
at different times and one can find a fixed coordinate
system in which $\bb$ is parallel to the $z$ axis.  In this
basis, the anisotropic term Eq.~(\ref{a}) can be expressed as
\begin{equation}
{\cal A} = \beta (S_1^x S_2^y - S_1^y S_2^x) + \delta S_1^z S_2^z ,
\end{equation}
with $\delta=\gamma\beta^2$.
In the singlet-triplet basis with basis vectors $\{\ket{T_{+}}=\spupup,\ket{S} = 
(\spupdown-\spdownup)/\sqrt{2},
\ket{T_0}=(\spupdown+\spdownup)/\sqrt{2},\ket{T_{-}}=\spdowndown\}$ 
the gate sequence Eq.~(\ref{cpf_circuit}), including the anisotropy Eq.~(\ref{a}),
yields the unitary operation
\begin{equation}
   \label{umatrix}
   U_g={\rm diag}(ie^{-i\varphi(1+\delta)},1,1,-ie^{-i\varphi(1+\delta)}),
\end{equation}
where ${\rm diag}(x_1,\ldots , x_4)$ denotes the diagonal matrix with
diagonal entries $x_1,\ldots , x_4$.
Note that the pulse strength $\varphi$ and the spin-orbit parameters only
enter $U_{g}$ in the $S^z=\pm 1$ subspaces.
Moreover, the terms linear in $\bb$ have canceled out exactly in $U_g$.
With the choice $\varphi=\pi/2(1+\delta)$, one obtains
the conditional phase flip gate
$U_g=U_{\sc CPF}={\rm diag}(1,1,1,-1)$,
being equivalent to the {\sc XOR} up to the basis change,
Eq.~(\ref{basis_change_unitary}).
Therefore, the anisoptropic terms ${\cal A}={\rm const.}$ in the spin 
Hamiltonian cancel exactly in the gate sequence Eq.~(\ref{basis_change})
for the quantum {\sc XOR}.

We briefly discuss what happens if, as can be expected
in real systems, the anisotropic terms in the
Hamiltonian $H$ are not exactly proportional to $J(t)$, 
i.e.\ if ${\cal A}(t)$ is time-dependent.  
Generally, both $\bb$ and $\gamma$ depend on time.  
In this more general case, we cannot exactly
eliminate the effect of the anisotropy
because of the time-ordering in Eq.~(\ref{t-exp}) and
since the Hamiltonian cannot be expected to commute with 
itself at different times, $[H(t),H(t')]\neq 0$.
The estimated gate errors $\epsilon=||U_g-U_{\sc CPF}||^2$ due
to the anisotropy in the case where
${\cal A}(t)$ is only weakly time-dependent are
$\epsilon\le\Delta^2$ where
we use $\Delta\beta(t) = \beta(t)-\beta_0$ and
\begin{equation}
\Delta  = (|\varphi | \beta_0 /2) \max_{|t|\le \tau_s/2}
|(J(t)/J_0)(\beta(t)/\beta_0-1)|,
\end{equation}
where $J_0$ stands for the average exchange coupling,
$J_0 = \varphi/\tau_s\neq 0$.
It can be shown \cite{burkard02a} that for tunnel-coupled quantum dots,
it is possible to choose a weakly time-dependent ${\cal A}$
by using Eq.~(\ref{J})
for the exchange coupling and the result
\begin{equation}
b(d,q) \equiv |J(d,q) \bb(d,q)|=b_0  \sqrt{q} d \exp(-2 q d^2),
\end{equation}
where $b_0=a/a_B^0$, $a_B^0=\sqrt{\hbar/m\omega_0}$, and where $a$
is a constant depending on the
spin-orbit parameter (for a $5\:{\rm nm}$ wide [100] GaAs quantum well
$a \approx 2\:{\rm meV}\,{\rm nm}$), $q=\omega/\omega_0$.
The minimal value of the quantum dot confinement energy $\omega$
is denoted by $\omega_0$.
A possible model for the switching process is
the use of a time-dependent confinement strength
$q(t) = \omega(t)/\omega_0 = \cosh ^2 (\alpha t/\tau_s )$
(where $alpha$ is a number of order $1$, e.g., $\alpha=4$).
This pulse shape has favorable adiabatic properties
\cite{burkard99a,schliemann01a}, as detailed in Sec.~\ref{sec-adiab},
and leads to a pulsed exchange interaction $J(t)=J(d,q(t))$
and spin-orbit field $b(t)=b(d,q(t))$,
where $-\tau_s/2\le t\le \tau_s/2$.
The resulting error was estimates in \cite{burkard02a}
as $\Delta\approx  7\cdot 10^{-3}$, leading to
gate errors occurring at a rate
$\epsilon \approx 4  \Delta^2 \approx  2\cdot 10^{-4}$
being around the currently known threshold
for fault tolerant quantum computation \cite{preskill98a,preskill98b}.
The error $\epsilon$ can be further reduced by performing the
gates more slowly, with a long period of constant ${\cal A}$ 
between the rise and fall of the pulses.

\subsection{Universal QC with the exchange coupling}
\label{ch-exchange}

In some situations, a local controllable field ${\bf B}_i$ or 
g-factor $g_i$  in the Hamiltonian
Eq.~(\ref{spin-Hamiltonian}) and thus the single-qubit operations 
$SU(2)$ in the universal set Eq.~(\ref{eq:universal2}) may be more 
costly to implement than the tunable exchange coupling generating the
spin-spin coupling $U_{(2)}$ (note, however, that there exist
all-electric switching schemes using g-factor modulation, see Sec.~\ref{ch-gfactor}).
A scheme has been developed in which the Heisenberg interaction alone
suffices to exactly implement any quantum computer circuit, at a price
of a factor of three in additional qubits and about a factor of ten in
additional two-qubit operations.  
However, the Heisenberg interaction by itself is not
a universal gate \cite{barenco95b}, in the sense that it cannot generate
any arbitrary unitary transformation on a collection of spin-1/2
qubits.  This is why in Eq.~(\ref{eq:universal2}), the Heisenberg
interaction needs to be combined with some other means of applying 
independent one-qubit gates.  
The Heisenberg interaction alone does not give a
universal quantum gate because it has too much symmetry: it commutes
with the operators $S^2$ and $S_z$, where the total spin is defined as
\begin{equation}
  \label{totalspin}
  {\bf S} = \sum_{i=1}^n {\bf S}_i,
\end{equation}
and therefore it
can only rotate among states with the same $S,\ S_z$ quantum numbers.

\subsubsection{Encoding}\label{encoding}
\begin{figure}
\begin{minipage}{8.3cm}
\includegraphics[width=8cm]{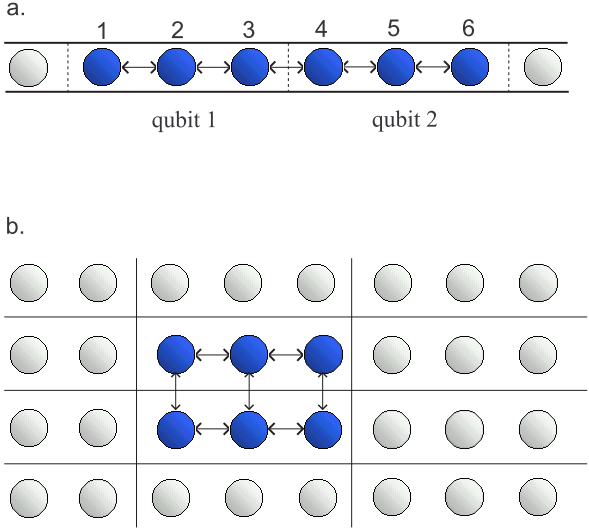}
\end{minipage}
\hfill
\begin{minipage}{6cm}
\caption{\small
Possible layouts of spin-1/2 devices.  a) One-dimensional
layout.  We consider two different assumptions about how the exchange
interactions can be turned on and off in this layout: 1) At any given
time each spin can be exchange-coupled to at most one other spin (we
refer to this as ``serial operation'' in the text), 2) All exchange
interactions can be turned on simultaneously between any neighboring
pair of spins in the line shown (``1D parallel operation'').  b)
Possible two-dimensional layout with interactions in a rectangular
array.  We imagine that any exchange interaction can be turned on
between neighboring spins in this array (``2D parallel operation'').
Of course other arrangements are possible, but these should be
representative of the constraints that will be faced in actual device
layouts.
\label{exchange-fig1}}
\end{minipage}
\end{figure}

The exchange coupling is thus not universal in the full Hilbert space;
but, by working exclusively in one symmetry sector of the Hilbert space
with fixed $S,\ S_z$ quantum numbers, the exchange coupling can be made
universal.  This restriction is achieved 
by defining coded qubit states, ones for which the spin quantum
numbers always remain the same \cite{bacon00,viola00,kempe00}.
The smallest number of spins 1/2 for which two orthogonal states 
with identical $S,\ S_z$ exist is three.  The space of three-spin states 
with spin quantum numbers $S=1/2$, $S_z=+1/2$ is two-dimensional and 
will serve to represent our coded qubit.  
An explicit choice for the basis states of this qubit are
\begin{eqnarray}
|0_L\rangle &=& |S\rangle|\uparrow\rangle,\label{code0}\\
|1_L\rangle &=& \sqrt{2/3}|T_+\rangle|\downarrow\rangle-\sqrt{1/3}|T_0\rangle
|\uparrow\rangle,\label{code1}
\end{eqnarray}
where $|S\rangle=\sqrt{1/2}(|\uparrow\downarrow\rangle-
|\downarrow\uparrow\rangle)$ is the singlet state of spins 1 and 2
(see Fig.~\ref{exchange-fig1}a) of the three-spin block, and
$|T_+\rangle=|\uparrow\uparrow\rangle$ and
$|T_0\rangle=\sqrt{1/2}(|\uparrow\downarrow\rangle+
|\downarrow\uparrow\rangle)$ are triplet states of these two spins.
While in principle this solves the problem of exchange-only QC,
in practice we would like to know what the overhead in terms of
qubits (for coding) and gates (for operating on encoded qubits
with the exchange interaction) will be, and how a universal set
of operations on the encoded qubits can be achieved \cite{divincenzo00a}.
It has also been found that the anisotropic
XY interaction (\ref{XY}) alone is sufficient for quantum
computation \cite{kempe02}, a result which was later generalized
to large class of anisotropic exchange Hamiltonians \cite{vala02}.
An encoding involving two spins per qubit has also been demonstrated
for universal quantum logic starting from locally alternating g-factors
\cite{levy02} and from a homogeneous magnetic field combined with
anisotropic exchange interactions \cite{wu02a,wu02b,wu04}.

\subsubsection{One-qubit gates}
A one-qubit gate on a single three-spin block is performed as
follows.  The Hamiltonian $H_{12}$ generates a
rotation $U_{12}=\exp(i/\hbar\int J{\vec S}_1\cdot{\vec S}_2 dt)$
which is just a $z$-axis rotation (in Bloch-sphere notation) on the
coded qubit, while $H_{23}$ produces a rotation about an axis in the
{\em x-z} plane, at an angle of 120$^o$ from the $z$-axis.  Since
simultaneous application of $H_{12}$ and $H_{23}$ can generate a
rotation around the $x$-axis, three steps of 1D parallel operation
(defined in Fig.~\ref{exchange-fig1}a) permit any one-qubit rotation,
using the classic Euler-angle construction.  
In serial operation, it can be found numerically
that four steps are always adequate when only nearest-neighbor
interactions are possible (e.g.\ the sequence
$H_{12}$-$H_{23}$-$H_{12}$-$H_{23}$ shown in
Fig.~\ref{exchange-fig2}a, with suitable
interaction strengths), while three steps suffice if interactions can
be turned on between any pair of spins (e.g.\
$H_{12}$-$H_{23}$-$H_{13}$, see Fig.~\ref{exchange-fig2}b).

\subsubsection{Two-qubit gates}
\begin{figure}
\begin{minipage}{8.3cm}
\includegraphics[width=7cm]{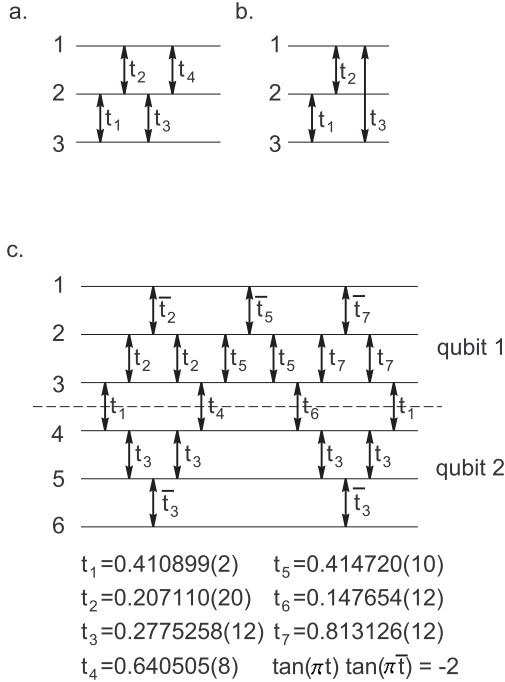}
\end{minipage}
\hfill
\begin{minipage}{9cm}
\caption{\small
Circuits for implementing single-qubit and two-qubit
rotations using serial operations.  a) Single-qubit rotations by
nearest-neighbor interactions.  Four exchanges (double-headed arrows)
with variable time parameters $\tau_i$ are always enough to perform
any such rotation, one of the two possible layouts is shown.  b)
Non-nearest neighbor interactions.  Only three interactions are
needed, one of the possible layouts is shown.  c) Circuit of 19
interactions that produce a cNOT between two coded qubits (up to
one-qubit gates before and after).  The durations of each interaction
are given in units such that for $t=1/2$ the rotation
$U_{ij}=\exp(iJt{\vec S}_i\cdot{\vec S}_j/\hbar)$ is a SWAP,
interchanging the quantum states of the two spins $i$, $j$.  The
$\bar{t}_i$ parameters are not independent, they are related to the
$t_i$s as indicated.  The uncertainty of the final digits of these
times are indicated in parentheses.  With these uncertainties, the
absolute inaccuracy of the matrix elements of the two-qubit gate
rotations achieved is no greater than $6\times 10^{-5}$.  Further fine
tuning of these time parameters would give the cNOT to any desired
accuracy.  In a practical implementation, the exchange couplings
$J(t)$ would be turned on and off smoothly; then the time values given
here provide a specification for the integrated value $\int J(t)dt$.
The functional form of $J(t)$ is irrelevant, but its integral must be
controlled to the precision indicated.
\label{exchange-fig2}}
\end{minipage}
\end{figure}

The implementation of two-qubit gates for universal QC with the
exchange interaction on two three-spin code blocks is less
intuitive that the corresponding task for one-qubit gates.
Much of the
difficulty of these searches arises from the fact that while the four
basis states $|0_L,1_L\rangle|0_L,1_L\rangle$ have total spin quantum
numbers $S=1$, $S_z=+1$, the complete space with these quantum numbers
for six spins has nine states, and exchanges involving these spins
perform rotations in this full nine-dimensional space.
Numerical searches for the implementation of two-qubit gates using a 
simple minimization algorithm \cite{divincenzo00a} aided by 
the two-qubit gates invariants \cite{makhlin02}  have resulted in
a sequence for an encoded CNOT operation that is depicted
in Fig.~\ref{exchange-fig2}.
The solution shown in Fig.~\ref{exchange-fig2}c
appears to be optimal for serial operation and
happens to involve only nearest neighbors in the 1D
arrangement of Fig.~\ref{exchange-fig1}a.
There also are (apparently) optimal numerical solutions for parallel
operation mode.  
For the 1D layout of Fig.~\ref{exchange-fig1}a, the simplest solution found
involves 8 clock cycles
with just 8*4 different interaction-time parameters ($H_{12}$ can
always be zero in this implementation).  For the 2D parallel mode of
Fig.~\ref{exchange-fig1}b, a solution was found using just 7 clock 
cycles (7*7 interaction times).

In the present scheme, quantum computation would proceed as follows.  
In the beginning, all the computational qubits would be set to the 
$|0_L\rangle$ state which is easily obtained using the exchange 
interaction: if a strong $H_{12}$ is turned on in each coded block 
and the temperature made lower than the strength $J$ of the interaction, 
these two spins will equilibrate to their ground state, which is the 
singlet state.  The third spin in the block should be in the 
$|\uparrow\rangle$ state,
which can be achieved by also placing the entire system in a
moderately strong magnetic field $B$, such that $k_B T<<g\mu_B B<J$
(it can be shown that in a slightly more general scheme involving
both the $S_z=+1$ and $S_z=-1$ subspaces, the last step can be
omitted).  After the computation, with the one- and two-qubit gates
 implemented according to the schemes mentioned above, the final
qubit measurement, we note that determining whether the spins 1 and 2
of the block are in a singlet or a triplet suffices to perfectly
distinguish \cite{divincenzo99b} $|0_L\rangle$ from $|1_L\rangle$ (again, the
state of the third spin does not enter).

\subsubsection{Protection against errors}
Codes of the type of Eqs.~(\ref{code0}) and (\ref{code1}) have first
been introduced as a computational basis in \textit{decoherence-free subspaces},
i.e., subspaces of a Hilbert space which are protected against errors with 
a certain type of symmetry \cite{zanardi97,lidar98,lidar99,bacon00,kempe00}.
Moreover, it has been suggested that the logical subspace may be energetically
separated from the remaining Hilbert space and thus protected against errors
in a system where a certain combination of exchange couplings always remains
switched on \cite{bacon01,weinstein04}.

\subsubsection{Related ideas}
Encoded qubits of a different kind, so-called spin-cluster qubits
\cite{meier03a,meier03b}, have been proposed in order to relax the requirements for 
control on the single-spin level while inheriting the favorable
single-spin properties such as long decoherence time and fast gate
operating time.  Spin cluster qubits are finite spin chains with 
Heisenberg or anisotropic (XY and Ising-like) antiferromagnetic 
exchange interaction that can have uniform or nonuniform 
interaction constants.

The use of many-electron QDs for
exchange-based quantum computations has been analyzed in \cite{vorojtsov04}.
A particular implementation of three-spin QDs encoding one qubit has
been put forward in \cite{kyriakidis04}.

\subsection{Optimization of quantum circuits}
\label{ch-qec}

A quantum gate operating on $n$ qubits can be represented as a $2^n\times 2^n$ 
unitary matrix.
Any quantum computation or algorithm can be split up into a series of elementary
gate operations drawn from a universal set involving only one or two qubits, 
as in Eq.~(\ref{eq:universal2}).  This is the quantum circuit representation
of quantum algorithms (or, unitary operations).  For a simple example, 
see Fig.~\ref{cpf_fig} 
for a circuit representations of $\cpf$ in terms of sqrt-of-SWAP gates.  
However, quantum circuits are 
in general not the most efficient way of implementing a quantum computation \cite{burkard99b}.
There are a number of related but different approaches
using, e.g., genetic algorithms and chirped Gaussian pulses \cite{sanders99}
or control theory \cite{khaneja01}.

If one is interested in optimizing the switching time $\tau_s$ for a desired unitary $U$,
with a given Hamiltonian, e.g., the spin Hamiltonian Eq.~(\ref{spin-Hamiltonian}),
one can depart from the circuit representation of the unitary by allowing arbitrary time
dependent parameters $\vec{p}(t)$ in the Hamiltonian.  In the case of the spin
Hamiltonian (\ref{spin-Hamiltonian}), we have $\vec{p}=(J, {\bf B}_1, {\bf B}_2, \ldots)$.
We will only demonstrate this optimization in the case of a simple two-qubit unitary,
the $\xor$ (CNOT) gate.  The optimization method can in principle be applied to unitaries of any
size; note, however, that the optimization as an arbitrary classical computational task 
is typically a hard computation in itself.

\subsubsection{Serial pulse mode}\label{serial_pulse}
We first restrict ourselves to a special class of parameter functions $\vec{p}(t)$,
in which at every time $t$, only one component of $\vec{p}(t)$ is non-zero.
If we further restrict ourselves to parameter functions in which the duration of the
$J$-pulses with $\vec{p}=(J, 0, 0, \ldots)$ are $\pi/2$ pulses generating the
sqrt-of-swap $S$, then we are back to the circuit model with the universal set
Eq.~(\ref{eq:universal2}) and $U_{(2)}=S$.  In this case we can optimize 
circuits, e.g., to have as few instances of $S$ as possible.  E.g., it turns
out that the use of two $S$ for a $\cpf$ as in the
sequence Eq.~(\ref{cpf_circuit}), and therefore, for $\xor$, is optimal.
Such minimal requirements for the implementation of a unitary $U$ can be proven by 
analyzing the set $\p(U)$ of product states
$\left\{|\Psi\rangle\in{\cal H}={\cal H}_2^{\otimes M}\Big||\Psi\rangle = |\phi_1\rangle\otimes\cdots\otimes|\phi_M\rangle ;|\phi_i\rangle\in{\cal H}_2\right\}$ which are mapped back onto 
product states by $U$ \cite{burkard99b}.
An alternative method for determining whether a Hamiltonian generates
a gate in a single pulse involves the invariants under addition of single-qubit
gates \cite{makhlin02}.

\subsubsection{Parallel pulse mode}\label{parallel_pulse}

In the case where several or all parameters $\vec{p}$ can be changed simultaneously, we expect that a given quantum gate, say $\xor$, can be performed faster than by changing only one parameter at a time as in the serial pulse mode.
The unitary time evolution operator after time $t$ is the following functional in $\vec{p}$, 
\begin{equation}
U_t[\vec{p}(\tau)] = T\exp\left(\frac{i}{\hbar}\int_0^t H(\vec{p}(\tau))\,d\tau\right),\label{propagator_general}
\end{equation}
where $T$ denotes the time-ordering.
For a given quantum gate $U_g$, the integral equation $U_t[\vec{p}(\tau)]=U_g$ has to be solved for the functions $\vec{p}(\tau)$.   An \textit{optimal} solution is given by a set of bounded functions $|p_i(\tau)|<M_i$ requiring minimal time $t$ for a fixed set of bounds $M_i$.  In order to simplify the problem, one can restrict the problem to piecewise-constant functions,
\begin{eqnarray}
U_N(\vec{p}^{\,(1)},...,\vec{p}^{\,(N)}; \phi) &=& e^{i\phi} U_N(\vec{p}^{\,(N)})\cdots U_2(\vec{p}^{\,(2)}) U_1(\vec{p}^{\,(1)}),\nonumber\\
U_k(\vec{p}^{\,(k)})  &=& \exp\left\{ i t H(\vec{p}^{\,(k)})\right\}.\label{propagator_discrete}
\end{eqnarray}
For each of the $N$ time intervals, one has the freedom to choose a new set of parameters $\vec{p}^{\,(k)}=(J, {\bf B}_1, {\bf B}_2)$.  The discretized problem can now be treated both analytically and numerically \cite{burkard99b}.

One finds analytically that $\cpf$ can be implemented in a single step
by fixing $N=1$, i.e., all parameters in Eq.~(\ref{spin-Hamiltonian}) simultaneously non-zero but constant,
\begin{equation}
  \label{CPF_solution}
  U_{\sc CPF} = \exp\left[ it  H(J,{\bf B}_1,{\bf B}_2)\right],
\end{equation}
The parameters are (in units of $2\pi \hbar/t$),
\begin{eqnarray}
J = k-n-2m-\frac{1}{2}  &,&  \quad \quad \phi = -\pi (n+\frac{1}{2}) \nonumber\\
{\bf B}_1 = \frac{1}{2}(0,0,n+\frac{1}{2}+\sqrt{k^2-J^2}) &,& \quad \quad 
{\bf B}_2 = \frac{1}{2}(0,0,n+\frac{1}{2}-\sqrt{k^2-J^2}),
\label{CPF_solution_param}
\end{eqnarray}
where $n$, and $m$ are arbitrary integers, and $k$ is an integer satisfying $2|k|\ge |n+2m+\frac{1}{2}|$.
In the specific case where all constraints are equal to $M$, we find that the solution for $k=1$, $m=n=0$,
\begin{equation}\label{cpf_parallel}
J=\frac{1}{2},\quad B_1^z=\frac{1}{4}(1+\sqrt{3}),\quad B_2^z=\frac{1}{4}(1-\sqrt{3}), \quad \phi = - \frac{\pi}{2}
\end{equation}
has the shortest switching time,
\begin{equation}
t_{\cpf,p}=\frac{2\pi\hbar}{4 M}(1+\sqrt{3})=0.683 \frac{2\pi\hbar}{M},
\end{equation}
less than half the time which is used for the serial pulse quantum circuit Eq.~(\ref{cpf_circuit}), $t_{\cpf,s}=1.5\cdot 2\pi\hbar/M$.
Numerically, one finds that $\xor$ requires at least $N=2$ steps,
\begin{equation}
U_{\xor}=e^{i\phi}e^{i\phi}\exp\left[ it  H(\vec{p}^{\,(2)})\right]\exp\left[ it H(\vec{p}^{\,(1)})\right],
\end{equation}
with the parameter values (in units of $2\pi \hbar/t$)
\begin{equation}\label{xor_parallel}
\begin{array}{| r | r r r r r r r |}
\hline
k & J^{(k)} & B_{1x}^{(k)} & B_{2x}^{(k)} & B_{1y}^{(k)} & B_{2y}^{(k)} & B_{1z}^{(k)} & B_{2z}^{(k)}\\ \hline
1 &     0.187       &  -0.025 &       {0.464} &  0.205   & 0.195        & -0.420      & 0.395\\ 
2 &     {0.617} &  -0.220 &       0.345 & -0.384         & 0.244        & 0.353       & 0.108\\
\hline
\end{array}
\end{equation}
and the global phase $\phi=-0.8481 \cdot \pi$,
where the time $t$ has to be chosen such that none of the parameters exceeds the bound $M$.
The total switching time for equal bounds is in this case $t_{\xor,p}=(0.4643+0.6170)2\pi\hbar/M=1.0813\cdot 2\pi\hbar/M$, compared to $t_{\xor,s}=2\cdot 2\pi\hbar/M$ for the serial switching.

\subsubsection{Anisotropic systems}\label{anisotropic}

Parallel switching is also possible with the XY dynamics Eq.~(\ref{XY}).
It can be shown that $U_{\cpf}$ requires two pulses,
\begin{equation}
U_{CPF} = e^{i\phi}\,U_2 U_1,
\end{equation}
\[ {\rm where}\quad\quad U_k=\exp\left[2\pi i H_{XY,B}(J^{(k)},B_x^{(k)},B_z^{(k)})\right],\quad
k=1,2.\]
Note that all magnetic fields can be chosen homogeneous (${\bf
B}_1^{(k)}={\bf B}_2^{(k)}\equiv{\bf B}^{(k)}$) and perpendicular to the
$y$-axis ($B_y=0$). 
Here we give one possible realization which is found numerically ($\phi=-3\pi/4$):
\begin{equation}\label{cpf_parallel_xy}
\begin{array}{| r | r r r |}
\hline
k &  J^{(k)}  &  B_x^{(k)} &  B_z^{(k)}\\ \hline
1 &  0.7500 &  {\bf 0.7906}  &  0.5728\\
2 &  {\bf 0.5000} &  0.0000  &  0.2500\\ \hline
\end{array}
\end{equation}
The total switching time for $\cpf$, assuming equal bounds $M_J=M_B\equiv M$ for $J$ and $B$, is $t_{\cpf,p}^{XY}=1.291\cdot 2\pi\hbar/M$, compared to $t_{\cpf,s}^{XY}=2.167\cdot 2\pi\hbar/M$ for the serial pulse sequence defined in Eq.~(\ref{CPF-XY}).

In order to produce the XOR gate Eq.~(\ref{eq:XOR})
we can implement the basis change Eq.~(\ref{basis_change})
using the single-qubit rotation $V$.
This procedure requires a total of four steps for the XOR gate.
Another way of achieving XOR is the following sequence which
we found numerically and which takes only three steps:
\begin{equation}
U_{\sc XOR} = \exp(3i\pi/4)U_3  U_2  U_1,
\end{equation}
with the following parameters:
\begin{equation}\label{xor_parallel_xy}
\begin{array}{| r | r r r r r r r |}
\hline
  k &   J^{(k)}    &    B_{1x}^{(k)} & B_{2x}^{(k)} &   B_{1y}^{(k)} & B_{2y}^{(k)} & B_{1z}^{(k)} & B_{2z}^{(k)}\\
\hline
  1 &   1.802   &    0.615     &  {\bf 2.045}    &   0.020     &  0.316    &   0.794   &    0.130 \\
  2 &   {\bf 3.344}   &    0.348     &  0.718    &   0.259     &  0.493    &   1.583   &    1.062 \\
  3 &   {\bf 1.903}   &    1.193     &  0.705    &   0.413     &  -0.305   &   0.589   &    0.604 \\\hline
\end{array}
\end{equation}
The total switching time of $t_{\xor,p}^{XY}=7.29\cdot 2\pi\hbar/M$ (compared to $2.67\cdot 2\pi\hbar/M$ using $\cpf$ and a basis change) indicates that Eq.~(\ref{xor_parallel_xy}) is not an optimal solution.

\subsection{Adiabaticity}
\label{sec-adiab}
Quantum gates are generated by controlling the parameters 
in the Hamiltonian Eq.~(\ref{spin-Hamiltonian}),
$J_{ij}(t)$ and ${\bf B}_i(t)$ (or $g_i(t)$), as a function
of time.  E.g., the exchange coupling $J$ depends on time
via some physically controlled quantity, such as an
electric gate voltage $v(t)$, i.e., $J(t)=J(v(t))$
(similarly for the effective g-factor $g(t)$).
According to Eq.~(\ref{Jpluse}), 
only the time integral $\int_0^{\tau} J(v(t)) dt$
needs to assume a certain value (modulo $2\pi$)
in order to generate the correct quantum gate and the pulse form
of $v(t)$ does not matter.
However, the exchange interaction $J(t)$ needs to be switched \textit{adiabatically}
in order to avoid unwanted excitations in the system.
The adiabaticity condition is \cite{burkard99a,burkard99b,burkard00c}
 $|\dot{v}/v| \ll \delta\varepsilon/\hbar$,
where $\delta\varepsilon$ is the energy scale on which excitations may
occur.   Here, $\delta\varepsilon$ denotes
 the energy-level separation of a single dot, i.e.,
the smaller of either the single-electron level spacing or
the on-site Coulomb energy $U$ required to add a second electron
to a dot.
A rectangular pulse leads to excitation of higher levels,
 whereas an adiabatic pulse with amplitude $v_0$ is e.g. given by
 $v(t) = v_0\,{\rm sech}(t/\Delta t)$
 where $\Delta t$ controls the width of the pulse.
We need to use a switching time $\tau_s > \Delta t$,
  such that $v(t\!=\!\tau_s/2)/v_0$ becomes vanishingly small.
We then have $|\dot v/v|=|{\rm tanh}(t/\Delta t)|/\Delta t \leq 1/\Delta t$,
 so we need $1/\Delta t \ll \delta\varepsilon/\hbar$ for adiabatic
 switching.
The Fourier transform
 $v(\omega) = \Delta t v_0 \pi \, {\rm sech} (\pi\omega\Delta t)$
 has the same shape as $v(t)$ but width $2/\pi\Delta t$.
In particular,
 $v(\omega)$ decays exponentially in the frequency $\omega$,
 whereas it decays only with $1/\omega$ for a rectangular pulse.

Adiabatic switching of the exchange coupling in two coupled
quantum dots and the error probability for different pulse
forms have been studied numerically in \cite{schliemann01a}.
Furthermore, corrections to the fully adiabatic result have
been investigated \cite{requist04}.



\section{Electron spins}
\label{spin}

Being a natural two-level system, the spin 1/2 of the electron represents  
an ideal candidate for a qubit.  On the one hand, the electron spin
is typically quite well isolated from charge degrees of freedom (not completely,
though, due to, e.g., the spin-orbit coupling).
In some situations, electron spin decoherence times in solids appear to be relatively
long, exceeding microseconds \cite{kikkawa97,kikkawa98,awschalom99}.   
On the other hand, single spins in solid-state
structures are not readily available and controllable.  However, large
experimental efforts are currently made to isolate and control single spins in solid-state
structures.  The spin-based proposals for quantum information processing which will 
be discussed below are all based on artificial nano- or micrometer-scale semiconductor 
structures, such as quantum dots (QDs) or microcavities.

\subsection{Quantum Dots}
\begin{figure}[h]
\begin{minipage}{8.3cm}
\includegraphics[width=8cm]{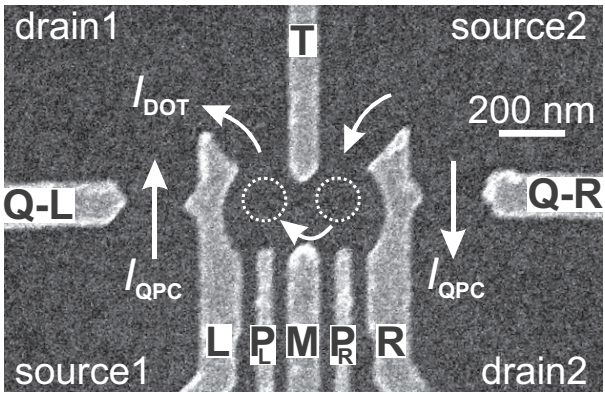}
\end{minipage}
\hfill
\begin{minipage}{8.5cm}
\caption{\label{delft-dots}
Electron micrograph of a structure comprising two QDs, defined by metal electrodes (bright structures)
on the surface of a GaAs/AlGaAs heterostructure (Courtesy of J. Elzerman, TU Delft).
The charge on the dots is controlled in steps of single electron charges, down to one electron per dot,
by tuning the voltage applied to the plunger gates ${\rm P}_{\rm L,R}$
and is monitored by measuring the conductance of (i.e., the currents $I_{\rm QPC}$ through) 
the quantum point contacts (QPCs) Q-R and Q-L.
Conductance spectroscopy was performed by measuring the current $I_{\rm dot}$
\cite{elzerman03}.}
\end{minipage}
\end{figure}

In \cite{loss98}, a quantum register is proposed in which single electrons are 
trapped in quantum dots (QDs) that are arranged in an array or lattice in a semiconductor
structure, e.g., as in Fig.~\ref{qd-array}.   Electrically defined QDs
in two-dimensional semiconductor heterostructures (typically, GaAs) are well-studied
objects \cite{kouwenhoven01} in which charge transport has attracted much
attention \cite{averin92,kouwenhoven97b,vanderwiel03}.
In recent years, the controlled storage of a \textit{single} electron---and thus a 
spin 1/2 or qubit---in a QD has been achieved \cite{ciorga00,elzerman03}.
Structures in which two QDs, each containing a well-controlled number
of electrons (down to a single electron), are adjacent and tunnel-coupled, have
been fabricated and studied \cite{elzerman03}.
In Fig.~\ref{delft-dots}, we show an electron micrograph of a structure of the type
that was used in \cite{elzerman03}.  The tunneling of electrons between the two dots
is predicted to give rise to the spin exchange coupling $J{\bf S}_1\cdot{\bf S}_2$
in Eq.~(\ref{spin-Hamiltonian}).  In the next section, we are going to outline a 
theory of this spin exchange mechanism.

\subsection{Exchange in laterally coupled QDs}
\label{lateral}

Due to the Coulomb interaction and the Pauli exclusion principle,
the ground state of two coupled electron sites (atoms, QDs) 
in the absence of a magnetic field is a spin singlet (a highly 
entangled spin state), while the spin triplet states (one of them entangled)
are typically separated by some energy gap $J$.
This energy gap is called \textit{exchange coupling}, as it arises from virtual
electron exchange between the two sites due to the interaction.
The virtual electron exchanges are allowed for opposite spins (spin singlet, $S=0$)
but forbidden by the Pauli principle for parallel spins (spin triplet, $S=1$),
therefore the energy of the singlet is lowered by the interaction.
\begin{figure}
\begin{minipage}{9.3cm}
\includegraphics[width=9cm]{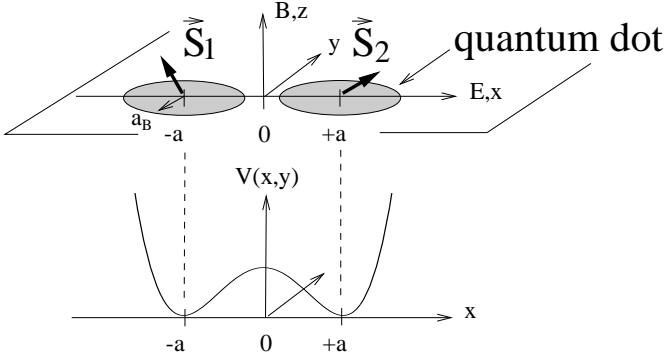}
\end{minipage}
\hfill
\begin{minipage}{5.5cm}
\caption{\label{qd-system}\small
Two coupled QDs with one valence electron per dot.
Each electron is confined to the $xy$ plane.
The spins of the electrons in
dots $1$ and $2$ are denoted by ${\bf S}_1$ and ${\bf S}_2$. The
magnetic field $B$ is perpendicular to the plane, i.e. along the $z$
axis, and the electric field $E$ is in-plane and along the $x$
axis. The quartic potential is given in Eq.~(\ref{potential}) and is
used to model the coupling of two harmonic wells centered at $(\pm
a,0,0)$. The exchange coupling $J$ between the spins is a function of
$B$, $E$, and the inter-dot distance $2a$.}
\end{minipage}
\end{figure}

We now introduce a model for the two laterally coupled QDs containing
one (conduction band) electron each \cite{burkard99a}.  The two-dot system is 
shown schematically in Fig.~\ref{qd-system}. 
It is essential that the electrons are allowed to tunnel between the dots,
and that the total wave function of the coupled system must be
antisymmetric.  It is this fact which introduces correlations
between the spins via the charge (orbital) degrees of freedom.
The electronic Hamiltonian in the effective-mass approximation
for the coupled system is then given by
\begin{equation}
H = \sum_{i=1,2} h({\bf r}_i, {\bf p}_i)+C+H_{\rm Z} = H_{\rm orb} + H_{\rm Z}, \label{hamiltonian}
\end{equation}
where the single-particle Hamiltonian,
\begin{equation}
h({\bf r}_i, {\bf p}_i) = \frac{1}{2m}\left({\bf p}_i-\frac{e}{c}{\bf A}(\br_i)
\right)^2+ex_iE+V(\br_i),\label{single-particle}
\end{equation}
describes the electron dynamics
confined to the $xy$-plane and
\begin{equation}
C = \frac{e^2}{\kappa\left| \br_1-\br_2\right|},\label{Coulomb}
\end{equation}
represents the Coulomb interaction
(unscreened in this case where the dot diameter is small or comparable to the 
screening length).
The electrons have an effective mass $m$ ($m=0.067\, m_e$
in GaAs) and carry a spin-1/2 ${\bf S}_i$.
The dielectric constant in GaAs is $\kappa = 13.1$. We allow for
a magnetic field ${\bf B}= (0,0,B)$ applied along the $z$-axis
and which couples to the electron charge via the
vector potential ${\bf A}(\br) = \frac{B}{2}(-y,x,0)$.
We also allow for an electric field $E$ applied in-plane along
the x-direction, i.e. along the line connecting the centers of the dots.
The coupling of the dots (which includes tunneling) can be
modeled by a quartic potential,
\begin{equation}\label{potential}
V(x,y)=\frac{m\omega_0^2}{2}\left(\frac{1}{4 a^2}\left(x^2-a^2
\right)^2+y^2\right)\, ,
\end{equation}
which separates (for $x$ around $\pm a$) into two harmonic wells of
frequency $\omega_0$, one for each dot, in the limit
of large inter-dot distance, i.e. for $2a\gg 2 a_{\rm B}$, where $a$ is
half the distance between the centers of the dots, and
$a_{\rm B}=\sqrt{\hbar/m\omega_0}$
is the effective Bohr radius of a single isolated harmonic well.
This choice for the potential is motivated by the experimental
fact \cite{tarucha96,kouwenhoven97a} that
the spectrum of single dots in GaAs is well described by a parabolic
confinement potential,
e.g. with $\hbar\omega_0 =3\,{\rm meV}$ \cite{tarucha96,kouwenhoven97a}.
We note that in this simplified model, increasing (decreasing) 
the inter-dot distance is physically equivalent to
raising (lowering) the inter-dot barrier, which can be
achieved experimentally by e.g.
applying a gate voltage between the dots \cite{waugh95,livermore96}.
Thus, the effect of such gate voltages is described in this model
simply by a change of the inter-dot distance $2a$.

The magnetic field $B$ also couples to the electron spins via the
Zeeman term $H_{\rm Z}=g\mu_{\rm B} \sum_i{\bf B}_i\cdot{\bf S}_i$,
where $g$ is the effective g-factor ($g\approx -0.44$ for GaAs), and
$\mu_{\rm B}$ the Bohr magneton.  The ratio between the Zeeman splitting and
the relevant orbital energies is small for all $B$-values of interest
here; indeed, $g\mu_{\rm B} B/\hbar\omega_0\lesssim 0.03$, for $B\ll
B_0=(\hbar\omega_0/\mu_{\rm B})(m/m_e)\approx 3.5\,{\rm T}$,
and $g\mu_{\rm B} B/\hbar\omega_{\rm L}\lesssim 0.03$, for $B\gg B_0$,
where $\omega_{\rm L}=eB/2mc$ is the Larmor
frequency, and where we used $\hbar\omega_0 =3\,{\rm meV}$. Thus, we
can safely ignore the Zeeman splitting when we discuss the orbital
degrees of freedom and include it later into the effective spin
Hamiltonian.

\subsubsection{The Heitler-London approach}\label{HL}

We consider first the Heitler-London (HL) approximation (also known as
valence orbit approximation), and then refine this
approach by including hybridization as well as double occupancy
in a Hund-Mulliken approach, which will finally lead us to an extension of
the Hubbard description. We will see, however,  that
the qualitative features of $J$ as a  function of the control parameters
are already
captured by the simplest HL approximation for the artificial
hydrogen molecule described by Eq.~\ref{hamiltonian}.

The HL approximation is borrowed from molecular physics.  In the
present case, think of a hydrogen molecule ${\rm H}_2$.  The HL approach starts from
single-dot ground-state ($s$ wave) orbital wavefunctions $\varphi(\br)$ and combines
them into the (anti-) symmetric  two-particle orbital state vector
\begin{equation}
\label{HL-ansatz}
|\Psi_{\pm}\rangle = \frac{|12\rangle \pm |21\rangle}{\sqrt{2(1\pm S^2)}},
\end{equation}
the positive (negative) sign
corresponding to the spin singlet (triplet) state,
and $S=\int d^2r\varphi_{+a}^{*}(\br)\varphi_{-a}(\br)=\langle 2|1\rangle$
denoting the overlap of the right and left orbitals.
A non-vanishing overlap implies that the electrons tunnel
between the dots (see also Sec. \ref{HM}).
Here, $\varphi_{-a}(\br)=\langle \br| 1\rangle$ and $\varphi_{+a}(\br)
=\langle \br|2\rangle$ denote the one-particle orbitals
centered at $\br=(\mp a,0)$, and $|ij\rangle = |i\rangle |j\rangle$ are
two-particle product states. 
The exchange energy is then obtained
through $J = \epsilon_{\rm t}-\epsilon_{\rm s} =
\langle\Psi_{-}|H_{\rm orb}|\Psi_{-}\rangle -
\langle\Psi_{+}|H_{\rm orb}|\Psi_{+}\rangle$.
The single-dot orbitals for harmonic
confinement in two dimensions in a perpendicular magnetic field are the
Fock-Darwin states \cite{fock28,darwin30}, which are the usual harmonic oscillator
states, magnetically compressed by a factor $b=\omega/\omega_0 =
\sqrt{1+\omega_{\rm L}^2/\omega_0^2}$, where $\omega_{\rm L} = eB/2mc$
denotes the Larmor frequency. The ground state
(energy $\hbar\omega=b\hbar\omega_0$) centered at the origin is
\begin{equation}
\varphi(x,y) = \sqrt{\frac{m \omega}{\pi\hbar}} e^{-m\omega
\left(x^2+y^2\right)/2\hbar}.
\end{equation}
Shifting the single particle orbitals to $(\pm a,0)$ in the
presence of a magnetic field we obtain $\varphi_{\pm a}(x,y) = \exp(\pm
iya/2l_B^2)\varphi(x\mp a,y)$, where the phase factor involving the magnetic
length $l_B=\sqrt{\hbar c/eB}$ is due to the gauge transformation ${\bf
A}_{\pm a}=B(-y,x\mp a,0)/2\rightarrow {\bf A}=B(-y,x,0)/2$.
We obtain \cite{burkard99a}
\begin{equation}\label{Jformal}
J = \frac{2S^2}{1-S^4}\left(\langle 12|C+W|12\rangle-\frac{{\rm Re}
\langle 12|C+W|21\rangle}{S^2}\right),
\end{equation}
where the overlap becomes $S=\exp(-m\omega a^2/\hbar-a^2\hbar/4l_B^4m\omega)$.
Evaluation of the matrix elements of $C$ and $W$ yields
\begin{equation}
J =  \frac{\hbar\omega_0}{\sinh\!\left(2d^2(2b-\frac{1}{b})\right)}\Bigg[c\sqrt{b}
\Big(e^{-bd^2}{\rm I_0}(bd^2) - e^{d^2 (b-1/b)}{\rm I_0}(d^2
\{b-\frac{1}{b}\})\Big) +\frac{3}{4b}\left(1+bd^2\right)\Bigg],\label{J}
\end{equation}
where we introduce the dimensionless distance $d=a/a_{\rm B}$, and ${\rm
I_0}$ is the zeroth order Bessel function.  The first and second terms
in Eq.~(\ref{J}) are due to the Coulomb interaction $C$, where the
exchange term enters with a minus sign.  The parameter
$c=\sqrt{\pi/2}(e^2/\kappa a_{\rm B})/\hbar\omega_0$ ($\approx 2.4$, for
$\hbar\omega_0=3\,$meV) is the ratio between Coulomb and
confining energy. The last term comes from the confinement potential
$W$.  The result $J(B)$ is plotted in Fig.~\ref{exchange} (dashed
line).  Note that typically $|J/\hbar\omega_0|\lesssim 0.2$. Also, we
see that $J>0$ for $B=0$, which must be the case for a two-particle
system that is time-reversal invariant \cite{mattis88}.
The most remarkable feature of $J(B)$, however, is the change of sign
from positive to negative at $B=B_*^{\rm s}$, which occurs over a wide
range of parameters $c$ and $a$. This singlet-triplet crossing occurs
at about $B_*^{\rm s}=1.3\,{\rm T}$ for $\hbar\omega_0=3\,{\rm meV}$
 ($c=2.42$) and $d=0.7$. The transition from antiferromagnetic ($J>0$) to
ferromagnetic ($J<0$) spin-spin coupling with increasing magnetic
field is caused by the long-range Coulomb interaction, in particular
by the negative exchange term, the second term in Eq.~(\ref{J}).  As
$B\gg B_0$ ($\approx 3.5\,{\rm T}$ for $\hbar\omega_0=3\,$meV),
the magnetic field compresses the orbits by a factor $b\approx
B/B_0\gg 1$ and thereby reduces the overlap of the wavefunctions,
$S^2\approx\exp(-2 d^2(2b-1/b))$, exponentially strongly. Similarly, the
overlap decays exponentially for large inter-dot distances, $d\gg 1$.
Note however, that this exponential suppression is partly compensated
by the exponentially growing exchange term $\langle 12|C|21\rangle/S^2\propto
\exp(2d^2(b-1/b))$. As a result, the exchange coupling $J$ decays
exponentially as $\exp(-2d^2b)$ for large $b$ or $d$, as shown in
Fig.~\ref{plots}b for $B=0$ ($b=1$).
Thus, the exchange coupling $J$ can be tuned
through zero and then suppressed to zero by a magnetic field in a
very efficient way.

\subsubsection{Limitations and extensions of HL}
We  note that the HL approximation breaks down
explicitly (i.e. $J$ becomes negative even when $B=0$) for certain
inter-dot distances if the interaction becomes too strong ($c$ exceeds $\approx 2.8$).

\begin{figure}
\begin{minipage}{8.1cm}
\includegraphics[width=7.8cm]{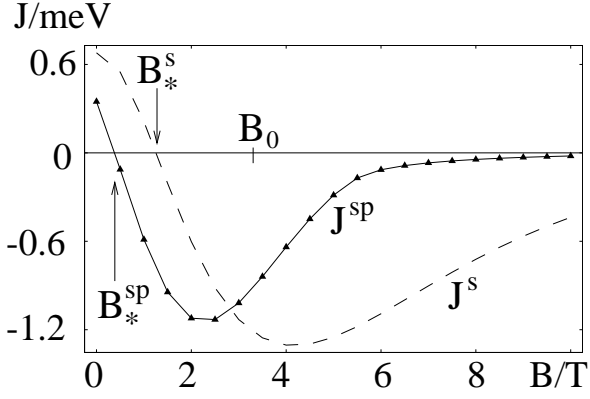}
\end{minipage}
\hfill
\begin{minipage}{9cm}
\caption{\label{exchange}\small
Exchange energy $J$ in units of
meV plotted against the magnetic field $B$ (in units of Tesla), as
obtained from the s-wave Heitler-London approximation (dashed line),
Eq.~(\ref{J}), and the  result from the improved
sp-hybridized Heitler-London approximation (triangles) which is obtained
numerically as explained in the text.
Note that the qualitative behavior
of the two curves is similar, i.e. they both
have zeroes, the s-wave approximation at $B_*^{\rm s}$,
and the sp-hybridized approximation at $B_*^{\rm sp}$, and also both curves
vanish exponentially for large fields.
$B_0=(\hbar\omega_0/\mu_{\rm B})(m/m_e)$ denotes the crossover
field to magnetically dominated confining ($B\gg B_0$).
The curves are given for a confinement energy $\hbar\omega_0=3\,{\rm meV}$
(implying for the Coulomb parameter $c=2.42$), and
inter-dot distance $a=0.7\, a_{\rm B}$.}
\end{minipage}
\end{figure}

The HL method can be improved by taking into account 
more than one single-dot orbital.  Admixture of higher orbitals can
be taken into account using a variational approach;  the orbitals
obtained in this way are termed \textit{hybridized} orbitals, in 
analogy to hybridized molecular orbitals in chemistry. Some
results obtained with sp-hybridized QD orbitals are plotted 
in Fig.~\ref{exchange}.

Another limitation of the HL approximation its restriction to quantum
dots that are occupied with a single electron.  Even with a single orbital,
the Pauli principle allows for the presence of a second electron with opposite
spin on a QD orbital.  While this admixture of double occupancy
is suppressed by the repulsive Coulomb interaction between electrons, it
nevertheless plays a relevant role.

\subsubsection{The Hund-Mulliken approach and the Hubbard Limit}
\label{HM}

\begin{figure}
\includegraphics[width=14cm]{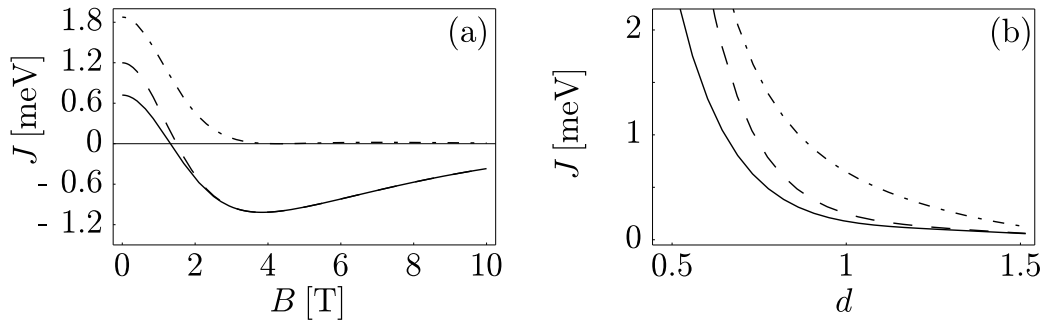}
\caption{\label{plots}\small
The exchange coupling $J$ obtained from
HM (full line), Eq.~(\ref{HMresult}), and from the extended
Hubbard approximation (dashed line), Eq.~(\ref{Hubbard}). For
comparison, we also plot the usual Hubbard approximation where the
long-range interaction term $V$ is omitted, i.e.  $J=4t_{\rm
H}^2/U_{\rm H}$ (dashed-dotted line).  In (a), $J$ is plotted as a
function of the magnetic field $B$ at fixed inter-dot distance
($d=a/a_{\rm B}=0.7$), and for $c=2.42$, in (b) as a function of
inter-dot distance $d=a/a_{\rm B}$ at zero field ($B=0$), and again
$c=2.42$.  For these parameter values, the s-wave Heitler-London $J$,
Eq.~(\ref{J}), and the HM $J$ (full line) are almost identical.}
\end{figure}

The Hund-Mulliken (HM) approximation (also known as molecular orbit
approximation \cite{mattis88}) extends the HL approach by 
including also the two doubly occupied states, which both are spin singlets
\cite{burkard99a}.
This extends the orbital Hilbert space from two to four dimensions.
First, the single particle states have to be orthonormalized,
leading to the states $\Phi _{\pm a}=(\varphi_{\pm a}-g\varphi_{\mp
a})/\sqrt{1-2Sg+g^2}$, where $S$ again denotes the overlap of
$\varphi_{-a}$ with $\varphi_{+a}$ and $g=(1-\sqrt{1-S^2})/S$. Then,
diagonalization of
\begin{equation}\label{matrix}
H_{\rm orb} = 2\epsilon + \left(\begin{array}{cccc}
U&X&-\sqrt{2}t_{\rm H}&0\\
X&U&-\sqrt{2}t_{\rm H}&0\\
-\sqrt{2}t_{\rm H}&-\sqrt{2}t_{\rm H}&V_+&0\\
0&0&0&V_-
\end{array}\right)
\end{equation}
in the space spanned by 
$\Psi^{\rm d}_{\pm a}(\br_1,\br_2)=\Phi_{\pm a}(\br_1)\Phi_{\pm
a}(\br_2)$, $\Psi^{\rm
s}_{\pm}(\br_1,\br_2)=[\Phi_{+a}(\br_1)\Phi_{-a}(\br_2)\pm
\Phi_{-a}(\br_1)\Phi_{+a}(\br_2)]/\sqrt{2}$
yields the eigenvalues
$\epsilon_{{\rm s}\pm} = 2\epsilon + U_{\rm H}/2+V_+\pm\sqrt{U_{\rm
H}^2/4+4t_{\rm H}^2}$, $\epsilon_{{\rm s} 0} = 2\epsilon + U_{\rm H} - 2X
+ V_+$ (singlet), and $\epsilon_{\rm t} = 2\epsilon + V_-$ (triplet),
where the quantities are given in \cite{burkard99a}.
The exchange energy then becomes
\begin{equation}
J = \epsilon_{\rm t}-\epsilon_{{\rm s}-}=V - \frac{U_{\rm H}}{2} +
\frac{1}{2}\sqrt{U_{\rm H}^2 + 16t_{\rm H}^2}.
\label{HMresult}
\end{equation}
In the standard Hubbard approach for short-range Coulomb interactions (and
without
$B$-field) \cite{mattis88} $J$ reduces to $-U/2 + \sqrt{U^2+ 16t^2}/2$,
where $t$ denotes the hopping matrix element,
and $U$ the on-site repulsion.
Thus, $t_{\rm H}$ and $U_{\rm H}$ are the extended hopping matrix element
and the on-site repulsion, resp., renormalized by  long-range Coulomb
interactions. The remaining two singlet energies $\epsilon_{{\rm s}+}$ and
$\epsilon_{{\rm s}0}$
are separated from $\epsilon_{\rm t}$ and $\epsilon_{{\rm s}-}$ by a gap of
order $U_{\rm H}$ and are therefore neglected for the study of
low-energy properties.
Typically, the ``Hubbard ratio'' $t_{\rm H}/U_{\rm H}$ is less than
$1$, e.g., if $d=0.7$, $\hbar\omega_0=3\,$meV, and $B=0$,
we obtain $t_{\rm H}/U_{\rm H}=0.34$, and it decreases with
increasing $B$. Therefore, we are in an \textit{extended} Hubbard
limit, where $J$ takes the form
\begin{equation}
J = \frac{4t_{\rm H}^2}{U_{\rm H}}+V.
\label{Hubbard}
\end{equation}
The first term has the form of the standard Hubbard
approximation \cite{fradkin91} but with $t_{\rm H}$ and $U_{\rm H}$ 
being renormalized by long-range Coulomb interactions.
The second term $V$ is new and
accounts for the difference in Coulomb energy between the
singly occupied singlet and triplet states $\Psi^{\rm s}_{\pm}$. It is
precisely this $V$
that makes $J$ negative for high magnetic fields, whereas $t_{\rm
H}^2/U_{\rm H}>0$ for all values of $B$ (see Fig.~\ref{plots}a).
Thus, the usual Hubbard approximation (i.e. without $V$) would not give
reliable
results, neither for the $B$-dependence (Fig.~\ref{plots}a)
nor for the dependence on
the inter-dot distance $a$ (Fig.~\ref{plots}b).
Since only the singlet space has
been enlarged, it is clear that we obtain a lower singlet energy
$\epsilon_{\rm s}$ than that from the s-wave Heitler-London calculation,
but the same
triplet energy $\epsilon_{\rm t}$, and therefore
$J=\epsilon_{\rm t}-\epsilon_{\rm s}$
exceeds the s-wave Heitler-London result, Eq. (\ref{J}).
However, the on-site Coulomb
repulsion
$U\propto c$ strongly suppresses the doubly occupied states $\Psi^{\rm
d}_{\pm a}$ and already for the value of $c=2.4$ (corresponding
to $\hbar \omega_0=3 $meV) we
obtain almost perfect agreement with the s-wave Heitler-London result
(Fig.~\ref{exchange}).
For large fields, i.e. $B\gg B_0$,
the suppression becomes even stronger ($U\propto \sqrt{B}$) because
the electron orbits become compressed with increasing $B$ and two electrons on
the same dot are confined to a smaller area leading to an increased Coulomb
energy.
Being a completely orbital effect, the exchange interaction
between spins of course competes with the Zeeman coupling $H_Z$ of the
spins to the magnetic field. In our case, however, the Zeeman energy $H_Z$
is small and exceeds the exchange energy (polarizing the spins) only in a
narrow window (about $0.1\,{\rm T}$ wide) around $B_*^{\rm sp}$ and again
for high fields ($B>4\,{\rm T}$).

\subsubsection{Numerical work}
While the calculation discussed in Sec.~\ref{HM} above take only the 
ground-state orbital in each QD into account, the HM, like
the HL, approximation can be refined to include more orbital levels
of the QDs.  Such extended calculations are usually done
numerically, and are very closely related to Hartree-Fock (HF)
calculations.  Note, however, that HF is not sufficient for
the purpose of calculating a spin exchange coupling, since it
is not capable of including entangled (quantum correlated) states
such as the spin singlet or $m=0$ triplet.  This is typically
remedied by invoking the so-called configuration-interaction
method which includes linear superpositions of HF states.
Numerical studies of the double-dot system with one
\cite{hu00} and three \cite{hu01} electrons per QD
showed good agreement with the somewhat more crude
approximations discussed above.

At finite magnetic field, the exchange coupling Eq.~(\ref{J})
can be tuned through zero by changing electrostatic properties
(QD size, distance, electric field).  
It has been confirmed both numerically and
in actual experiment \cite{kyriakidis02} that singlet-triplet
crossings can be induced in a single QD by changes in the dot
potential at constant magnetic field.

\subsubsection{Measurements of QD exchange}
\begin{figure}
\begin{minipage}{8.3cm}
\includegraphics[width=8cm]{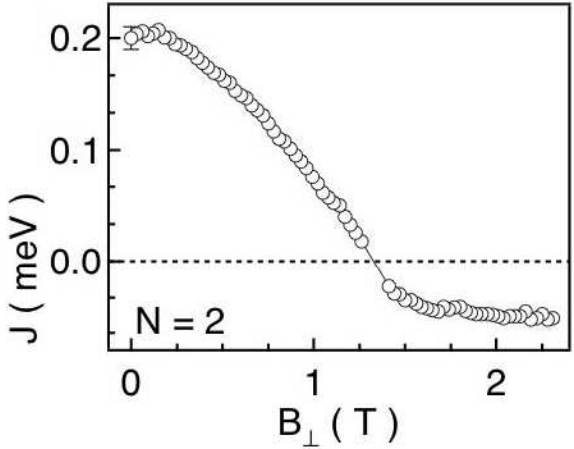}
\end{minipage}
\hfill
\begin{minipage}{8cm}
\caption{The exchange coupling $J$ measured as a function of the
applied magnetic field $B_\perp$ using conductance spectroscopy
in a two-electron dot system defined in a GaAs/AlGaAs heterostructure.
There are signatures that a double has been formed although
a single dot structure was used in the experiment \cite{zumbuhl04}.
The shape of the dot is not circular, but somewhat elongated.
The dot spectra appear to be consistent with a parabolic 
potential with harmonic energies $\hbar \omega_a = 1.2\,{\rm meV}$
and $\hbar\omega_b = 3.3\,{\rm meV}$, corresponding to a 
spatial elongation of $\sqrt{\omega_b/\omega_a} \sim 1.6$.
(Figure courtesy of D. M. Zumb\"uhl, Harvard University).
\label{fig-zumbuhl}}
\end{minipage}
\end{figure}

Signatures of singlet-triplet crossings
have been observed using transport spectroscopy
in lateral GaAs quantum dot structures \cite{zumbuhl04} 
(see Fig.~\ref{fig-zumbuhl}).
Although a single dot structure was used, there are
signatures that a double dot was formed in the experiment
\cite{engel04}.

These data seem to be in rather good qualitative agreement
with theory \cite{burkard99a},
bearing in mind that the absolute magnitude of the exchange 
coupling $J$ strongly depends on the inter-dot distance which
is a free parameter of the theory.
Similar double-dot experiments with the double-dot systems
shown in Fig.~\ref{delft-dots} are in preparation.

\subsection{Exchange in vertically coupled QDs}
\label{ch-ver}
\begin{figure}[b]
\begin{minipage}{8.3cm}
\includegraphics[width=8cm]{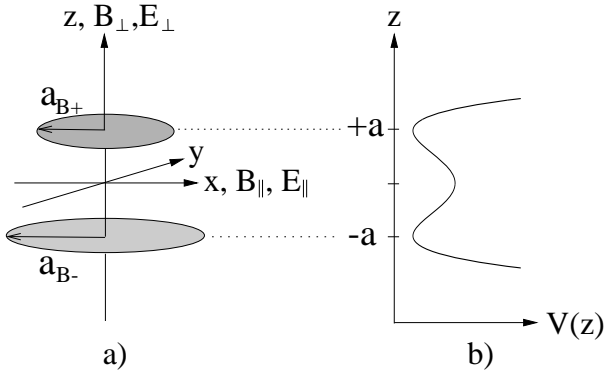}
\end{minipage}
\hfill
\begin{minipage}{8cm}
\caption{\label{potential-v}\small
(a) Sketch of the vertically coupled double quantum-dot
system. The two dots may have different lateral diameters,
$a_{B+}$ and $a_{B-}$. We consider magnetic and electric fields applied
either in-plane ($B_\parallel$, $E_\parallel$) or perpendicularly
($B_\perp$, $E_\perp$).
(b) The model potential for the vertical confinement
is a double well, which is obtained by combining two
harmonic wells at $z=\pm a$.}
\end{minipage}
\end{figure}

While lateral QDs are adjacent to each other in a two-dimensional electron gas,
vertically coupled QDs are stacked on top of each other in a three-dimensional 
semiconductor structure.
Vertical coupling occurs both in QDs etched out of
multilayer structures and electrically gated \cite{austing98} 
and in self-assembled QDs originating from the Stran\-ski-Krastanov
growth \cite{fafard99,luyken98,fricke96}.
A system of vertically coupled QDs is shown schematically
in Fig.~\ref{potential-v}.
There are a number of works on the exchange coupling between spins located 
in this type of coupled QDs \cite{oh96,imamura96,imamura98,imamura99,tokura99,burkard00a}.
This Section in devoted to the exchange coupling between spins in
vertically coupled dots under the influence of both in-plane 
magnetic and electric fields, $B_\parallel$
and $E_\parallel$, and perpendicular fields $B_\perp$, $E_\perp$ \cite{burkard00a}.
Electronic spectra and charge densities for two electrons in a system
of vertically tunnel-coupled QDs at zero magnetic field, ${\bf B}=0$, were
calculated in \cite{bryant93}.
Singlet-triplet crossings in the ground state of single \cite{wagner92,wojs96} and
coupled dots with two \cite{oh96} to four \cite{imamura96,imamura98,imamura99,tokura99} electrons in
vertically coupled dots in the presence of a magnetic field perpendicular
to the growth direction ($B_\perp$ in Fig.~\ref{potential-v}) have been
predicted.
In order to apply the HL and HM methods introduced in Sec.~\ref{lateral} to
vertically coupled QDs, a Hamiltonian of the same form as Eq.~(\ref{hamiltonian})
can be used, but with the single-particle Hamiltonian replaced by \cite{burkard00a}
\begin{equation}
h({\bf r}, {\bf p}) = \frac{1}{2m}\left({\bf p}-\frac{e}{c}{\bf A}(\br) 
\right)^2+ezE+V_l(\br)+V_v(\br).\label{hamiltonian-v}
\end{equation}
Also note that ${\bf r}$ is a three-dimensional vector here, as opposed
to a two-dimensional vector in Sec.~\ref{lateral}.
The potential $V_l$ in $h$ describes the lateral 
confinement, whereas $V_v$ models the vertical double-well structure.  For 
the lateral confinement the parabolic potential 
\begin{equation}\label{lateral-v}
V_l (x,y)=\frac{m}{2} \omega^2_{z}\left\{ 
\begin{array}{ll} \alpha^2_{0+} (x^2 + y^2), & z>0,\\ \alpha^2_{0-} (x^2 + 
y^2), & z<0,\\ \end{array} \right.
\end{equation}
is chosen, where we have introduced the anisotropy parameters
$\alpha_{0\pm}$ determining the strength of the 
vertical relative to the lateral confinement.
In the presence of a magnetic field $B_\perp$ perpendicular to the 2DES,
the one-particle problem has the Fock-Darwin states \cite{fock28,darwin30} as an
exact solution. Furthermore, it has been shown 
experimentally  \cite{fricke96} and theoretically \cite{wojs96} that a 
two-dimensional harmonic confinement potential is a reasonable 
approximation to the real confinement potential in a lens-shaped SAD.
In describing the  confinement $V_v$ along the inter-dot axis,
a (locally harmonic) double well potential of the form (see Fig.~\ref{potential-v}b)
\begin{equation}\label{VOsc}
V_v = \frac{m\omega_z^2}{8a^2}\left(z^2-a^2 \right)^2,
\end{equation}
can be used;
in the limit of large inter-dot distance $a\gg a_{\rm B}$, the potential $V_v$
in the vicinity of $z\approx \pm a$ becomes a harmonic well of frequency $\omega_z$. 
Here $a$ is half the distance between the centers
of the dots and $a_B=\sqrt{\hbar/(m\omega_z})$ is the 
vertical effective Bohr radius.
For most vertically coupled dots, the vertical 
confinement is determined by the conduction band offset between different 
semiconductor layers; therefore in principle a square-well potential would
be a more accurate description of the real potential than the harmonic
double well (note however, that the required conduction-band offsets
are not always known exactly).
There is no qualitative difference between the results presented below
obtained with harmonic potentials and the corresponding
results obtained using square-well potentials \cite{seelig99}.

\subsubsection{Perpendicular Magnetic Field $B_\perp$}\label{perpendicular}

For a magnetic field $B=B_\perp$ (cf. Fig.~\ref{potential-v})
and $E=0$, one obtains the ground-state Fock-Darwin \cite{fock28,darwin30} solution
\begin{equation}\label{oneparticle}
\varphi_{\pm a}(x,y,z) = \left(\frac{m \omega_z}{\pi\hbar}\right)^{3/4}
\sqrt{\alpha_\pm}
e^{-m\omega_z \left (\alpha_\pm \left(x^2+ y^2\right)
+(z \mp a)^2\right)/2\hbar}, 
\end{equation}
corresponding to the ground-state energy
$\epsilon_{\pm}=\hbar\omega_z(1+2\alpha_{\pm})/2$.
In Eq.~(\ref{oneparticle}) we have introduced
$\alpha_{\pm}({\rm B})=\sqrt{\alpha^2_{0\pm}+\omega_L(B)^2/\omega^2_z}
=\sqrt{\alpha^2_{0\pm}+B^2/B^2_0}$,
with $\omega_L(B) =eB/2mc$ the Larmor frequency and $B_0=2mc\omega_z/e$ 
the magnetic field for which $\omega_z=\omega_L$.  The parameters 
$\alpha_{\pm}(B)$ describe the compression of the one-particle wave 
function perpendicular to the magnetic field.
The HL with this Hamiltonian yields
\begin{eqnarray}\label{JHLBz}
J &=&\frac{2S^2}{1-S^4}\hbar\omega_z \bigg[ c \sqrt{\mu}\,e^{2\mu 
d^{2}}\left(1-{\rm erf}\left(d \sqrt{2\mu}\right)\right)
 -  \frac{c}{\pi} \frac{\alpha_+ + \alpha_-}{\sqrt{1-(\alpha_+ + \alpha_- 
-1)^{2}}}\, {\rm arccos}(\alpha_+ + \alpha_- -1) \nonumber\\
& & +  \frac{1}{4} \left(\alpha^2_{0+}-\alpha^2_{0-}\right)\left(\frac{\alpha_+ - 
\alpha_-}{\alpha_+ \alpha_-}\right)\, \left(1-{\rm 
erf(d)}\right)+\frac{3}{4} \left(1+d^2\right) \bigg],
\end{eqnarray}
where ${\rm erf}(x)$ denotes the error function.
We have introduced the dimensionless parameters $d=a/a_B$ for the inter-dot 
distance, and $c=\sqrt{\pi/2}(e^2/\kappa a_{\rm B})/\hbar\omega_z$ for the
Coulomb interaction. Note that
$\alpha_{\pm}$, $\mu=2\alpha_+ \alpha_-/\left(\alpha_+ +\alpha_-\right)$,
and the overlap
\begin{equation}\label{overlaposc} 
S = 2\frac{\sqrt{\alpha_{+}\alpha_{-}}}{\alpha_{+} 
+\alpha_{-}}\,{\rm exp}(-d^2),
\end{equation}
depend on the magnetic field $B$.
The first term in the 
square brackets in Eq.~(\ref{JHLBz}) is an approximate evaluation of the 
direct Coulomb integral $\langle 12|C|12\rangle$ for $d\gtrsim 0.7$ and
for magnetic fields $B\lesssim B_0$.
The second term in Eq.~(\ref{JHLBz}) is the (exact) exchange Coulomb integral
$\langle 12|C|21\rangle/S^2$, while the last two terms stem from the potential 
integrals, which were also evaluated exactly.  If the two dots have the 
same size, the expression for the exchange energy Eq.~(\ref{JHLBz}) can be 
simplified considerably. 

For two vertically coupled dots of equal size, we set
$\alpha_{0+}=\alpha_{0-}\equiv\alpha_0$ in 
Eq.~(\ref{JHLBz}) and using Eq.~(\ref{overlaposc}), we obtain 
\begin{equation}\label{Jequal}
J = \frac{\hbar\omega_z}{{\rm sinh}(2d^2)}
\Bigg[ c \sqrt{\alpha} e^{2\alpha d^{2}}
\left(1-{\rm erf}\left(d \sqrt{2 \alpha}\right)\right)
- \frac{c}{\pi} \frac{2\alpha}{\sqrt{1-(2\alpha -1)^{2}}}
\arccos (2\alpha -1)+\frac{3}{4} \left(1+d^2\right) \Bigg],
\end{equation} 
where $\alpha=\sqrt{\alpha^2_0+B^2/B^2_0}$.  
As before, the first term in Eq.~(\ref{Jequal}) is the direct Coulomb term,
while the second term (appearing with a negative sign) is the exchange
Coulomb term. Finally, the potential term in this case equals $W=(3/4)(1+d^2)$
and is due to the vertical confinement only.
For two dots of equal size neither the prefactor $2S^2/(1-S^4)$ nor 
the potential term depends on the magnetic field.  Since the direct Coulomb
term depends on $B_\perp$ only weakly, the field dependence 
of the exchange energy is mostly determined by the exchange Coulomb term.  
The exchange coupling can also be calculated using the HM method \cite{burkard00a}.

The dependence of the exchange energy $J$ on an electric field
$E_\perp$ applied in parallel to the magnetic field, i.e. perpendicular to the
$xy$ plane, withing the HL approximation, was found to be
\begin{equation}\label{electric}
J(B,E_\perp) = J(B,0)+ 
\hbar\omega_z\frac{2S^2}{1-S^4}\frac{3}{2}\left(\frac{E_\perp}{E_0}\right)^2,
\end{equation}
where $E_0=m\omega^2_z/e a_B$.  
The growth of $J$ is thus proportional to the square of the electric 
field $E_\perp$, if the field is not too large. 
This result is supported by a HM calculation, yielding 
the same field dependence at small electric fields, whereas if $eE_\perp a$ is 
larger than $U_H$, double occupancy must be taken into account.
The electric field causes the exchange $J$ at a constant magnetic field $B$
to cross through zero from $J(E=0,B)<0$ to $J>0$.
This effect is signaled by a change in the magnetization $M$
\cite{burkard00a}.
\begin{figure}
\begin{tabular}{r r r}
\includegraphics[scale=0.45]{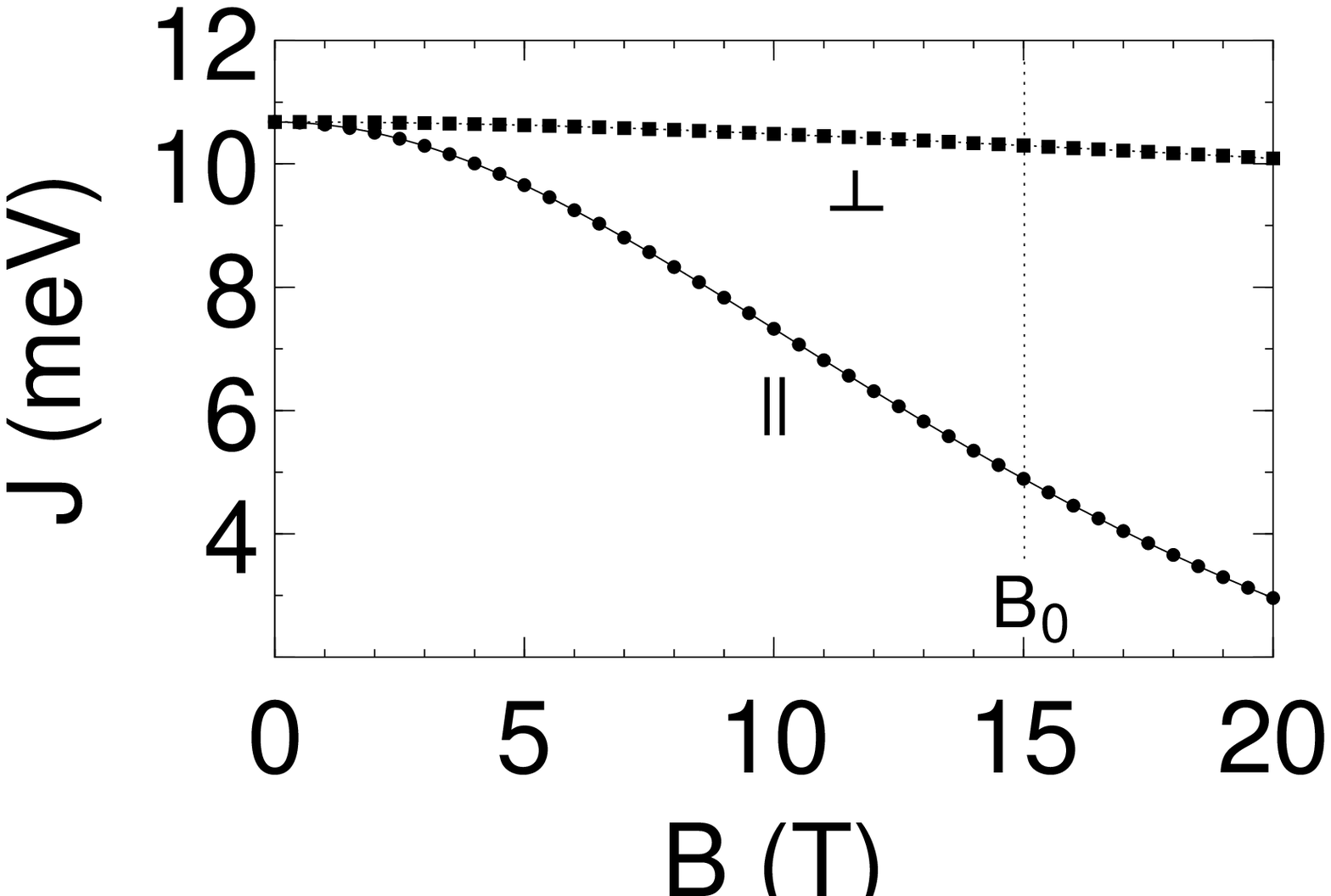} & &
\includegraphics[scale=0.45]{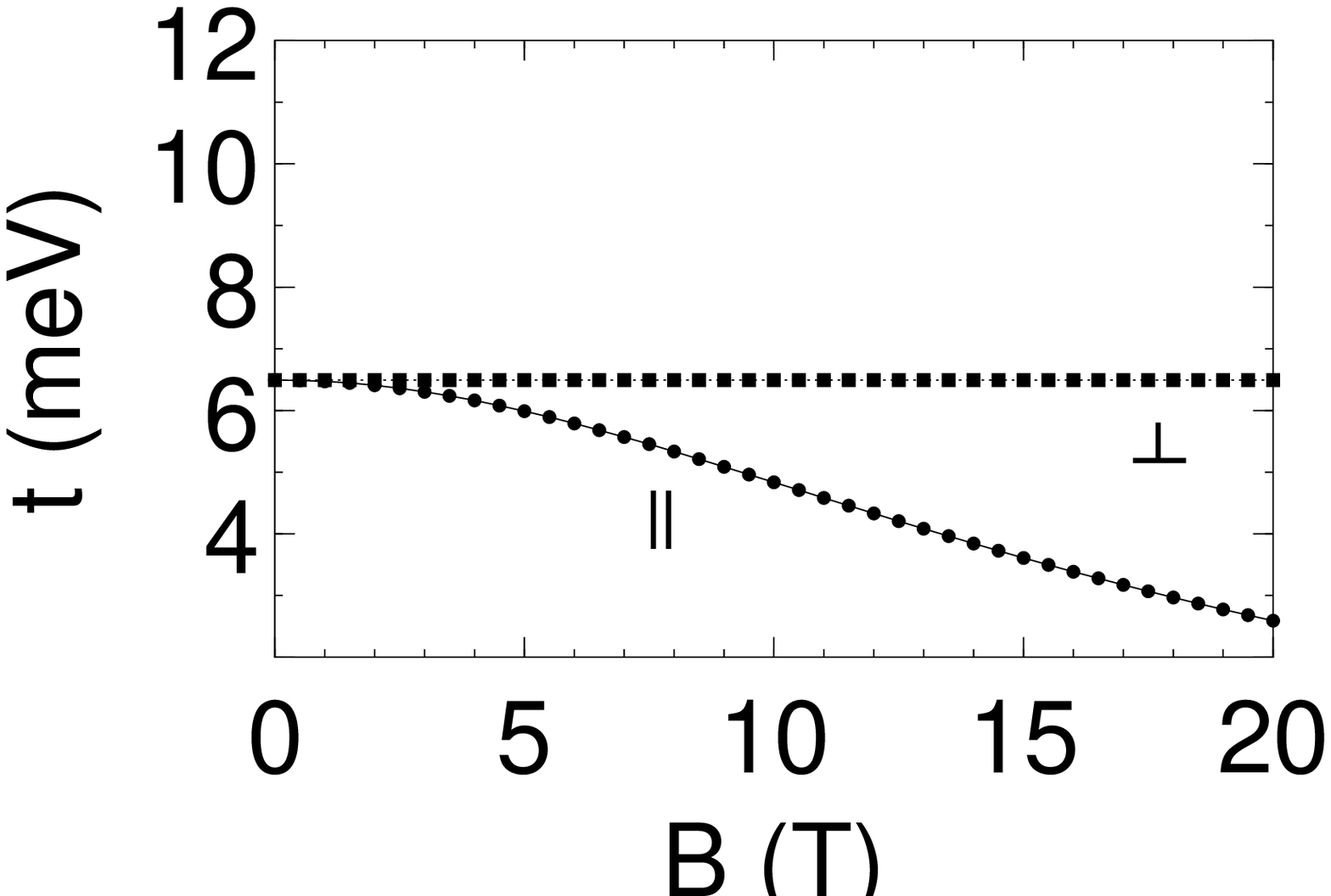}
\end{tabular}
\caption{\label{osc_fig_1b}\small
Exchange energy $J$ (left graph) and single-electron tunneling amplitude
$t$ (right graph) as a function of the applied magnetic field for
two vertically coupled small (height $6\,{\rm nm}$, width $12\,{\rm nm}$)
InAs ($m=0.08m_e$, $\kappa=14.6$)
self-assembled QDs in a center-to-center
distance of $9\,{\rm nm}$ ($d=1.5$). The box-shaped symbols correspond
to the magnetic field $B_\perp$ applied in $z$ direction, the circle
symbols to the field $B_\parallel$ in $x$ direction.
The plotted results were obtained using the HM method and
are reliable up to a field $B_0\approx 15\,{\rm T}$ where
higher levels start to become important.
}
\end{figure}

\subsubsection{In-plane magnetic field $B_\parallel$}\label{parallel}

In this section we consider two dots of equal size in a magnetic field 
$B_\parallel$ which is applied along the $x$-axis, i.e. \textit{in-plane}
(see Fig.~\ref{potential-v}). Since the two dots have the same 
size, the lateral confining potential Eq.~(\ref{lateral}) reduces to 
$V(x,y)=m\omega^2_{z}\alpha^2_0(x^2+y^2)/2$, where the parameter $\alpha_0$
describes the ratio between the lateral and the vertical confinement energy.
The vertical double-dot structure is modeled using the potential
Eq.~(\ref{VOsc}).
The situation for an in-plane field is a bit more complicated than for
a perpendicular field, because the planar and vertical motion do
not separate.
In order to find the ground-state wave function of the one-particle
Hamiltonian $h^0_{\pm a}$, an approximate variational method can be applied
\cite{burkard00a} with the result
\begin{equation}
\varphi_{\pm a} = 
\left(\frac{m \omega_z}{\pi\hbar}\right)^{\frac{3}{4}}
\left(\alpha_{0} \alpha\beta \right)^{\frac{1}{4}}
\,\exp \left[-\frac{m\omega_z}{2\hbar} \left(\alpha_{0} x^2+\alpha 
y^2+\beta(z \mp a)^2\right) \pm i\frac{ya}{2l_B^2}\right],
\label{var}
\end{equation}
where the parameters $\alpha(B)=\sqrt{\alpha^2_0+(B/B_0)^2}$ and
$\beta(B)=\sqrt{1+(B/B_0)^2}$, describing the wave-function compression in 
$y$ and $z$ direction, respectively, have been introduced.

The resulting exchange coupling $J$ in this case is
\begin{equation}\label{JformalBx} 
J(B,d)=J_0(B,d)-\hbar\omega_z\frac{4S^2}{1-S^4}
\frac{\beta -\alpha }{\alpha }d^2\left(\frac{B}{B_0}\right)^2,
\end{equation} 
where $J_0$ denotes the expression from Eq.~(\ref{Jformal}).
The variation of the exchange energy $J$ as a function of the
magnetic field $B$ is, through the prefactor $2S^2/(1-S^4)$,
determined by the overlap
$S(B,d)=\exp[-d^2(\beta (B)+(B/B_0)^2)/\alpha (B)]$,
depending exponentially on the in-plane field, while for a perpendicular
field the overlap is independent of the field (for two dots of 
equal size), see Eq.~(\ref{overlaposc}).
The HL result can again be improved by performing a
molecular-orbital (HM) calculation of the exchange energy, which
we plot in Fig.~\ref{osc_fig_1b} (left graph, circle symbols).

\subsubsection{Electrical switching of the interaction}
\label{e_switch}

\begin{figure}
\begin{minipage}{8.3cm}
\includegraphics[width=8cm]{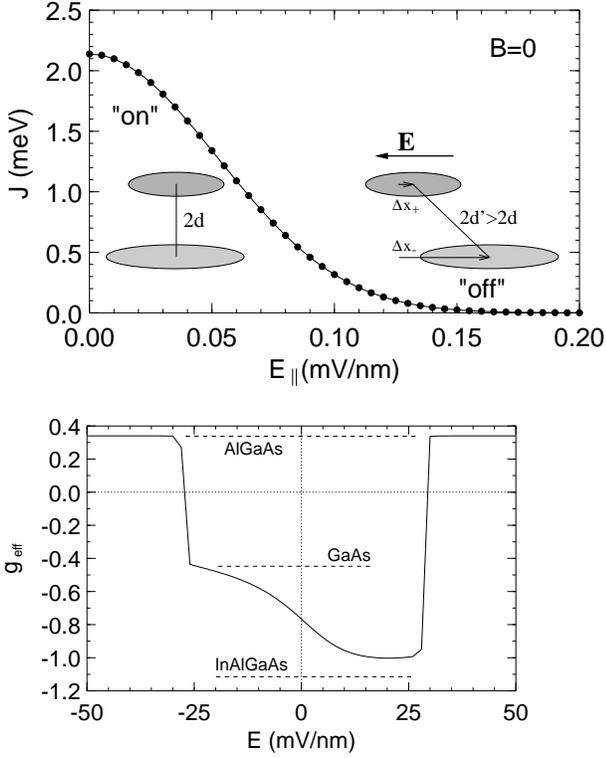}
\end{minipage}
\hfill
\begin{minipage}{9.5cm}
\caption{\label{switch}\small
Controlled switching of the exchange coupling $J$ between dots of different 
size by means of an in-plane electric field $E_\parallel$ 
at zero magnetic field, $B=0$.
The coupling is ``on'' at $E=0$. When $E_\parallel$ is applied, the larger
dot is shifted to the right by $\Delta x_-$,
whereas the smaller dot is shifted by $\Delta x_+<\Delta x_-$,
where $\Delta x_\pm =E_\parallel/E_0\alpha_{0\pm}^2$ and
$E_0=\hbar\omega_z/ea_B$.
With $E_\parallel$ increasing, $J$ decreases exponentially,
$J\approx S^2\approx \exp[-2A^2(E_\parallel/E_0)^2]$.
The parameters used for this plot are 
$\hbar\omega_z=7\,{\rm meV}$, $d=1$,
$\alpha_{0+}=1/2$ and $\alpha_{0-}=1/4$,
yielding
$E_0=\hbar\omega_z/ea_B=0.56\,{\rm mV/nm}$ and
$A=(\alpha_{0+}^2-\alpha_{0-}^2)/2\alpha_{0+}^2\alpha_{0-}^2=6$.
The coupling $J$ decreases exponentially on the scale
$E_0/2A = 0.047 \,{\rm mV/nm}$ for the electric field.}
\end{minipage}
\end{figure}

Operating a coupled quantum
dot as a quantum gate requires the ability to switch on and off
the interaction between the electron spins on neighboring dots.
A simple method of achieving a high-sensitivity
switch for vertically coupled dots involves a horizontally
applied electric field $E_\parallel$.
The idea is to use a pair of QDs with different lateral
sizes, e.g. a small dot on top of a large dot
($\alpha_{0+}>\alpha_{0-}$, see Fig.~\ref{potential-v}).
Note that only the radius in the $xy$ plane has to be different,
while it can be  assumed that the dots have the same height.
Applying an in-plane electric field $E_\parallel$ in this case
causes a shift of the single-dot orbitals by
$\Delta x_\pm = eE_\parallel/m\omega_z^2\alpha_{0\pm}^2
=E_\parallel/E_0\alpha_{0\pm}^2$, where $E_0=\hbar\omega_z/ea_B$, see
Fig.~\ref{switch}.
It is clear that the electron in the larger dot moves further in the
(reversed) direction of the electric field ($\Delta x_- >\Delta x_+$),
since its confinement potential is weaker. As a result, the mean
distance between the two electrons changes from $2d$ to $2d'$, where
\begin{equation}
  \label{d_electric}
  d' = \sqrt{d^2 + \frac{1}{4}\left(\Delta x_- - \Delta x_+\right)^2}
     = \sqrt{d^2 + A^2\left(\frac{E_\parallel}{E_0}\right) ^2},
\end{equation}
with $A=(1/\alpha_{0-}^2-1/\alpha_{0+}^2)/2$.
Using Eq.~(\ref{overlaposc}), we
find that the wavefunction overlap scales as
$S\propto \exp(-{d'}^2)\propto \exp[-A^2(E_\parallel/E_0)^2]$.
Due to this high sensitivity,
the electric field is an ideal ``switch'' for the exchange coupling
$J$ which is (asymptotically) proportional to $S^2$ and thus decreases
exponentially on the scale $E_0/2A$.
Note that if
the dots have exactly the same size, then $A=0$ and the effect vanishes.
An estimate of $J$ as a function of $E_\parallel$ can be obtained
by substituting $d'$ from Eq.~(\ref{d_electric}) into the
HL result, Eq.~(\ref{JHLBz}). A plot of $J(E_\parallel)$
obtained in this way is shown in Fig.~\ref{switch} for a
specific choice of GaAs dots. Note that this procedure is not
exact, since it neglects the tilt of the orbitals with respect to their
connecting line. 
Exponential switching is highly desirable for quantum computation,
because in the ``off'' state of the switch, fluctuations in the
external control parameter (e.g. the electric field $E_\parallel$)
or charge fluctuations
cause only exponentially small fluctuations in the coupling $J$.
If this were not the case, the fluctuations in $J$ would lead to
uncontrolled coupling between qubits and therefore to multiple-qubit
errors. Such correlated errors cannot be corrected by known
error-correction schemes, which are designed for
uncorrelated errors \cite{preskill98a}.

\subsection{Single-qubit operations}
\label{ch-gfactor}

Single-qubit operations with the Hamiltonian Eq.~(\ref{spin-Hamiltonian})
require a time-varying Zeeman coupling
$(g\mu_B {\bf S}\cdot {\bf B})(t)$ \cite{loss98,burkard99a},
which can be controlled by changing the magnetic field
${\bf B}$ or the g-factor $g$. Effective
magnetic fields/g-factors can be produced by coupling the spin via
exchange to a ferromagnet \cite{loss98} or to polarized nuclear
spins \cite{burkard99a}.
We review here how the g-factor of an electron in a semiconductor
heterostructure can be modulated by shifting its orbital between
layers of host material with different g-factors \cite{divincenzo99a,divincenzo99b}.

The spin-orbit coupling can leads to large deviations of the Land\'e g-factor
(both in the positive and negative direction)
in bulk semiconductors from the free-electron value $g_0=2.0023$.
The effective g-factors in these materials range from large negative
to large positive numbers. In confined structures such as
quantum wells, wires, and dots, the g-factor is modified
 with respect to the bulk value and sensitive
to an external bias voltage \cite{ivchenko97}. 
In the case of a layered structure, the effective
g-factor of electrons can be varied by electrically shifting their
equilibrium position from one layer (with g-factor $g_1$) to
another (with another g-factor $g_2\neq g_1$).
The bulk g-factors of the layer materials and linear interpolations between
them, have been used here as an approximation which becomes
increasingly inaccurate as the layers become thinner \cite{kiselev98}.

\begin{figure}
\begin{minipage}{7.3cm}
\includegraphics[width=7cm]{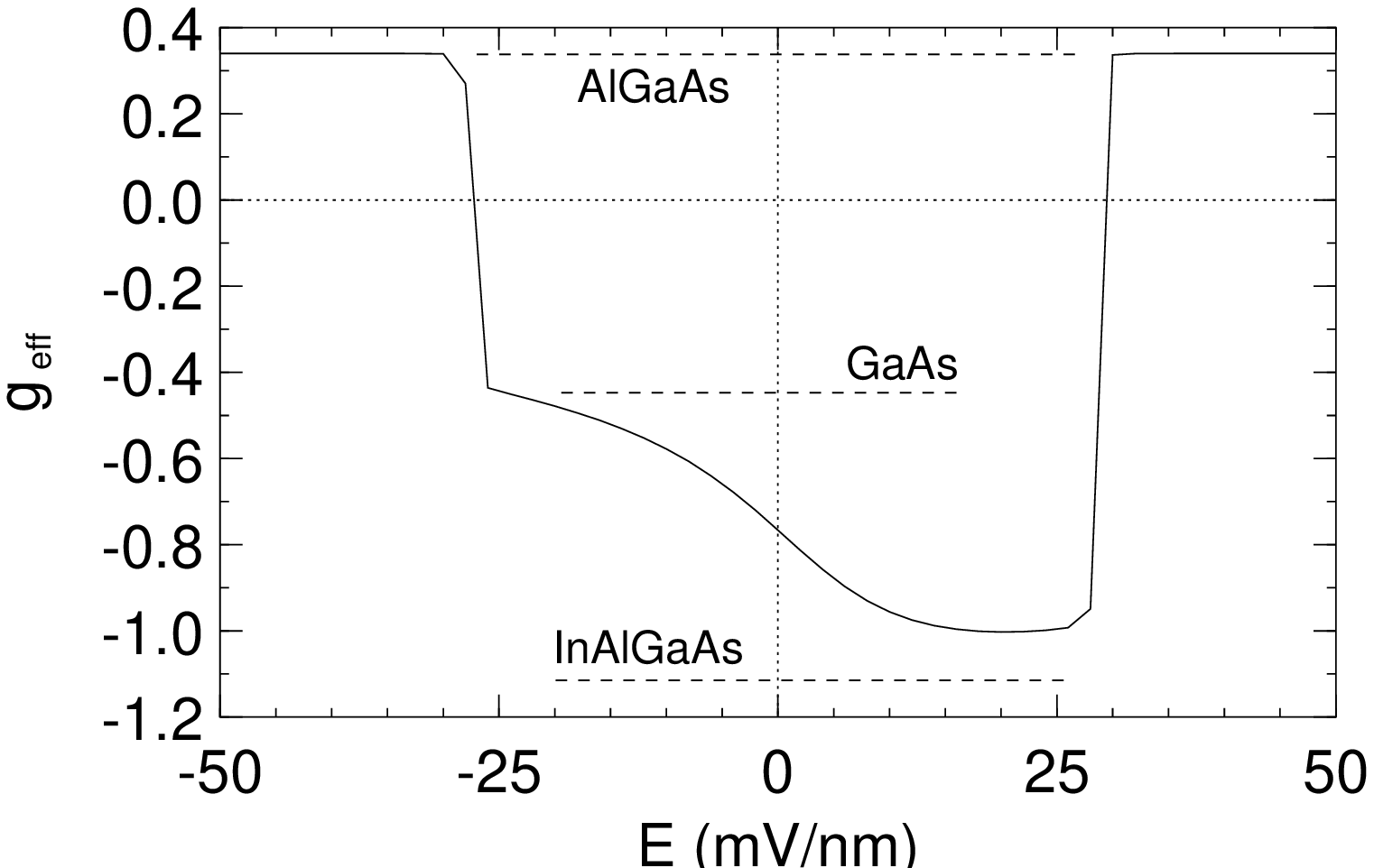}
\end{minipage}
\hfill
\begin{minipage}{8cm}
\caption{\small
Effective g-factor $g_{\rm eff}$ of electrons
confined in a
Al$_{x}$Ga$_{1-x}$As--GaAs--In$_{y}$Al$_{x}$Ga$_{1-x-y}$As--Al$_{x}$Ga$_{1-x}$As
heterostructure ($x=0.3$, $y=0.1$)
as a function of the applied electric field $E$ in growth direction. 
The widths of the quantum well and the barriers are
$w=w_{\rm B}=10\,{\rm nm}$.
The g-factors which are used for the materials are indicated
with dashed horizontal lines.}
\label{gfig}
\end{minipage}
\end{figure}

Let us assume that by replacing some fraction $y$ of Ga atoms in the upper half of a
AlGaAs-GaAs-AlGaAs quantum well by In atoms (we have used $y=0.1$)
we obtain the following layered heterostructure:
\begin{center}
Al$_{x}$Ga$_{1-x}$As--GaAs--In$_{y}$Al$_{x}$Ga$_{1-x-y}$As--Al$_{x}$Ga$_{1-x}$As,
\end{center}
where $x$ denotes the Al content in the barriers (typically around 30\%).
In such a structure, the effective g-factor can be modified
by changing the vertical position of the electrons via top or back gates.
If the electron is mostly in a pure GaAs environment, then its effective
g-factor will be around the GaAs bulk value ($g_{\rm GaAs}=-0.44$) whereas
if the electron is in the InAlGaAs region, the g-factor will be more
negative due to the large negative InAs value ($g_{\rm InAs}=-15$).
The one-dimensional problem of one electron in such a structure has been analyzed
numerically. When the effective mass $m(z)$ is spatially
varying, the Hamiltonian in the effective mass approximation can be written as
\begin{equation}
\left[ -\frac{d}{dz}\frac{\hbar^2}{2m(z)}\frac{d}{dz}+V(z)
\right]\Psi(z) = E\Psi(z).
\end{equation}
This problem can be discretized in real space and subsequently
diagonalized numerically \cite{divincenzo99b}.
Finally, the effective g-factor is
calculated by averaging the local g-factor $g(z)$ over the electronic density
in the ground-state (see Fig.~\ref{gfig}),
\begin{equation}
g_{\rm eff} = \int dz g(z) |\Psi (z)|^2.
\end{equation}
The option of performing single-qubit rotations by electrostatically
controlling the g-factor makes all-electric control
of a spin-based quantum computer (an array of QDs as in Fig.~\ref{qd-array})
possible and thus offers a way around the problematic local magnetic field
implementation  of single-qubit gates.  Another method to circumvent
single-spin operations completely (however, at a higher cost of gates
and exchange operations) is the exchange-only architecture
outlined in Sec.~\ref{ch-exchange}.

\subsection{Semiconductor microcavities}
\label{qed}

Here, we present a modification of the Loss-DiVincenzo 
scheme (Sec.~\ref{LD}) for QC based on QD electron spins.
In contrast to the original scheme \cite{loss98},
where the spins are coupled via direct exchange, this
coupling which is mediated through a single microcavity mode
and uses laser fields to mediate coherent interactions between
distant QD spins  \cite{imamoglu99}.

\begin{figure}
\begin{minipage}{6.3cm}
\includegraphics[width=6cm]{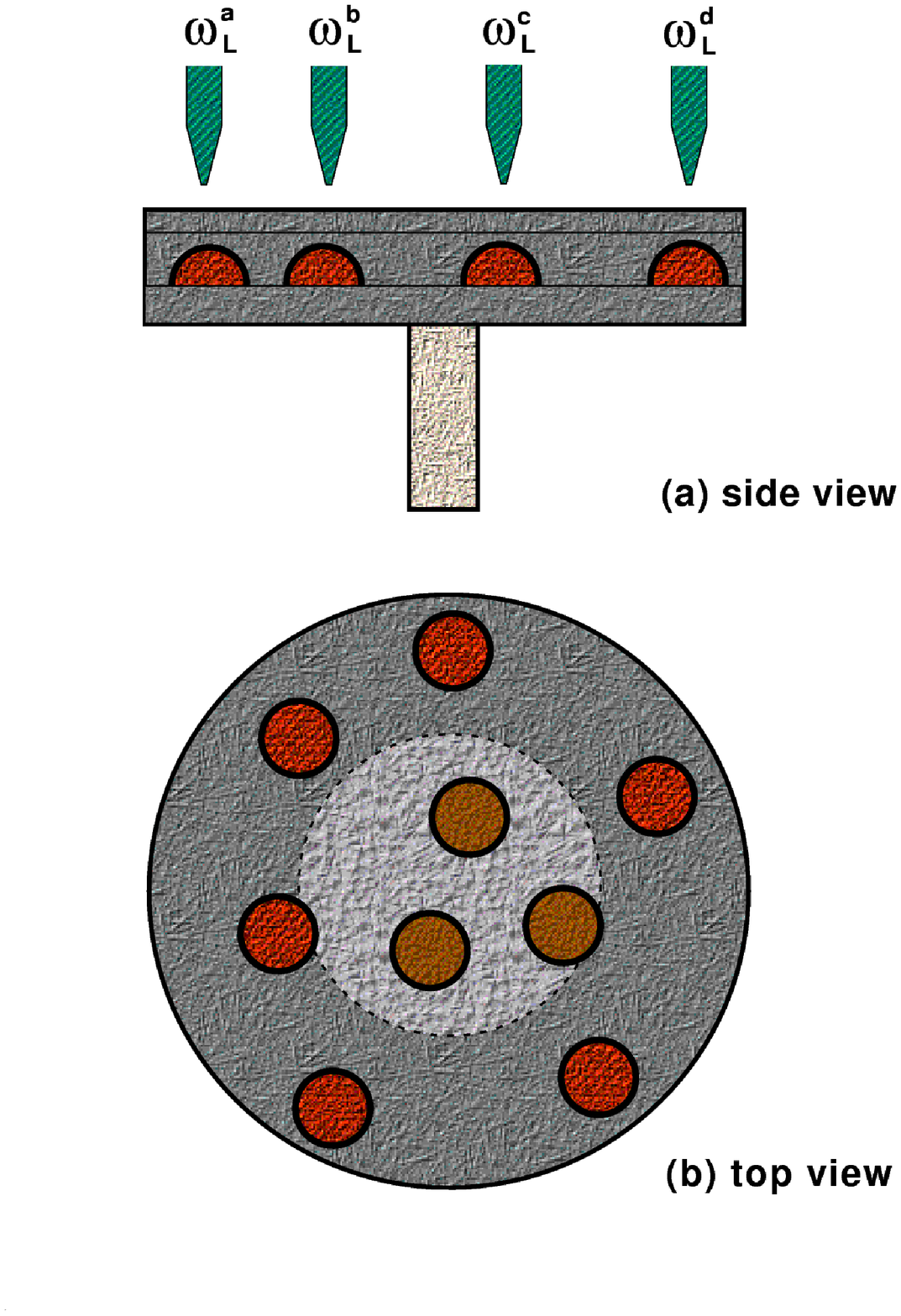}
\end{minipage}
\begin{minipage}{6.3cm}
\includegraphics[width=6cm]{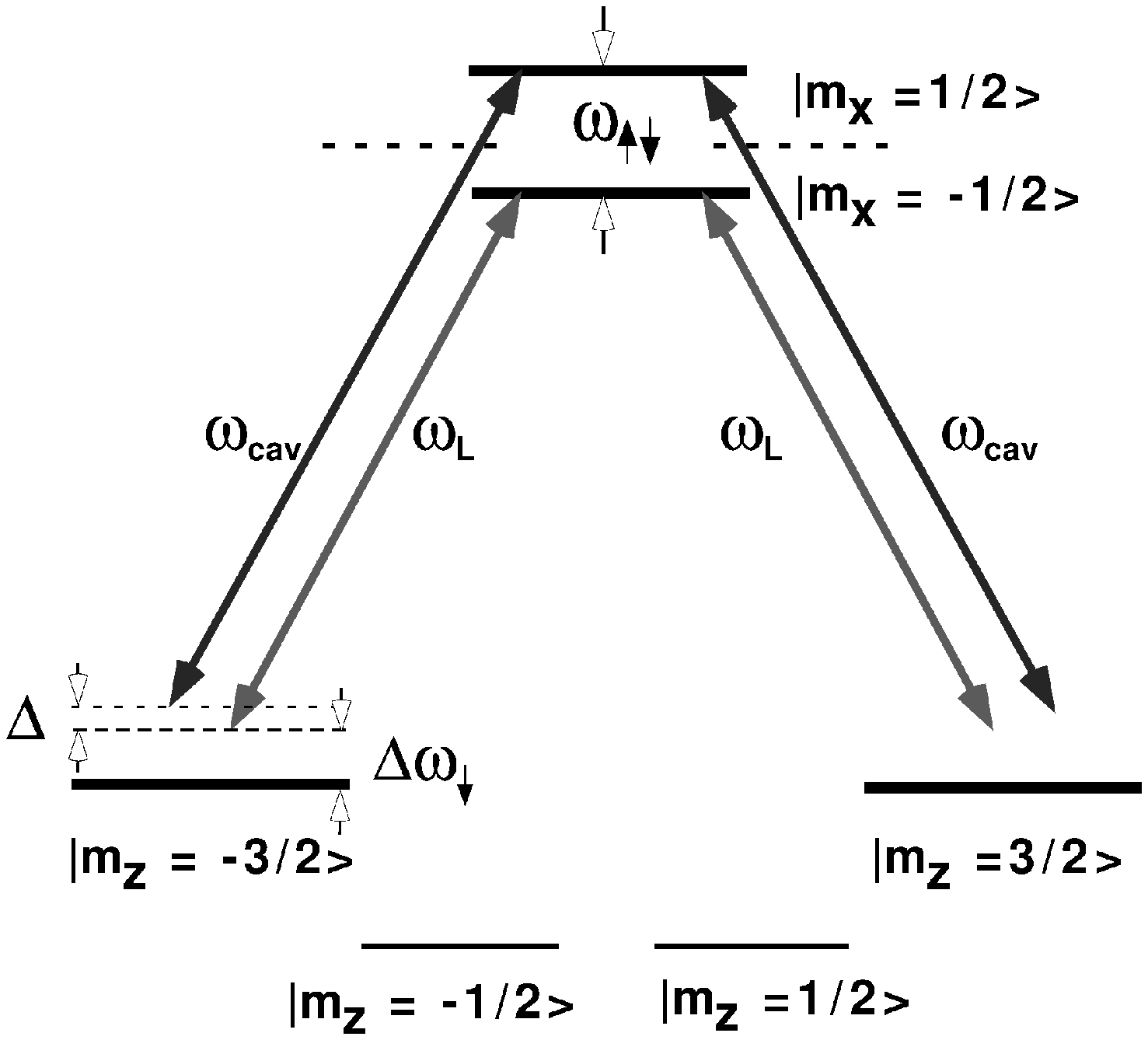}
\end{minipage}
\hfill
\begin{minipage}{4cm}
\caption{\label{cavity-fig1}\small
Left: QDs embedded inside a microdisk structure, from \cite{imamoglu99}. 
Each QD is addressed selectively by a laser field from a fiber-tip. The laser
frequencies are chosen to select out the pair of QDs that will participate
in gate operation. All dots strongly couple to a single cavity-mode.
Right: Energy levels of a III-V (or II-VI) semiconductor QD.
It is assumed that confinement along the z-direction is strongest.
}
\end{minipage}
\end{figure}

The cavity scheme is shown in Fig.~\ref{cavity-fig1}: the doped QDs are
embedded in a microdisk structure with diameter $d \simeq 2 \,\mu{\rm m}$ and
thickness $d \simeq 0.1 \,\mu{\rm m}$.
Experiments have shown that InAs self-assembled QDs can be
embedded in microdisk structures with a cavity quality factor $Q \simeq 12000$
\cite{gerard99}.  It is assumed that the QDs are designed such that the quantum
confinement along the z-direction is the strongest. The
in-plane confinement is also assumed to be large enough to guarantee
that the electron will always be in the ground-state orbital.  Because
of the strong z-axis confinement, the lowest energy eigenstates of
such a III-V or II-VI semiconductor QD consist of $|m_z=\pm1/2\rangle$
conduction-band states and $|m_z=\pm3/2\rangle$ valence-band
states. The QDs are doped such that each QD has a full valence band
and a single conduction band electron: we assume that a uniform
magnetic field along the x-direction ($B_x$) is applied, so the QD
qubit is defined by the conduction-band states $|m_x=-1/2\rangle =
|\!\!\downarrow\rangle$ and $|m_x=1/2\rangle = |\!\!\uparrow\rangle$
(Fig.~\ref{cavity-fig1}, right).

\subsubsection{Single-qubit operations}
Single-bit operations are carried out in this scheme by applying two laser
fields $E_{L,x}(t)$ and $E_{L,y}(t)$ with Rabi frequencies 
$\Omega_{L,x}$ and $\Omega_{L,y}$, and frequencies $\omega_{L,x}$
and $\omega_{L,y}$ (polarized along the x and y directions,
respectively) that exactly satisfy the Raman-resonance condition
between $|\!\!\downarrow\rangle$ and $|\!\!\uparrow\rangle$.  The
laser fields are turned on for a short time duration that satisfies a
$\pi/r$-pulse condition, where $r$ is any real number. The process can
be best understood as a Raman $\pi/r$-pulse for the {\sl hole} in the
conduction band state. The laser field polarizations
should have non-parallel components in order to create a non-zero
Raman coupling (if there is no heavy-hole light-hole mixing). These
arbitrary single-bit rotations can naturally be carried out in parallel. 
In addition, the QDs that are not doped by a single electron never couple 
to the Raman fields and can safely be ignored.

\subsubsection{Two-qubit operations}
Two-qubit operations are mediated by virtual photons that are emitted to and reabsorbed
from the microcavity field. It is assumed that the x-polarized
cavity-mode with energy $\omega_{\rm cav}$ ($\hbar=1$) and a laser field
(assumed to be y-polarized) establish the Raman transition between the two
conduction-band states, in close analogy with the atomic cavity-QED schemes
\cite{pellizzari95}.
The Hamiltonian for a single QD is written as $H=H_0+H_{\rm int}$, with
\begin{equation}
  H_0 = \sum_{\sigma=\uparrow,\downarrow,\pm 3/2}
         \omega_\sigma e_\sigma^\dagger e_\sigma
      + \omega_{\rm cav} a_{\rm cav}^\dagger a_{\rm cav}
      + \omega_L a_L^\dagger a_L,
\label{cavity-h0}
\end{equation}
where $e_\uparrow$, $e_\downarrow$ annihilate an electron with spin $\uparrow$,
$\downarrow$ along the $x$ direction in the conduction band and $e_{\pm 3/2}$
annihilates an electron with spin $\pm 3/2$ along the $z$ direction
in the valence band, cf.\ Fig.~\ref{cavity-fig1} (right).
The interaction can be written as
\begin{equation}
  H_{\rm int}= g \left(a_+^\dagger e_{-3/2}^\dagger e_{-1/2} 
                 - a_-^\dagger e_{3/2}^\dagger e_{1/2} + h.c.\right),
  \label{cavity-hint}
\end{equation}
where the operators for the circularly polarized light are expressed
in terms of the x-polarized cavity mode and the y-polarized laser field,
$a_\pm = (a_{\rm cav}\pm i a_L)/\sqrt{2}$,
and the conduction-band operators in the $z$ basis can be expressed
in terms of those in the $x$ basis,
$e_{\pm 1/2} = (e_\downarrow \pm e_\uparrow)/\sqrt{2}$.
With $\omega_{-3/2}=\omega_{3/2}\equiv\omega_v$ 
and the definition $e_v=(e_{-3/2}-e_{3/2})/2$,
the following result for the cavity Hamiltonian is obtained,
\begin{equation}
  H_{\rm int} = g \left(a_{\rm cav}^\dagger e_v^\dagger e_\uparrow 
              + e_\uparrow^\dagger e_v a_{\rm cav}\right)
              - i g \left(a_L^\dagger e_v^\dagger e_\downarrow
              - e_\downarrow^\dagger e_v a_L \right).
\label{cavity-hamiltonian}
\end{equation}
The valence band states are eliminated by a Schrieffer-Wolff
transformation \cite{madelung78,schrieffer66}, $H_{\rm eff} = e^{-S} H e^S$, with
\begin{equation}
  S = \frac{g}{\Delta\omega_{\uparrow}}\left(a_{\rm cav}^\dagger e_v^\dagger e_\uparrow 
   - e_\uparrow^\dagger e_v a_{\rm cav}\right)
  - i\frac{g}{\Delta \omega_{\downarrow}}\left(a_L^\dagger e_v^\dagger e_\downarrow
   + e_\downarrow^\dagger e_v a_L \right),
\end{equation}
where $\Delta\omega_{\uparrow} =\omega_{\uparrow} - \omega_v - \omega_{\rm cav}$ 
and $\Delta \omega_{\downarrow}  = \omega_{\downarrow} - \omega_v - \omega_L$.
Neglecting all terms $O(g^3)$ and replacing
$e_v^\dagger e_v$ by its expectation value $\langle e_v^\dagger e_v\rangle =1$
and $g a_L$ by $\Omega_L \exp(-i\omega_L t)$
one obtains the effective Hamiltonian
\begin{eqnarray}
H_{\rm eff} & = & \omega_{\rm cav} a_{\rm cav}^{\dagger}
a_{\rm cav} + \sum_i \left[\omega_{\uparrow \downarrow}^i
\sigma_{\uparrow \uparrow}^i
- \frac{g_{\rm cav}^2}{ \Delta \omega_{\uparrow}^i}
 \, \sigma_{\downarrow \downarrow}^i \, a_{\rm cav}^{\dagger}
a_{\rm cav} \right.
 -   \left. \frac{(\Omega_{L,y}^i)^2}{\Delta \omega_{\downarrow}^i}
\sigma_{\uparrow \uparrow}^i \, + \, i  g_{\rm eff}^i \left[
a_{\rm cav}^{\dagger}
\sigma_{\downarrow \uparrow}^i e^{-i \omega_{L,y}^i t} - h.c.\right]
\right], \label{cavity-eq1} \\
g_{\rm eff}^i(t) & = & \frac{g_{\rm cav}
\Omega_{L,y}^i(t)}{2} \left(\frac{1}{\Delta \omega_{\uparrow}^i}  +
\frac{1}{\Delta \omega_{\downarrow}^i}\right),
\label{cavity-eq2}
\end{eqnarray}
where the sum runs over all QDs of the system, 
$g_{\rm eff}^i$ is the effective 2-photon coupling coefficient,
$\sigma_{\uparrow \downarrow}^i = |\!\!\uparrow\rangle \langle \downarrow \!\!|$ 
the spin projection operator for the $i$-th QD, 
and $\omega_{\uparrow \downarrow}^i = \omega_{\uparrow}^i -\omega_{\downarrow}^i$. 
The exact two-photon-resonance condition would be $\Delta
\omega_{\uparrow}^i =\omega_{\uparrow}^i - \omega_v^i - \omega_{\rm cav} = \Delta
\omega_{\downarrow}^i  = \omega_{\downarrow}^i - \omega_v^i - \omega_L^i$.
The derivation of $H_{\rm eff}$ assumes
$\Delta\omega_{\uparrow,\downarrow}^i \gg
g_{\rm cav}$, $\omega_{\uparrow \downarrow}^i \gg k_B T$,
and $\omega_{\uparrow\downarrow}^{i,j} \gg g_{\rm eff}^i > \Gamma_{\rm cav}$,
where $\Gamma_{\rm cav}$ denotes the cavity decay rate (not included in
Eq.~(\ref{cavity-eq1})). The third and fourth terms of
Eq.~(\ref{cavity-eq1}) describe the ac-Stark-effect caused by the cavity and
laser fields, respectively.

In order to implement a CNOT quantum gate, one would
turn on laser fields $\omega_L^i$ and $\omega_L^j$ to establish near
two-photon resonance condition for both the control (i) and the target
(j) qubits,
\begin{eqnarray}
\Delta_i & = & \omega_{\uparrow \downarrow}^i -
\omega_{cav} + \omega_L^i \; =  \; \Delta_j  \ll
\omega_{\uparrow \downarrow}^{i,j} \;\;\;.
\label{cavity-eq3}
\end{eqnarray}
If the two-photon detunings $\Delta_i$ are chosen large compared to the
cavity linewidth and $g_{\rm eff}^i(t)$,
the cavity modes can be eliminated 
with a second Schrieffer-Wolff transformation
to obtain an effective two-qubit interaction Hamiltonian in the rotating
frame (interaction picture with $H_0 = \sum_i
\omega^i_{\uparrow\downarrow}\sigma^i_{\uparrow\uparrow}$),
\begin{eqnarray}
H_{\rm int}^{(2)} & = & \sum_{i \neq j} \tilde{g}_{ij}(t)
\left[ \sigma_{\uparrow
\downarrow}^i \sigma_{\downarrow \uparrow}^j e^{i\Delta_{ij}t} \, + \,
\sigma_{\uparrow
\downarrow}^j \sigma_{\downarrow \uparrow}^i e^{-i\Delta_{ij}t} \right] \;\;\;,
\label{cavity-eq4}
\end{eqnarray}
where $\tilde{g}_{ij}(t) = g_{\rm eff}^i(t) g_{\rm eff}^j(t)/ \Delta_i$ and $\Delta_{ij} =
\Delta_i - \Delta_j$.
The implementation of the conditional phase-flip (CPF) and the CNOT or quantum XOR
gates between two spins $i$ and $j$ from a transversal (XY) spin coupling
of the form Eq.~(\ref{cavity-eq4}) has been discussed in Sec.~\ref{QC-anisotropic}.

The interaction Hamiltonian $H_{\rm int}^{(2)}$ describes the coupling of the QD
spins via the following virtual process.  One of the QDs emits a virtual photon
into the cavity while absorbing a laser photon.  The cavity photon is then
reabsorbed by the other QD while a laser photon is emitted.
Due to the spin splitting in the QD spectrum, Fig.~\ref{cavity-fig1} (right),
this process is spin
sensitive and leads to the spin-spin coupling $H_{\rm int}^{(2)}$ between the QDs.

\subsubsection{Measurement}
In the cavity QED scheme, measurement of a single QD spin can be achieved by
applying a laser field $E_{L,y}$ to the QD to be measured, in order to realize
exact two-photon resonance with the cavity mode. If the QD spin is in state
$|\!\!\downarrow\rangle$, there is no Raman coupling and no photons will be
detected. If on the other hand, the spin state is $|\!\!\uparrow\rangle$, the
electron will exchange energy with the cavity mode and eventually a single
photon will be emitted from the cavity. A single photon
detection capability is thus sufficient for detecting a single spin. 

\subsubsection{Related proposals}
A related proposal is to use optically controlled virtual excitations
of delocalized excitons as mediators of RKKY type spin interaction between 
electrons localized in neighboring QDs \cite{piermarocchi02b}.  
The spin interaction in this case is isotropic, $H = J{\bf S}_i\cdot{\bf S}_j$.

\subsection{Decoherence}
The spin coherence time in semiconductors---the time over 
which the phase of a superposition of spin-up and spin-down states, 
Eq.~(\ref{qubitstate}), is well-defined---can be much longer than the charge
coherence time (a few nanoseconds).   In fact it is known from
experiment that they can be orders of magnitude longer.  
This is of course one of the reasons for using spin as a
qubit \cite{loss98} rather than charge.  In bulk GaAs and in
CdSe quantum dots, the ensemble spin coherence time $T_2^{\star}$,
being a lower bound on the single-spin decoherence time $T_2$,
was measured using a technique called time-resolved Faraday 
rotation \cite{kikkawa97,gupta99}.  For a detailed account
of these experiments, we refer the reader to Chaps.~4 and 5
of \cite{awschalom02}. 
The spin relaxation time $T_1$ in a single-electron QD in a GaAs 
heterostructure was probed via transport measurements and
found to approach one microsecond \cite{hanson03,hanson04}.
It has been proposed to also measure the single-spin $T_2$ 
in such a structure in a transport experiment by
applying electron spin resonance (ESR) techniques \cite{engel01}. 
In this scheme, the stationary current exhibits a resonance whose 
line width is determined by the single-spin decoherence time $T_2$.

Below, a number of decoherence mechanisms for spin in semiconductor
nanostructures will be listed.  It Should be emphasized, though,
that it is usually hard for \textit{theory} to predict which
mechanism is dominant.  Nevertheless, the understanding of the 
underlying mechanisms for a list of possible causes can be a
very valuable tool for the purpose of achieving long coherent
operation in a future quantum device.

\subsubsection{Phonons and the spin-orbit coupling}

Phonon-assisted transitions between different discrete energy levels (or Zeeman
sublevels) in GaAs quantum dots can cause spin flips
and therefore spin decoherence
\cite{khaetskii00,khaetskii01a,khaetskii01b}.   
There are various mechanisms originating from the spin-orbit coupling which lead to such
spin flip processes;  the most effective mechanisms in 2D have to do with the broken 
inversion symmetry, either in the elementary crystal cell or at the heterointerface.
The spin-orbit Hamiltonian for the electron in such a structure is given by Eq.~(\ref{s-o}).
The relaxation rates $\Gamma = T_{1}^{-1}$ are evaluated in 
leading perturbation order in this coupling, with and without a
magnetic field. 
The spin-orbit coupling $H_{\rm so}$  mixes the spin-up and
spin-down states of the electron and leads to a non-vanishing matrix element of
the phonon-assisted transition between two states with opposite
spins.  However, one of the main findings of \cite{khaetskii00,khaetskii01a,khaetskii01b}
is that the spin relaxation of the electrons localized in
the dots differs strongly from that of delocalized electrons. 
It turns out that in quantum dots (in contrast to extended 2D states), 
the contributions to the spin-flip rate
  proportional to $\beta^2$ are  absent in general.
This greatly reduces the spin-flip rates of electrons confined to dots.
The finite Zeeman splitting in the
energy spectrum also leads to contributions $\propto \beta^2$,
\begin{equation}
\Gamma \simeq \Gamma_0(B) \left (\frac{m\beta^2}{\hbar \omega_0}\right)
\left(\frac{g\mu_B B}{\hbar \omega_0}\right)^2,
\label{7.0}
\end{equation}
where $\hbar \omega_0$ is the orbital energy level splitting in the 
QD and $\Gamma_0(B)$ is the inelastic rate without spin flip for the
transition between neighboring orbital levels. 

Spin-flip transitions between Zeeman sublevels occur with a rate
that is proportional to the fifth power of the  Zeeman splitting,
  \begin{equation}
\Gamma_z \simeq \frac{(g\mu_B B)^5}{\hbar(\hbar \omega_0)^4}\Lambda_p.
\label{6}
\end{equation}
The dimensionless constant $\Lambda_p\propto \beta^2$
characterizes the strength of the effective spin--piezo-phonon
coupling in the heterostructure and ranges from $\approx 7\cdot
10^{-3}$ to $\approx 6\cdot 10^{-2}$ depending on $\beta$. To give
a number, $\Gamma _{z} \approx 1.5\cdot 10^3 \,{\rm s}^{-1}$ for
$\hbar \omega_0 = 10 \,{\rm K}$ and at a magnetic field $B=1\,{\rm T}$.

It can be shown that under realistic conditions, a general symmetry
argument leads to the conclusion that the spin
decoherence time $T_2$ does not have a transverse contribution (in leading order),
in other words, $T_2=2T_1$ for spin-orbit (phonon) related processes
\cite{golovach04}.

\subsubsection{Nuclear spins}

The nuclear spins of the host material can cause decoherence
via spin flips that are caused by the hyperfine interaction.
A rough perturbative estimate of this effect \cite{burkard99a}
suggests that the rate of such processes
can be suppressed by either polarizing the nuclear spins or by 
applying an external magnetic field.  The suppression factor is
$( B_{\rm n}^{*} / B )^2/N$, where $B_{\rm n}^{*}= AI/g\mu_{\rm B}$ 
is the maximal magnitude of the
effective nuclear field (Overhauser field), $N$ the number of nuclear
spins in the vicinity of the electron, and $A$ the hyperfine coupling
constant.  In GaAs, the nuclear spin of both Ga and As is $I=3/2$.
The field $B$ denotes either the external field,
or, in the absence of an external field, the Overhauser field
$B=p B_{\rm n}^{*}$ due to the nuclear spin polarization $p$, which
can be obtained e.g.\ by optical pumping \cite{dobers88} or
by spin-polarized currents at the edge of a 2DEG \cite{dixon97}.
In the latter case, the suppression of the spin flip rate becomes
$1/p^2N$.

A more detailed analysis treats a single electron
confined to an isolated QD under the influence of the hyperfine
interaction with the surrounding nuclei \cite{khaetskii02}.
It turns out that the electron spin decoherence time $T_2$ is shorter than
the nuclear spin relaxation time $T_{n2}$ determined by the 
dipole-dipole interaction between nuclei, and
therefore the problem can be considered in the absence of the
nuclear dipole-dipole interaction.
Since the hyperfine interaction depends on the position via 
a factor $|\psi({\bf r})|^2$ where $\psi({\bf r})$ is the
electron wavefunction, the value of the hyperfine interaction
varies spatially.  It turns out that this is the relevant cause of decoherence.
The analysis is complicated by the fact that in a weak external
Zeeman field (smaller than a typical fluctuating Overhauser field seen 
by the electron, $\sim 100$ Gauss in a GaAs QD), the
perturbative treatment of the electron spin decoherence breaks
down and the decay of the spin precession amplitude
is not exponential in time, but either described by a power law, 
$1/t^{d/2}$ (for finite Zeeman fields) or an inverse logarithm, 
$1/ (\ln t)^{d/2}$ (for vanishing fields). 

The decay rate $1/T_2$ is thus roughly given by $A/\hbar N$, where $A$ is the
hyperfine interaction constant, and $N$ is the number of nuclei
within the dot, with $N$ typically $10^5$. This time is of the
order of several $\mu {\rm s}$.   However, it needs to be stressed
that there is no simple exponential decay which, strictly speaking,
means that decoherence cannot simply be characterized by the decay
times $T_1$ and $T_2$ in this case.  
The case of a fully polarized nuclear spin state was solved 
exactly in \cite{khaetskii02}.  The amplitude of the precession
which is approached after the decay, is of
order one, while the decaying part is $1/N$, in agreement with
earlier results \cite{burkard99a}, see above.  A large difference 
between the values of $T_2$ (decoherence time for a single dot) and 
$T_2^{\star}$ (dephasing time for an ensemble of dots), 
i.e. $T_2^{\star} \ll T_2$ is found and indicates 
that it is desirable to have direct
experimental access to single spin decoherence times.

The non-Markovian dynamics of a localized electron spin interacting with an environment of nuclear spins with arbitrary polarization $p$ was calculated in \cite{coish04} from a perturbative analysis of the generalized master equation for the longitudinal and transverse components of the electron spin.



\section{Superconducting micro-circuits}
\label{sc}

\subsection{Overview}
\begin{figure}[b]
\begin{minipage}{8.3cm}
\includegraphics[width=8cm]{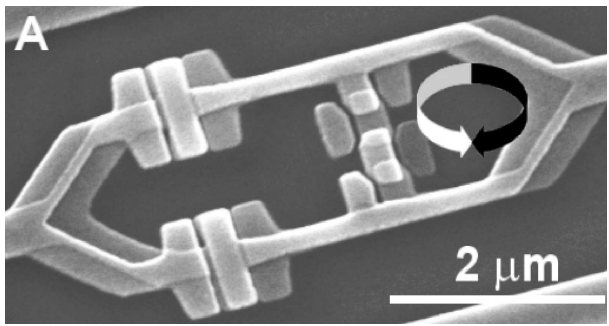}
\end{minipage}
\hfill
\begin{minipage}{8cm}
\caption{Electron micrograph of the SC flux qubit circuit studied
in \cite{chiorescu03}.  
The logical qubit basis states correspond to circulating
SC currents in the smaller loop as indicated.
The bright areas are the Al wires;
the double-layer structure from the shadow evaporation
deposition is clearly visible.
(Figure courtesy of I. Chiorescu and J. E. Mooij, TU Delft).
\label{fig-delft-qubit}}
\end{minipage}
\end{figure}

Roughly speaking, three prototypes of superconducting (SC) qubits are 
studied experimentally.
We only briefly review them here, and refer the reader
to \cite{makhlin01} for a comprehensive review.
The charge qubit 
\cite{shnirman97,averin98,makhlin99,nakamura99,vion02,pashkin03},
operating in the regime $E_C\gg E_J$,
and the flux qubit \cite{mooij99,orlando99,vanderwal00,chiorescu03},
operating in the regime $E_J\gg E_C$,
are distinguished by their Josephson junctions' relative 
magnitude of charging energy $E_C$ and Josephson energy $E_J$.
A third type, the phase qubit \cite{ioffe99,martinis02},
operates in the same regime as the flux qubit, but
is represented purely in the SC phase and is not associated with 
any magnetic flux or circulating current.  
The Josephson phase qubit consists of a single
Josephson junction \cite{martinis02}.
In flux qubits, the quantum state
of the SC phase differences across the Josephson
junctions in the circuit contain the quantum information, 
i.e., the state of the qubit.  A micrograph of the circuit
for a SC flux qubit studied in \cite{chiorescu03} is shown in 
Fig.~\ref{fig-delft-qubit}. In charge qubits, the quantum
state of the charge on SC islands contains the
quantum information.

Both charge and flux qubits have been described by an approximate
pseudo-spin Hamiltonian of the type \cite{makhlin01},
\begin{equation}
  \label{sc-pseudospin}
  H = \frac{\Delta}{2}\sigma_x + \frac{\epsilon}{2}\sigma_z,
\end{equation}
where $\Delta$ denotes the tunnel coupling between the two
qubit states $|0\rangle$ and $|1\rangle$ 
(eigenstates of $\sigma_z$) and $\epsilon$ the bias (asymmetry).
In Sec.~\ref{sec-circuit}, a more general model, including the 
full Hilbert space of a SC circuit, will be discussed.

\subsection{Decoherence, visibility, and leakage}

\subsubsection{Decoherence}
\begin{figure}
\begin{minipage}{10.3cm}
\includegraphics[width=10cm]{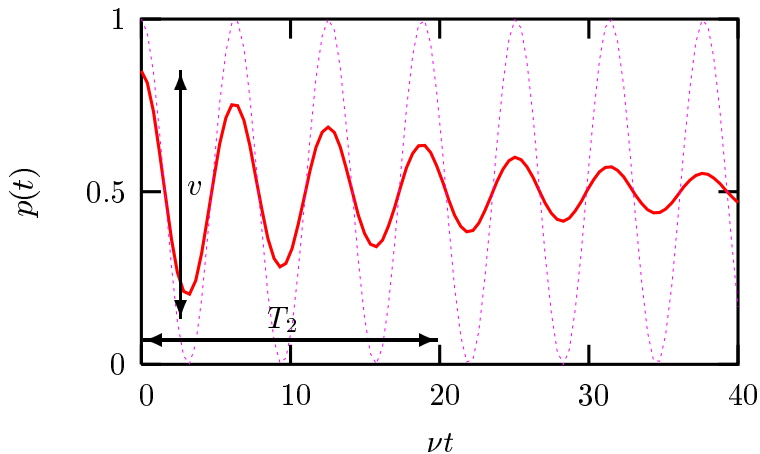}
\end{minipage}
\hfill
\begin{minipage}{7cm}
\caption{Theoretical Larmor precession (Ramsey fringe) curve with 
decoherence time $T_2<\infty$ and limited visibility $v<1$ (solid thick line),
compared to the ideal curve (dotted thin line).
The probability $p(t)$ to find the qubit in state $|1\rangle$
is plotted as a function of the free evolution time $t$.
The Larmor frequency of the coherent oscillations is denoted
with $\nu$.  The \textit{visibility} $v$ is the maximum range
of $p(0)$ whereas the \textit{decoherence time} $T_2$ is the time over
which the oscillations are damped out (in the case of an exponential decay).  
For this plot, we have chosen $T_2=20/\nu$ and $v=70\,\%$.
\label{fig-decoherence}}
\end{minipage}
\end{figure}

Decoherence within the model Eq.~(\ref{sc-pseudospin}) can be
understood phenomenologically as follows.  
If the qubit is initially prepared in state $|1\rangle$, 
then it will undergo free Larmor oscillations with frequency
$\nu = h^{-1}\sqrt{\Delta^2+\epsilon^2}$.  Ideally, the probability for
finding the qubit in state $|1\rangle$ after time $t$ would be
a perfect cosine function of $t$.  This ideal Larmor precession
is shown as a thin dotted line in Fig.~\ref{fig-decoherence}.
Such a Larmor precession experiment (also known as Ramsey fringe 
experiment) determines
how well the qubit satisfies item \ref{ss-decoherence}
of DiVincenzo's five criteria.
Decoherence is a process in which the amplitude of the 
oscillations decays over time, as shown by the thick solid
line in Fig.~\ref{fig-decoherence}.  This decay is often (but not
always) exponential with a characteristic decoherence time $T_2$.

All types of SC qubits suffer from decoherence that is caused
by a several sources.  Decoherence in charge qubits has been investigated
using the spin-boson model in \cite{makhlin01,makhlin04}.
In flux qubits, the Johnson-Nyquist noise from lossy
circuit elements (e.g., current sources) has been identified as one important
cause of decoherence \cite{tian99,tian02,vanderwal03,wilhelm03}.  
A systematic theory of decoherence of a qubit from such
dissipative elements, based on the network graph analysis \cite{devoret97} 
of the underlying SC circuit, was developed for SC flux qubits \cite{burkard04a},
and applied to study the effect of asymmetries in a persistent-current
qubit \cite{burkard04b}.  The circuit theory for SC qubits will be discussed
further below in Sec.~\ref{sec-circuit}. 
For the Josephson phase qubit \cite{martinis02},
decoherence due to bias noise and junction resonators 
was studied in \cite{martinis03,simmonds04}.

\subsubsection{Visibility}
A different type of imperfection that typically affects SC qubits
is a limited visibility $v$.  This effect means that the maximum range $v$
of the read-out probability of the qubit being in state $|1\rangle$ 
is smaller than one.
This means, e.g., that the probability $p(0)$ of measuring the qubit
in state $|1\rangle$ right after preparation in this state is
less than one.  In the case of a symmetric reduction of the visibility,
the relation is $p(0)=(1+v)/2$.
This effect is schematically shown in Fig.~\ref{fig-decoherence}.

\subsubsection{Leakage}
Limitations of the visibility are often attributed to
a mechanism called leakage.
Since the SC phase is a continuous variable as, e.g.,
the position of a particle, superconducting qubits (two-level systems)
have to be obtained by truncation of an infinite-dimensional Hilbert 
space.  This truncation is only approximate for various reasons;
(i) because it may not be possible to prepare the initial state 
with perfect fidelity in the lowest two states, (ii) because
of erroneous transitions to higher levels (leakage effects) 
due to imperfect gate operations on the system, and (iii) because 
of erroneous transitions to higher levels due to the unavoidable
interaction of the system with the environment.
Apparent leakage effects may occur if the read-out process is
not $100\%$ accurate.
Leakage effects due to the non-adiabaticity of externally applied
fields were studied in \cite{fazio99}.  Recent work 
\cite{meier04} shows that leakage in microwave-driven Josephson 
phase qubits leading to a reduced visibility can occur, even if the
microwave source is pulsed slowly.

\subsection{Circuit theory}
\label{sec-circuit}
A recently developed method for deriving the
Hamiltonian of SC circuits from their classical dynamics,
combined with the theory of dissipative quantum systems,
can be utilized to describe decoherence in arbitrary
SC circuits \cite{burkard04a}.

There exists a variety of theoretical models for dissipative
environments in general, and dissipative electrical circuit
elements (impedances) in particular.  A dissipative (resistive)
element can be modeled as a transmission line \cite{yurke94,yurke87,werner91},
i.e.\ an infinite set of dissipation-free elements
(capacitors and inductors), or, alternatively, within 
the widely known Caldeira-Leggett model \cite{caldeira83,leggett87,weiss99} 
as a continuum of harmonic oscillators that is
coupled to the degrees of freedom of the system (in this case, the SC circuit).
In the following, the Caldeira-Legget approach will be used.

The systematic derivation of the dynamical equations
for a general (classical) electric circuit is a well-known problem in
electric engineering that has been tackled using the elegant
and convenient network graph analysis methods \cite{peikari74}.
It has been suggested early on that these methods may also be
used for a description of the dissipative quantum dynamics of
superconducting circuits \cite{devoret97}.
We shall now explain the circuit graph analysis applied to SC qubits,
both of the flux (Sec.\ref{ssec-flux}) 
and charge (Sec.\ref{ssec-charge}) type.

\subsection{Flux qubits}
\label{ssec-flux}
In this Section, the results of the circuit theory for flux qubits
are presented; for a derivation, see \cite{burkard04a}.
The IBM qubit \cite{koch03} will be used as a first example, 
and then followed by other examples.  The IBM qubit is described by the 
electrical circuit drawn in Fig.~\ref{ibm-graph}.

\subsubsection{The network graph}
As a first step in the circuit analysis of a SC flux qubit, 
the \textit{network graph} of the SC circuit is drawn and labeled.
In the graph, each two-terminal element (Josephson junction,
capacitor, inductor, external impedance, current source)
is represented as a branch connecting two nodes.
In Fig.~\ref{ibm-graph} (left panel), the IBM qubit is represented
as a network graph, where thick lines are used as a shorthand 
for resistively shunted Josephson junctions, or RSJ
(see Fig.~\ref{ibm-graph}, right panel).
A convention for the direction of all branches has to be chosen--in
Fig.~\ref{ibm-graph}, the direction of branches
is represented by an arrow.

\subsubsection{The tree of the network graph}
\begin{figure}[b]
\begin{minipage}{6.8cm}
\includegraphics[width=6.5cm]{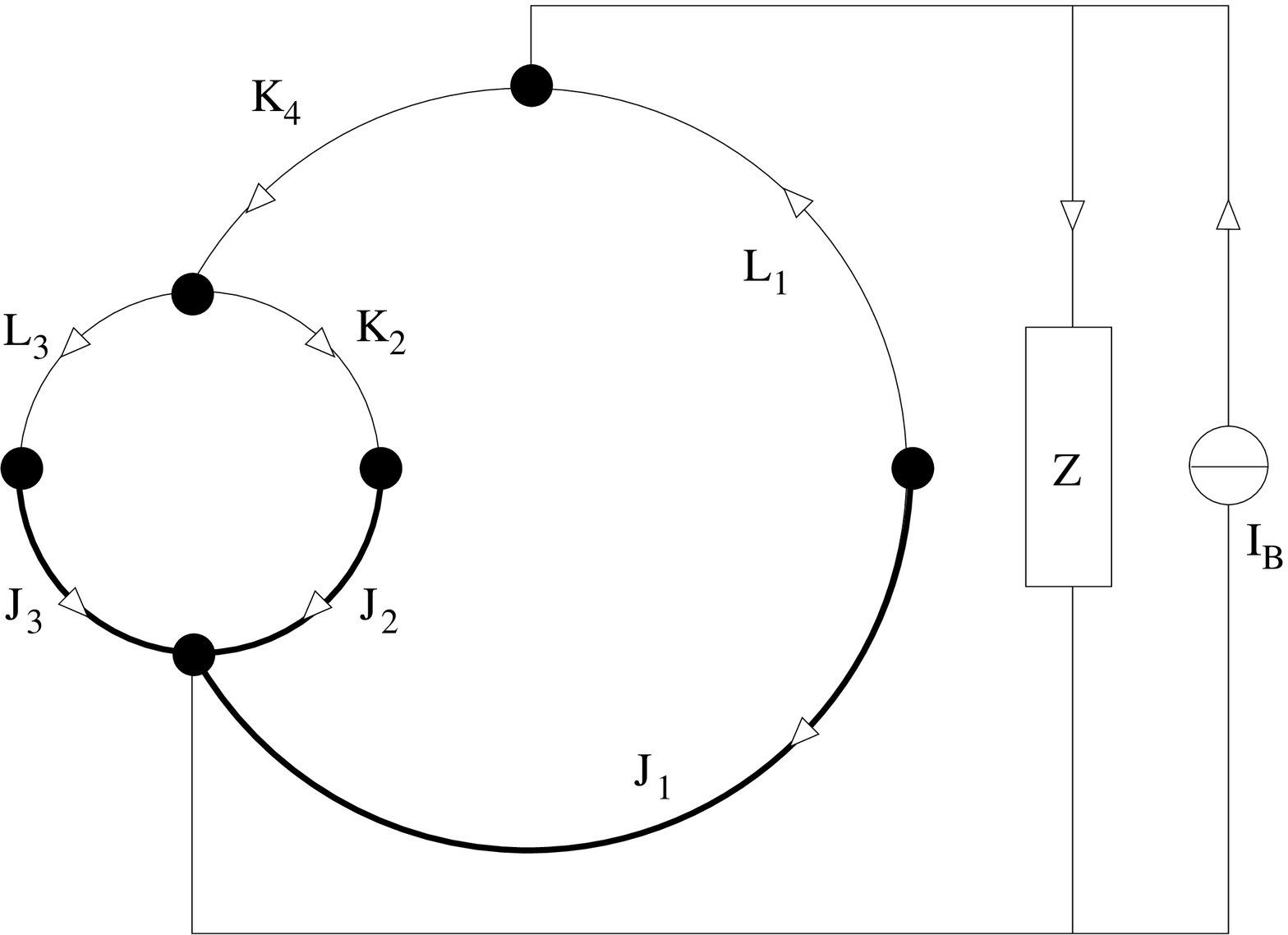}
\end{minipage}
\hfill
\begin{minipage}{4.8cm}
\includegraphics[width=4.5cm]{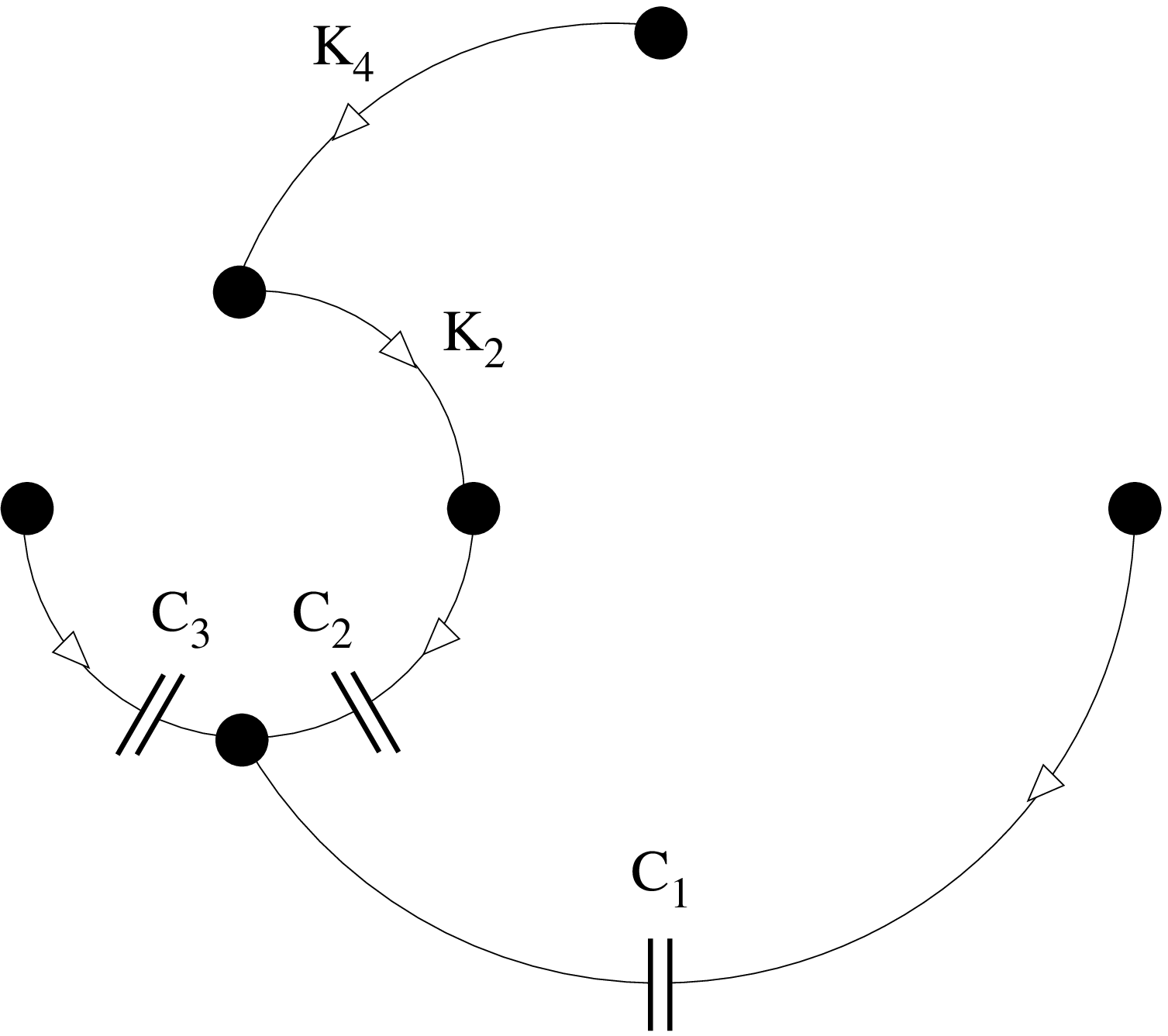}
\end{minipage}
\hfill
\begin{minipage}{3.8cm}
\includegraphics[width=3.5cm]{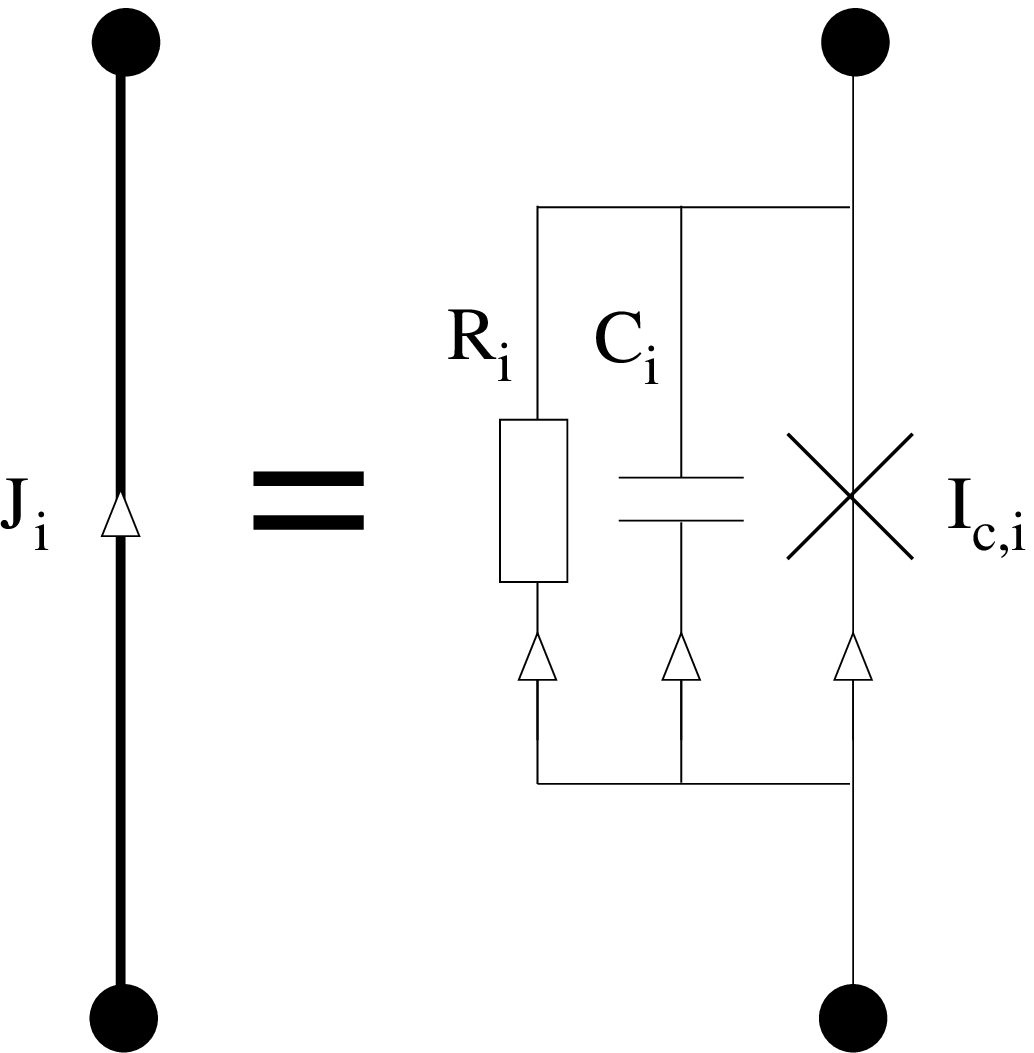}
\end{minipage}
\caption{\label{ibm-graph}
Left: 
The IBM qubit is an example of a network graph.  Each thick line represents a Josephson element, i.e.\ three branches in parallel, see right panel. Thin lines represent simple two-terminal elements, such as linear inductors (L, K), external impedances (Z), and current sources ($I_B$).
Center:
A Josephson subgraph (thick line) consists of three branches; a Josephson junction (cross), 
a shunt capacitor (C), a shunt resistor (R), and no extra nodes.
Right:
A tree for the circuit shown on the right.
A tree is a subgraph containing all nodes and no loop.  
Here, a tree was chosen that contains all capacitors (C), some inductors (K), 
but no current sources ($I_B$) or external impedances (Z).}
\end{figure}

As a second step, a \textit{tree} of the network graph needs to be specified.   
A tree of a graph is a set of branches connecting all nodes without containing
any loops.  Here, the tree is chosen such that it contains
all capacitors, as few inductors as possible, and neither resistors 
(external impedances) nor current sources.  The conditions under
which such a choice can be made are discussed in \cite{burkard04a}.
The tree of Fig.~\ref{ibm-graph}(right) that will be used here is 
shown in Fig.~\ref{ibm-graph}(center).
The branches in the tree are called \textit{tree branches};  all other
branches are called \textit{chords}.  
Each chord is associated with exactly one so-called fundamental loop
that is obtained when adding the chord to the tree.

\subsubsection{The loop matrices}
In a next step, the loop sub-matrices ${\bf F}_{CL}$, ${\bf F}_{CZ}$,
${\bf F}_{CB}$, ${\bf F}_{KL}$, ${\bf F}_{KZ}$, and ${\bf F}_{KB}$
need to be found.
The loop sub-matrices ${\bf F}_{XY}$ have entries $+1$, $-1$, or $0$,
and hold the information about which tree branches of type $X$ belong to
which fundamental loop associated with the chords of type $Y$.  
E.g., for our example,
\begin{equation}
  {\bf F}_{CL} = \left(\begin{array}{r r}
      1 &  0 \\
     -1 &  1 \\
      0 & -1
\end{array}\right),
\end{equation}
where the first column determines that the capacitor $C_1$ (part of $J_1$)
belongs to the large loop (associated with $L_1$), 
capacitor $C_2$ (part of $J_2$) belongs to the large loop 
(with different orientation),
while capacitor $C_3$ (part of $J_3$) does not belong to the 
large loop at all.
Similarly, the second column of ${\bf F}_{CL}$ says which of
the capacitors are contained in the small loop (associated with $L_3$).

The loop matrices have the purpose of systematically incorporating 
Kirchhoff's laws of current and energy conservation in the circuit,
\begin{eqnarray}
  {\bf F}^{(C)}{\bf I} &=& \left( \openone \, |\, {\bf F}\right){\bf I} = 0, \label{Kirchhoff-Q}\\
  {\bf F}^{(L)}{\bf V} &=& \left(- {\bf F}^T \, |\, \openone \right){\bf V} 
                        =  {\bf \dot\Phi}, \label{Kirchhoff-B}
\end{eqnarray}
where the external magnetic fluxes are denoted with ${\bf \Phi}$, 
and where the loop sub-submatrices
\begin{eqnarray}
  \label{Fsplit-flux}
  {\bf F} = \left(\begin{array}{c c c c c}
      {\bf F}_{CJ} & {\bf F}_{CL} & {\bf F}_{CR} & {\bf F}_{CZ} & {\bf F}_{CB} \\
      {\bf F}_{KJ} & {\bf F}_{KL} & {\bf F}_{KR} & {\bf F}_{KZ} & {\bf F}_{KB}
\end{array}\right)
\end{eqnarray}
are related to the full fundamental loop and cutset matrices
${\bf F}^{(L)}$ and ${\bf F}^{(C)}$ via the
grouping of the branch currents and voltages into a tree and a chord part,
${\bf I} = ({\bf I}_{\rm tr}, {\bf I}_{\rm ch})$ and
${\bf V} = ({\bf V}_{\rm tr}, {\bf V}_{\rm ch})$.

\subsubsection{Current-voltage relations (CVRs)}
In order to derive the equations of motion and eventually, the Hamiltonian
of the SC circuit, Kirchhoff's laws, Eqs.~(\ref{Kirchhoff-Q}) and (\ref{Kirchhoff-B})
need to be combined with the CVRs of the various branch elements.
For this purpose, the tree and chord currents and voltages are divided up further,
according to the various branch types, ${\bf I}_{\rm tr}  =  ({\bf I}_C, {\bf I}_K)$, 
and ${\bf I}_{\rm ch}  =  ({\bf I}_J, {\bf I}_L, {\bf I}_R, {\bf I}_Z, {\bf I}_B)$,
and similarly for the voltages.
The tree current and voltage vectors contain a capacitor (C)
and tree inductor (K) part, whereas the chord current and voltage
vectors consist of parts for chord inductors, both non-linear (J) and linear (L),
shunt resistors (R) and other external impedances (Z), and bias current sources (B).
The branch charges and fluxes ($X=C, K, J, L, R, Z, B$) are formally defined as
\begin{eqnarray}
  {\bf I}_X(t) &=& \dot{\bf Q}_X(t),    \label{charges}\\
  {\bf V}_X(t) &=& \dot{\bf \Phi}_X(t). \label{fluxes}
\end{eqnarray}
Using the second Josephson relation and Eq.~(\ref{fluxes}),
we identify the formal fluxes associated with the Josephson
junctions as the superconducting phase differences $\bphi$ across the junctions,
\begin{equation}
  \frac{{\bf \Phi}_J}{\Phi_0} =  \frac{\bphi}{2\pi} ,
\end{equation}
where $\Phi_0=h/2e$ is the superconducting flux quantum.
Each branch type has its own current-voltage relation (CVR);
e.g., the Josephson junction branches follow the first Josephson relation,
\begin{equation}
    {\bf I}_J  =  {\bf I}_{\rm c} \,\mbox{\boldmath $\sin$} \bphi ,\label{CVR-J}\\
\end{equation}
with the critical current matrix ${\bf I}_{\rm c}$,
while the external impedances are described by the integral relation,
\begin{equation}
  \label{CVR-Z-time}
  {\bf V}_Z(t)=\int_{-\infty}^t {\bf Z}(t-\tau){\bf I}_Z(\tau) d\tau \equiv ({\bf Z}*{\bf I}_Z)(t).
\end{equation}
The total inductance matrix
\begin{equation}
  \label{inductance-mat}
{\bf L}_{\rm t}
  =\left(\begin{array}{l l}{\bf L}        & {\bf L}_{LK}\\
                           {\bf L}_{LK}^T & {\bf L}_{K} \end{array}\right),
\end{equation}
is used for the CVR of the chord (L) and tree (K) inductances,
\begin{equation}
  \label{inductance-1}
   \left(\begin{array}{c}{\bf \Phi}_L\\ {\bf \Phi}_K\end{array}\right)
  = {\bf L}_{\rm t} \left(\begin{array}{c}{\bf I}_L\\ {\bf I}_K\end{array}\right),
\end{equation}
where ${\bf L}$ and ${\bf L}_K$ are the self inductances of the chord and tree branch
inductors, resp., off-diagonal elements describing the mutual inductances among
chord inductors and tree inductors separately, and  ${\bf L}_{LK}$ is the
mutual inductance matrix between tree and chord inductors.

\subsubsection{The Hamiltonian}
The elements described above are sufficient to determine the Hamiltonian of the 
dissipation-free system,
\begin{eqnarray}
{\cal H}_S &=& \frac{1}{2}{\bf Q}_C^T {\bf C}^{-1}{\bf Q}_C 
            + \left(\frac{\Phi_0}{2\pi}\right)^2  U(\bphi),   \label{Hamiltonian-S}\\
U(\bphi) &=& -\sum_i \frac{2\pi I_{c;i}}{\Phi_0}\cos\varphi_i 
             + \frac{1}{2}\bphi^T {\bf M}_0 \bphi 
             + \frac{2\pi}{\Phi_0}\bphi^T \left({\bf N} {\bf \Phi}_x 
                                                  + {\bf S}{\bf I}_B \right),\label{U}
\end{eqnarray}
where ${\bf Q}_C$ are the charges conjugate to the fluxes ${\bf \Phi}_J=(\Phi_0/2\pi)\bphi$
and ${\bf C}$ is the capacitance matrix of the circuit.
The matrices ${\bf M}_0$, ${\bf N}$, and ${\bf S}$ are obtained from the 
inductance and loop matrices ${\bf L}_t$ and ${\bf F}$ \cite{burkard04a}.
The Hamiltonian Eq.~(\ref{Hamiltonian-S}) is quantized using the commutator relation
\begin{equation}
  \label{commutator}
  \left[\frac{\Phi_0}{2\pi}\varphi_i , Q_{C;j} \right] = i\hbar\delta_{ij} .
\end{equation}
The system including dissipation can be described using the Caldeita-Legget model,
\begin{eqnarray}
{\cal H}   &=& {\cal H}_S + {\cal H}_B + {\cal H}_{SB},\label{Hamiltonian}\\
{\cal H}_B &=& \frac{1}{2}\sum_\alpha\left(\frac{p_\alpha^2}{m_\alpha}+m_\alpha \omega_\alpha^2 x_\alpha^2\right),\label{Hamiltonian-B}\\
{\cal H}_{SB} &=&  {\bf m}\cdot\bphi \sum_\alpha c_\alpha x_\alpha  + \Delta U(\bphi),\label{Hamiltonian-SB}
\end{eqnarray}
where ${\cal H}_S$ is the quantized Hamiltonian Eq.~(\ref{Hamiltonian-S}),
${\cal H}_B$ is the Hamiltonian describing 
a bath of harmonic oscillators with (fictitious) position and momentum 
operators $x_\alpha$ and $p_\alpha$ with $[x_\alpha, p_\beta]=i\hbar\delta_{\alpha\beta}$, 
masses $m_\alpha$, and oscillator 
frequencies $\omega_\alpha$.  Finally, ${\cal H}_{SB}$ describes the coupling 
between the system and bath degrees of freedom, $\bphi$ and $x_\alpha$, where 
$c_\alpha$ is a coupling parameter and ${\bf m}$ are obtained from the 
inductance and loop matrices ${\bf L}_t$ and ${\bf F}$ \cite{burkard04a}.

The quantum dynamics of the entire system (qubit plus bath of oscillators)
is described by the 
Liouville equation $\dot{\rho}(t) = -i[{\cal H}, \rho(t)] \equiv  -i{\cal L}\rho(t)$
for the density matrix $\rho$.  The state of the system alone is described
by the reduced density matrix $\rho_S(t) = {\rm Tr}_B\,\rho (t)$.
In the Born-Markov approximation, the master equation for $\rho_S(t)$
can be written in the form of the Redfield equations \cite{redfield57},
\begin{equation}
  \dot{\rho}_{nm}(t) 
   = -i\omega_{nm}\rho_{nm}(t) -\sum_{kl}R_{nmkl}\rho_{kl}(t),\label{Redfield-equation}
\end{equation}
where ${\rho}_{nm}=\langle n|\rho_S|m\rangle$ are the matrix elements of $\rho$
in the eigenbasis $|n\rangle$ of ${\cal H}_S$ (eigenenergies $\omega_n$),
and $\omega_{nm}=\omega_n-\omega_m$,
and with the Redfield tensor,
\begin{eqnarray}
  R_{nmkl} &=& \delta_{lm}\!\sum_r \Gamma_{nrrk}^{(+)} + \delta_{nk}\!\sum_r \Gamma_{lrrm}^{(-)}
-\Gamma_{lmnk}^{(+)}-\Gamma_{lmnk}^{(-)},\label{RGamma}\\
{\rm Re}\Gamma_{lmnk}^{(+)} &=&  ({\bf m}\cdot\bphi)_{lm} ({\bf m}\cdot\bphi)_{nk} 
J(|\omega_{nk}|)\frac{e^{-\beta \omega_{nk}/2}}{\sinh \beta|\omega_{nk}|/2}\, ,\nonumber\\
{\rm Im}\Gamma_{lmnk}^{(+)} &=& -({\bf m}\cdot\bphi)_{lm} ({\bf m}\cdot\bphi)_{nk} 
\frac{2}{\pi} P\!\!\int_0^\infty \!\!\!\!\!\!d\omega \frac{J(\omega)}{\omega^2 \!-\!\omega_{nk}^2}\!\left(\!\omega\!-\!\omega_{nk}\coth \frac{\beta\omega}{2}\!\right). \label{Gp-flux}
\end{eqnarray}

In the two-dimensional qubit subspace, the Bloch vector 
${\bf p} = {\rm Tr}({\mbox{\boldmath $\sigma$}}\rho )$ can be introduced
where ${\mbox{\boldmath $\sigma$}}=(\sigma_x, \sigma_y, \sigma_z)$
are the Pauli matrices, and the Redfield equation (\ref{Redfield-equation}) 
takes the form of the Bloch equation $\dot{\bf p} = \bomega\times{\bf p}-R{\bf p}+{\bf p}_0$,
with $\bomega = (0,0,\omega_{01})^T$, where in the secular approximation, the
relaxation matrix $R$ is diagonal, $R = {\rm diag}(T_2^{-1}, T_2^{-1},T_1^{-1})$.
The relaxation and decoherence times $T_1$ and $T_2$ are then given by
\begin{eqnarray}
  \frac{1}{T_1} &=& 4|\langle 0|{\bf m}\cdot\bphi|1\rangle|^2 J(\omega_{01}) \coth\frac{\omega_{01}}{2k_B T}, \label{T1-flux}\\
  \frac{1}{T_2} &=& \frac{1}{2T_1} + \frac{1}{T_\phi},\label{T2-flux}\\
  \frac{1}{T_\phi} &=&  |\langle 0|{\bf m}\cdot\bphi|0\rangle-\langle 1|{\bf m}\cdot\bphi|1\rangle|^2 \left.\frac{J(\omega)}{\omega}\right|_{\omega\rightarrow 0} \!\!\!\!\!\!\!\!\! 2k_B T. \label{Tphi-flux}
\end{eqnarray}
In the semiclassical approximotion, $T_1$ and $T_\phi$ can be related to the
parameters $\Delta$ and $\epsilon$ in the Hamiltonian Eq.~(\ref{sc-pseudospin}),
\begin{eqnarray}
  \frac{1}{T_1} &=& \left(\frac{\Delta}{\omega_{01}}\right)^2\left| \Delta\bphi \cdot{\bf m}\right|^{2} J(\omega_{01}) \coth\frac{\omega_{01}}{2k_B T}, \label{T1-dw}\\
  \frac{1}{T_\phi} &=&  \left(\frac{\epsilon}{\omega_{01}}\right)^2 |\Delta\bphi \cdot {\bf m}|^2 \left.\frac{J(\omega)}{\omega}\right|_{\omega\rightarrow 0} \!\!\!\!\!\!\!\!\! 2k_B T. \label{Tphi-dw}
\end{eqnarray}

\subsubsection{Leakage}
\label{leakage}
The leakage rate from a qubit state $k=0,1$ into higher levels $n=2,3,\ldots$ outside
the qubit space can be quantified from the Redfield equation Eq.~(\ref{Redfield-equation})
by the sum
\begin{equation}
  \frac{1}{T_{L,k}} = 4\sum_n |\langle n|{\bf m}\cdot\bphi|k\rangle|^2 
                       J(\omega_{kn}) \coth\frac{\omega_{kn}}{2k_B T}. \label{Tleak-flux}
\end{equation}

\subsubsection{The Delft qubit}
\label{new-delft-qubit}
\begin{figure}
\begin{minipage}{7.3cm}
\includegraphics[width=7cm]{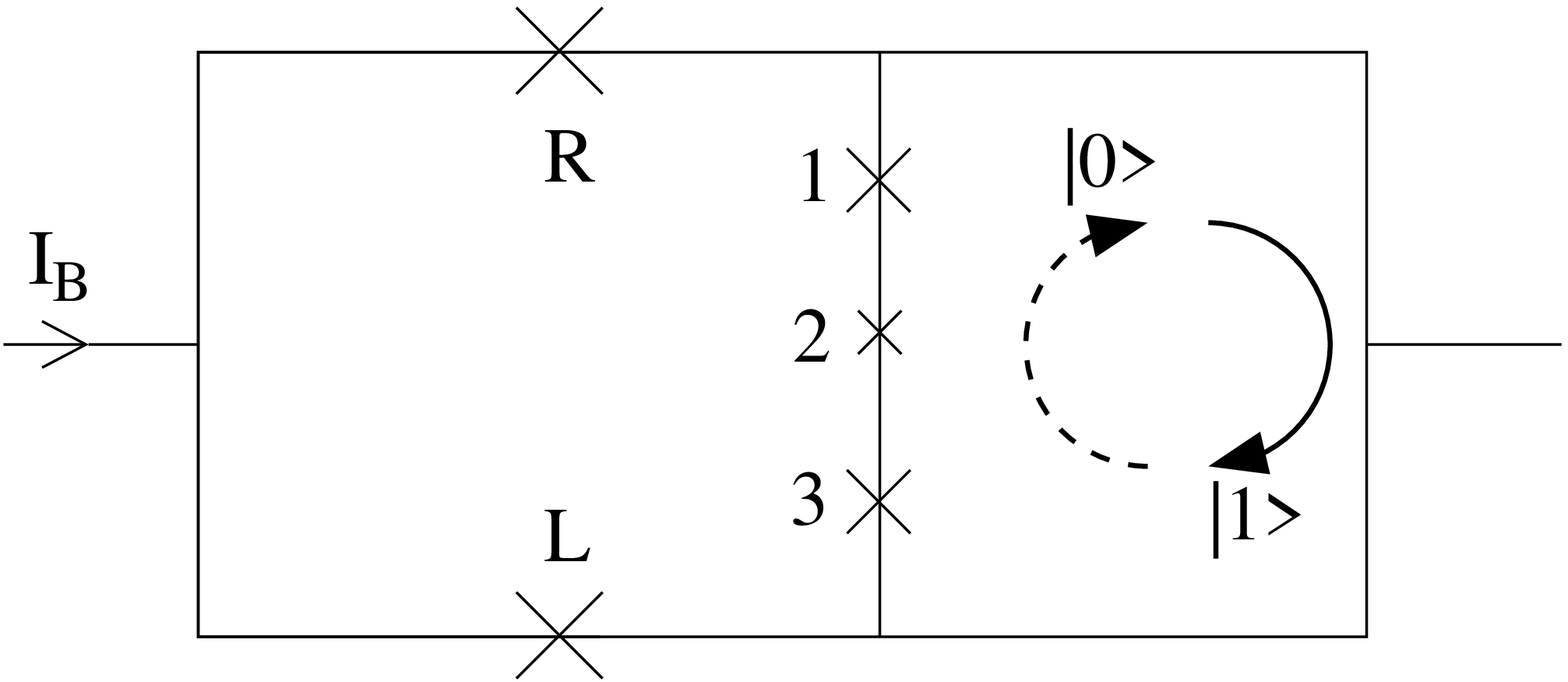}
\end{minipage}
\begin{minipage}{9cm}
\caption{\label{qubit-i}
Schematic of the Delft circuit, Fig.~\ref{fig-delft-qubit}, where the 
crosses denote Josephson junctions.
The outer loop with two junctions $L$ and $R$ forms a dc SQUID that
is used to read out the qubit.  The state of the qubit is determined 
by the orientation of the circulating current in the small loop,
comprising the junctions $1$, $2$, and $3$, one of which has a
slightly smaller critical current than the others.
A bias current $I_B$ can be applied as indicated for read-out.}
\end{minipage}
\end{figure}

A very successful qubit design is the Delft qubit \cite{chiorescu03} which is
depicted in  Fig.~\ref{fig-delft-qubit}, and which will be discussed in this Section.
A schematical drawing of the SC circuit for the Delft qubit is shown in Fig.~\ref{qubit-i}.
\begin{figure}[b]
\begin{minipage}{8.3cm}
\includegraphics[width=8cm]{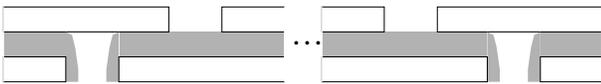}
\end{minipage}
\hfill
\begin{minipage}{9cm}
\caption{\label{shadow}
Schematics of Josephson junctions produced by the shadow 
evaporation technique, connecting the upper with the lower
aluminum layer.  Shaded regions represent the aluminum oxide.}
\end{minipage}
\end{figure}
This design is intended to be
immune to current fluctuations in the current bias $I_B$ 
due to its symmetry properties;  at zero dc bias, $I_B=0$,
and independent of the applied magnetic field,
a small fluctuating current $\delta I_B(t)$ caused by the
finite impedance of the external control circuit (the current 
source) is divided equally into the two arms of the 
SQUID loop and no net current flows through the
three-junction qubit line.
Hence, in the ideal circuit (Fig.~\ref{qubit-i}) the qubit is 
protected from decoherence due to current fluctuations in the bias current line.
However, asymmetries in the SQUID loop may spoil the 
protection of the qubit from decoherence.
In the case of an inductively coupled SQUID \cite{mooij99,orlando99,vanderwal00} 
neither a small geometrical asymmetry (imbalance of self- and mutual inductances in the SQUID loop),
nor thejunction (critical current) asymmetry of typically a few percent, 
would suffice to cause a relevant amount of decoherence
at zero bias current $I_B=0$ \cite{burkard04a}.
What turns out to be important here is that the circuit (Fig.~\ref{fig-delft-qubit})
contains another asymmetry, caused by its \textit{double layer structure}, 
being an artifact of the fabrication method used to produce SC circuits
with aluminum/aluminum oxide Josephson junctions, the
so-called shadow evaporation technique.  Junctions produced
with this technique will always connect
the top layer with the bottom layer, see Fig.~\ref{shadow}.   
\begin{figure}[t]
\begin{minipage}{7.9cm}
\includegraphics[width=7.6cm]{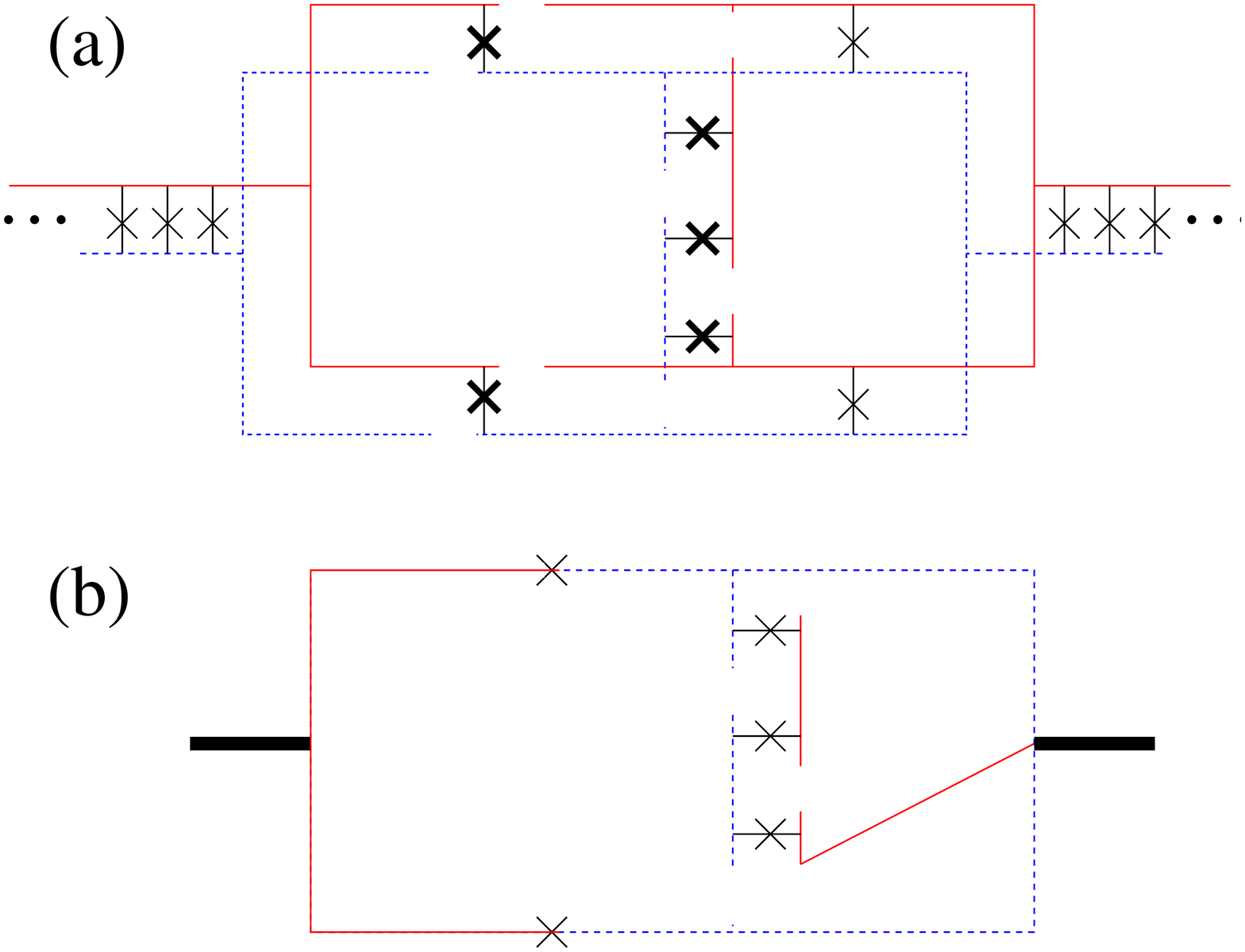}
\end{minipage}
\hfill
\begin{minipage}{9cm}
\caption{\label{qubit}
(a) Double layer structure.  Dashed blue lines represent the lower,
solid red lines the upper SC layer, and crosses indicate Josephson junctions.
The thick crosses are the intended junctions, while the thin
crosses are the unintended distributed junctions due to the double-layer
structure.
(b) Simplest circuit model of the double layer structure.
The symmetry between the upper and lower arms of the SQUID
has been broken by the qubit line comprising three junctions.
Thick black lines denote pieces of the SC in which the upper and lower
layer are connected by large area junctions.}
\end{minipage}
\end{figure}

Thus, while circuits like 
Fig.~\ref{qubit-i} can be produced with this technique,
strictly speaking, loops will always contain an
even number of junctions.
In order to analyze the implications of the double layer
structure for the circuit in Fig.~\ref{qubit-i},
the circuit is drawn again in Fig.~\ref{qubit}(a) with separate upper and 
lower layers.  Note that each piece of the upper layer is connected 
with the underlying piece of the lower layer via an ``unintentional'' Josephson junction.
\begin{figure}[t]
\begin{minipage}{7.3cm}
\includegraphics[width=7cm]{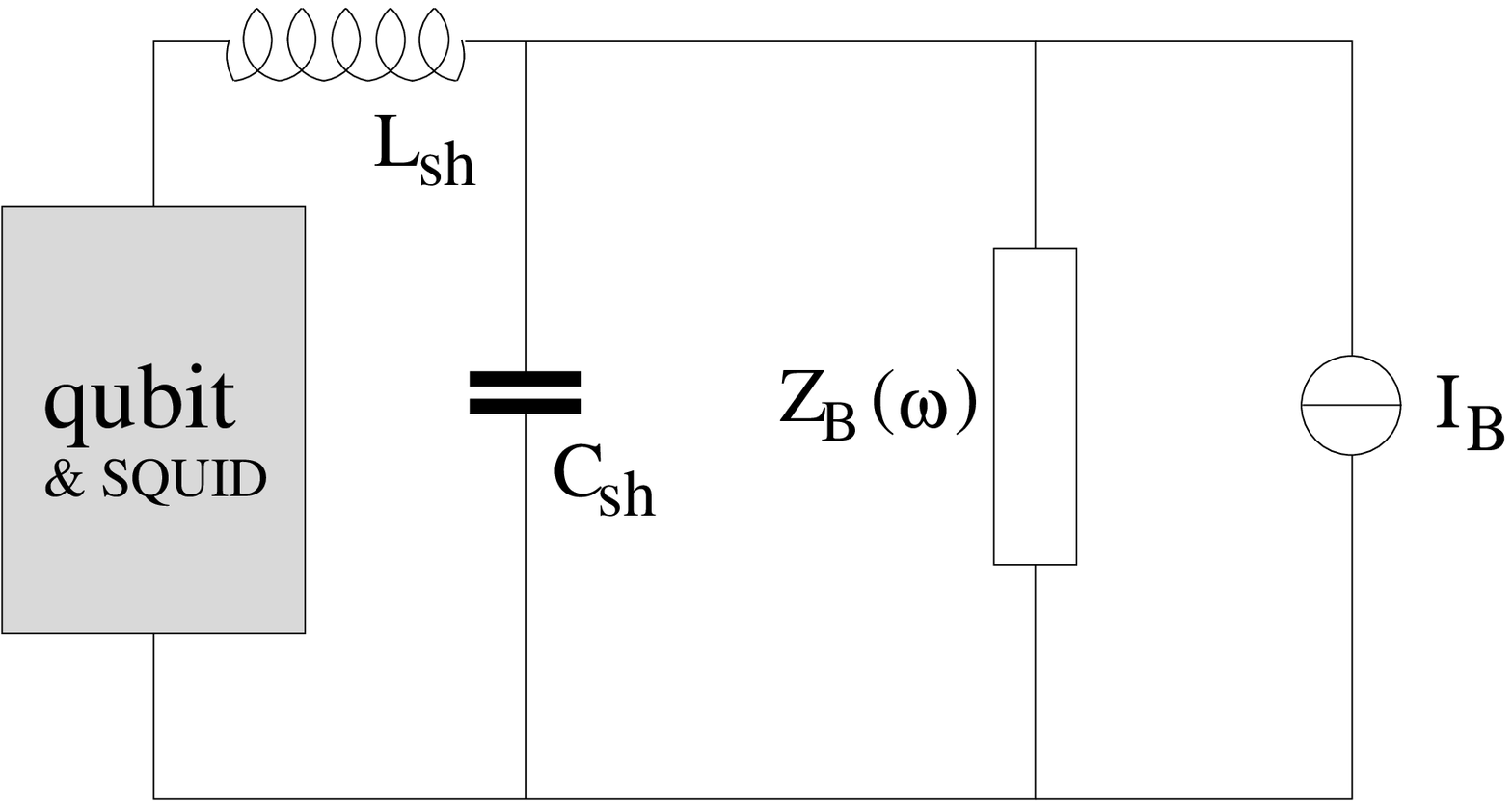}
\end{minipage}
\hfill
\begin{minipage}{9cm}
\caption{\label{sc-system}
External circuit attached to the qubit (Fig.~\ref{qubit-i})
that allows the application of a bias current $I_B$ for qubit read-out.
The inductance $L_{\rm sh}$ and capacitance $C_{\rm sh}$
form the \textit{shell circuit}, and $Z(\omega)$ is the
total impedance of the current source ($I_B$).  The case
where a voltage source is used to generate a current
can be reduced to this using Norton's theorem.}
\end{minipage}
\end{figure}
\begin{figure}[b]
\begin{minipage}{8.8cm}
\includegraphics[width=8.5cm]{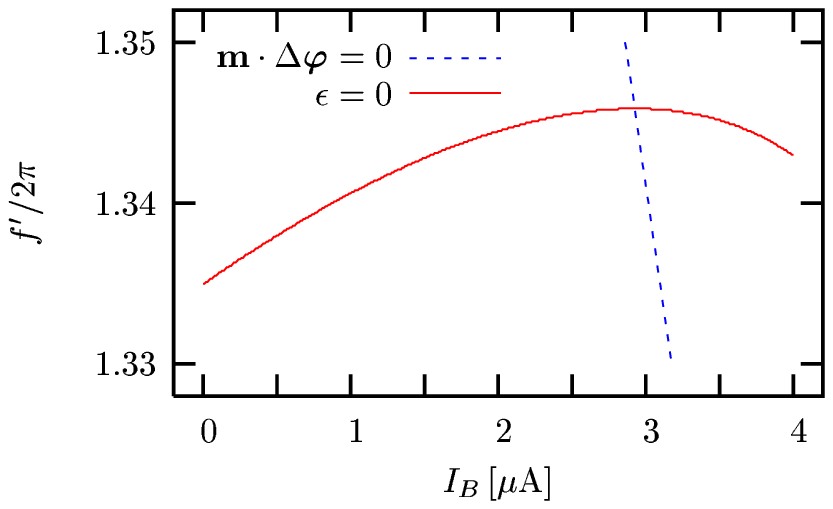}
\end{minipage}
\begin{minipage}{8.8cm}
\includegraphics[width=8.5cm]{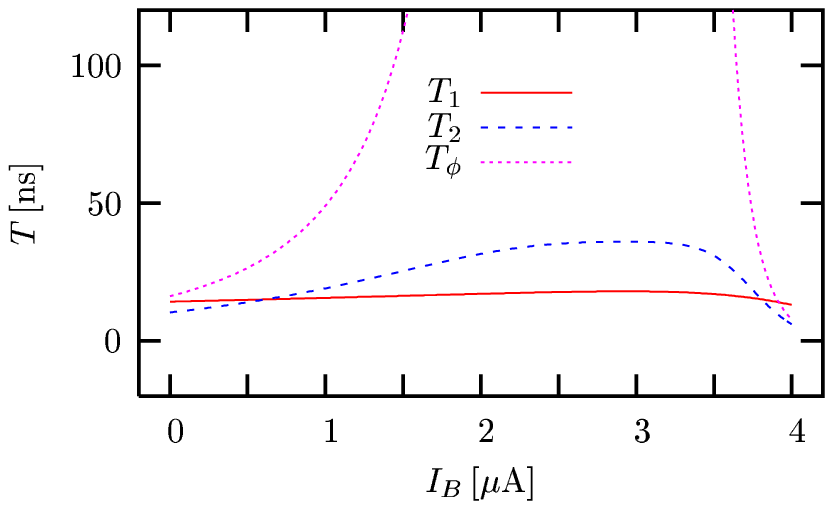}
\end{minipage}
\hfill
\caption{\label{decoh}
Left: Decoupling (red solid) and symmetric (blue dashed) curves in the $(I_B,f')$ plane,
where $I_B$ is the applied bias current and $f'=2\pi \Phi_x'/\Phi_0$
is the dimensionless externally applied magnetic flux threading the SQUID loop.
Both curves are obtained from the numerical minimization of the potential
Eq.~(\ref{U}).  The decoupling line is determined using the condition 
${\bf m}\cdot\Delta\bphi=0$, whereas the symmetric line follows from the
condiction $\epsilon=0$.
Right: Theoretical relaxation, pure dephasing, and decoherence times $T_1$, $T_\phi$, and $T_2$
as a function of applied bias current $I_B$, along the symmetric line (Fig.~\ref{decoh}, left).}
\end{figure}

These extra junctions typically have large areas 
and therefore large critical currents;  thus, their
Josephson energy can often be neglected.
In order to study the lowest-order effect of the
double layer structure, one can neglect all unintentional junctions in this
sense and arrive at the circuit Fig.~\ref{qubit}(b).  
It should be emphasized that the
resulting circuit is distinct from the 'ideal' circuit
Fig.~\ref{qubit-i} which does not reflect the double-layer
structure.  In the real circuit,  Fig.~\ref{qubit}(b), the
symmetry between the two arms of the dc SQUID is 
broken, and thus it can be expected that
bias current fluctuations cause decoherence of the
qubit at zero dc bias current, $I_B=0$.  

Starting from the circuit graph 
of the Delft qubit, the circuit theory can be used to find
the Hamiltonian of the circuit, which can subsequently be
analyzed numerically.
The double-well minima $\bphi_0$ and $\bphi_1$ was found
for a range of bias currents and applied external flux.  
The states localized at $\bphi_0$ and $\bphi_1$ are encoding the 
logical $|0\rangle$ and $|1\rangle$ states of the qubit. 
Two special lines in the plane spanned by the 
bias currents and applied external flux can now be determined,
see Fig.~\ref{decoh}.  (i) The line $f^*(I_B)$ on which
a \textit{symmetric} double well is predicted, 
$\epsilon \equiv U(\bphi_0)-U(\bphi_1)=0$.
On this line, the dephasing time $T_\phi$ diverges.
(ii) The line on which ${\bf m}\cdot\Delta\bphi=0$,
where $\Delta\bphi=\bphi_0\bphi_1$ is the vector
joining the two minima of the potential.
On this line, the environment is \textit{decoupled} from the
system, and both the relaxation and the decoherence times
diverge, $T_{1,2,\phi}\rightarrow\infty$.
The curve $f^*(I_B)$ agrees qualitatively with the experimentally measured
symmetry line \cite{bertet04}, but it underestimates the magnitude
of the variation in flux $f'$ as a function of $I_B$.
The point where the symmetric and the decoupling lines intersect
coincides with the maximum of the symmetric line, as can be understood
from the following argument.  Taking the total derivative with respect 
to $I_B$ of the relation 
$\epsilon = U(\bphi_0;f^*(I_B),I_B)-U(\bphi_1;f^*(I_B),I_B)=0$ 
on the symmetric line, and using that $\bphi_{0,1}$ are extremal
points of $U$, we obtain ${\bf n}\cdot\Delta\bphi \,\partial f^*/\partial I_B
+(2\pi/\Phi_0){\bf m}\cdot\Delta\bphi=0$ for some constant vector ${\bf n}$.
Therefore, ${\bf m}\cdot\Delta\bphi=0$ (decoupling line) and 
${\bf n}\cdot\Delta\bphi\neq 0$ implies $\partial f^*/\partial I_B =0$.

The relaxation and decoherence times $T_1$ and $T_2$ on the symmetric line have
been evaluated and are plotted (Fig.~\ref{decoh}, right)
where $\epsilon=0$, and therefore, $E=\Delta$.  The divergence in $T_\phi$ on the symmetric
line is cut off by higher-order effects, whereas the divergence of $T_1$ on the decoupling
line is cut off by residual impedances, e.g., due to the junctions' quasiparticle resistance
\cite{burkard04b}.  A peak in the relaxation and decoherence times where predicted
from theory can be observed experimentally \cite{bertet04}.

The symmetry breaking due to the double layer structure has 
another, very interesting, consequence.
It causes a coupling between the qubit and an external
harmonic oscillator, the plasmon mode formed by LC resonator in the SQUID (Fig.~\ref{sc-system}).
This coupling can be observed as resonances in the microwave spectrum of the system.
Moreover, it can be used to entangle the qubit with another degree of freedom \cite{chiorescu04}.
Here, the inductance $L_{\rm sh}$ and capacitance $C_{\rm sh}$ of the ``shell'' 
circuit (plasmon mode) are responsible for this resonator.
This coupling has been studied quantitatively in the framework of the circuit 
theory.
From the full Hamiltonian ${\cal H}_S$, a two-level Hamiltonian in the 
well-known Jaynes-Cummings form can be derived,
\begin{equation}
  {\cal H} =  \Delta \sigma_x + \epsilon \sigma_z
              + \hbar\omega_{\rm sh}\left(b^\dagger b +\frac{1}{2}\right) 
              + \lambda \sigma_z (b+b^\dagger), \label{Jaynes-Cummings}
\end{equation}
with the coupling parameter (Rabi frequency)
\begin{equation}
  \label{lambda}
  \lambda = -\sqrt{\pi}\left(\frac{\Phi_0}{2\pi}\right)^2 \sqrt{\frac{Z_{\rm sh}}{R_Q}}\frac{1}{M_{\rm sh}}{\bf m}\cdot\Delta\bphi.
\end{equation}
Note that this coupling vanishes along the decoupling line (Fig.~\ref{decoh}, left)
and also rapidly with the increase of $L_{\rm sh}$.
\begin{figure}
\begin{minipage}{8.3cm}
\includegraphics[width=7.9cm]{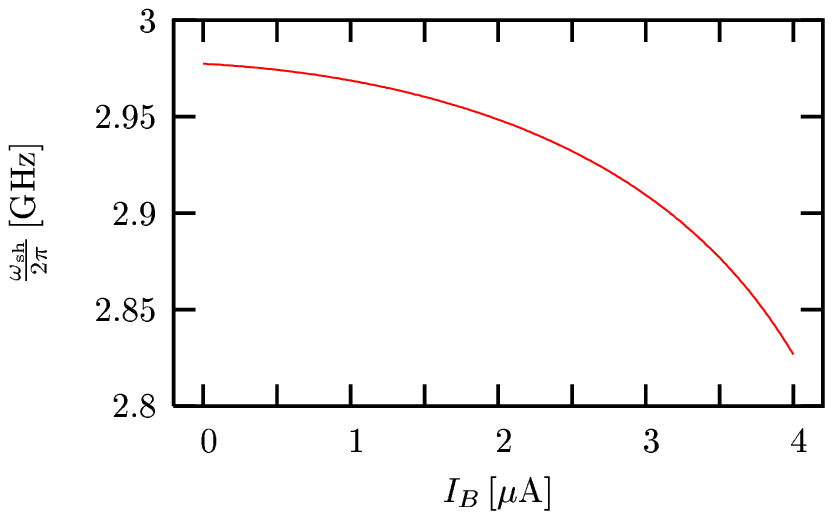}
\end{minipage}
\hfill
\begin{minipage}{8.3cm}
\vspace{1mm}
\includegraphics[width=8cm]{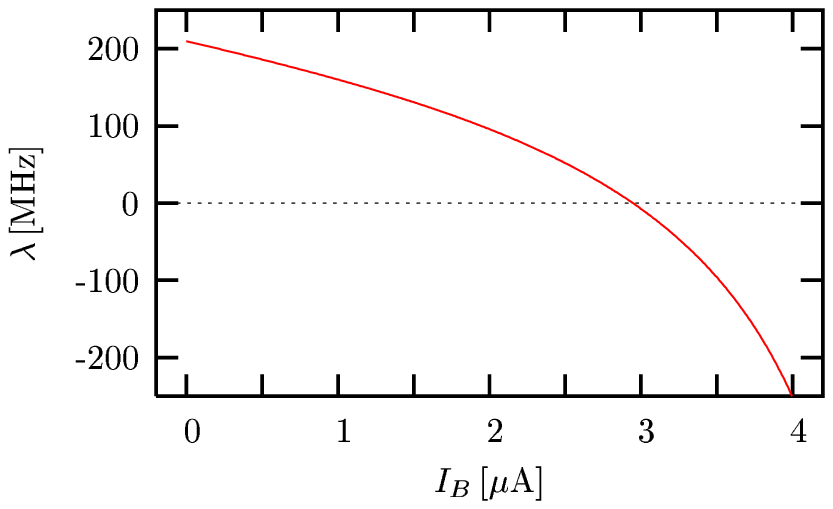}
\end{minipage}
\caption{\label{plasma-fig}
Left: Plasma frequency $\omega_{\rm sh}$ as a function of the applied bias current $I_B$.
The variation is due to the change the effective in Josephson inductances as
$I_B$ is varied.
Right: Rabi frequency of the coupling between the qubit and the plasmon mode.
The coupling disappears at the crossing with the decoupling line (Fig.~\ref{decoh}),
i.e., when ${\bf m}\cdot\Delta\bphi=0$.}
\end{figure}
The Rabi frequency at $I_B=0$ is predicted to be $\lambda\approx 210\,{\rm MHz}$.
The Rabi frequency as a function of the bias current $I_B$, together
with the plasma frequency, is plotted in Fig.~\ref{plasma-fig}.
Strong coupling between a \textit{charge} qubit and a quantum electromagnetic cavity
formed by a SC transmission line has been observed in \cite{wallraff04}.

\subsection{Charge qubits}
\label{ssec-charge}

In analogy to the circuit theory for flux qubits, a general circuit theory 
for charge qubits will be outlined and illustrated with examples in this Section
\cite{burkard04d}.
As in the case of the circuit theory for flux qubits,
we are not restricted to a Hilbert space of the SC device which is 
\textit{a priori} truncated to two levels only.
The role of the self and mutual inductances in SC charge qubits have been 
previously studied \cite{you01}, in particular as a means of coupling
two SC charge qubits \cite{makhlin01,makhlin04}.  Here, the
self and mutual inductances in the underlying SC circuit are fully taken
into account.

\subsubsection{Graph theory}
Note that the circuit theory for charge qubits is \textit{dual} to that
for flux qubits in the sense that the roles of chord and tree branches
are interchanged,
\begin{eqnarray}
  \label{IVsplit}
  {\bf I}_{\rm tr}  =  ({\bf I}_J, {\bf I}_L, {\bf I}_V, {\bf I}_Z), & & 
  {\bf I}_{\rm ch}  =  ({\bf I}_{C_J}, {\bf I}_C, {\bf I}_K), \\
  {\bf V}_{\rm tr}  =  ({\bf V}_J, {\bf V}_L, {\bf V}_V, {\bf V}_Z), & &
  {\bf V}_{\rm ch}  =  ({\bf V}_{C_J}, {\bf V}_C, {\bf V}_K).
\quad\quad
\end{eqnarray}
The loop matrix ${\bf F}$ now acquires the block form,
\begin{eqnarray}
  \label{Fsplit-charge}
  {\bf F} = \left(\begin{array}{c c c}
      \openone & {\bf F}_{JC} & {\bf F}_{JK} \\
      {\bf 0}  & {\bf F}_{LC} & {\bf F}_{LK} \\
      {\bf 0}  & {\bf F}_{VC} & {\bf F}_{VK} \\
      {\bf 0}  & {\bf F}_{ZC} & {\bf F}_{ZK} \\
\end{array}\right).
\end{eqnarray}
The Hamiltonian has the form
\begin{equation}
  \label{eq:HS}
  {\cal H}_S = \frac{1}{2}\left({\bf Q} - {\bf C}_V{\bf V}\right)^T{\cal C}^{-1}\left({\bf Q} 
                                  - {\bf C}_V{\bf V}\right) + U({\bf \Phi}),
\end{equation}
with the potential
\begin{equation}
  \label{eq:U}
  U({\bf \Phi})   = -{\bf L}_J^{-1} \,\mbox{\boldmath $\sin$} \bphi
                  +\frac{1}{2}{\bf \Phi}^T {\bf M}_0 {\bf \Phi}
                  +{\bf \Phi}^T {\bf N}{\bf \Phi}_x,
\end{equation}
where the Josephson and inductor flux variables are combined in
${\bf \Phi} = (\bphi,{\bf \Phi}_L)$, with the vector operator of conjugate 
charges denoted by ${\bf Q}$.
The coupling Hamiltonian in a Caldeira-Leggett description
${\cal H} = {\cal H}_S + {\cal H}_B + {\cal H}_{SB}$
now takes the form
\begin{equation}
  \label{eq:HSB}
  {\cal H}_{SB} = {\cal C}^{-1}\bar{\bf m}\cdot{\bf Q} \sum_\alpha c_\alpha x_\alpha
                = \bar{\bf m}\cdot  {\cal C}^{-1}{\bf Q} \sum_\alpha c_\alpha x_\alpha,
\end{equation}
where ${\cal C}$ is the total capacitance matric of the circuit.
The resulting Redfield equation takes the form Eq.~(\ref{Redfield-equation}) and
Eq.~(\ref{RGamma}), but with
\begin{eqnarray}
{\rm Re}\Gamma_{lmnk}^{(+)} &=&  \frac{1}{\hbar}({\bf m}\cdot{\bf Q})_{lm} ({\bf m}\cdot{\bf Q})_{nk} 
J(|\omega_{nk}|)\frac{e^{-\hbar \beta \omega_{nk}/2}}{\sinh \hbar \beta|\omega_{nk}|/2}\, ,\nonumber\\
{\rm Im}\Gamma_{lmnk}^{(+)} &=& -\frac{1}{\hbar}({\bf m}\cdot{\bf Q})_{lm} ({\bf m}\cdot{\bf Q})_{nk}\frac{2}{\pi} P\!\!\int_0^\infty \!\!\!\!\!\!d\omega \frac{J(\omega)}{\omega^2 \!-\!\omega_{nk}^2}\!\left(\!\omega\!-\!\omega_{nk}\coth \frac{\hbar \beta\omega}{2}\!\right), \label{Gp-charge}
\end{eqnarray}
and with ${\bf m}={\cal C}^{-1}\bar{\bf m}$.
Finally, the relaxation and decoherence times in a two-level description reduce to
\begin{eqnarray}
  \frac{1}{T_1} &=& \frac{4}{\hbar}|\langle 0|{\bf m}\cdot{\bf Q}|1\rangle|^2 J(\omega_{01}) \coth\frac{\hbar\omega_{01}}{2k_B T}, \label{T1-delft}\\
  \frac{1}{T_2} &=& \frac{1}{2 T_1} + \frac{1}{T_\phi},\label{T2-delft}\\\
  \frac{1}{T_\phi} &=&  \frac{1}{\hbar}|\langle 0|{\bf m}\cdot{\bf Q}|0\rangle-\langle 1|{\bf m}\cdot{\bf Q}|1\rangle|^2 \left.\frac{J(\omega)}{\hbar\omega}\right|_{\omega\rightarrow 0} \!\!\!\!\!\!\!\!\! 2k_B T. \label{Tphi-delft}
\end{eqnarray}
In the semiclassical limit, one finds
\begin{eqnarray}
  \frac{1}{T_1} &=& \frac{1}{\hbar}|{\bf m}\cdot \Delta{\bf Q}|^2 \left(\frac{\Delta}{\omega_{01}}\right)^2 J(\omega_{01}) \coth\frac{\hbar\omega_{01}}{2k_B T}, \label{T1-sc}\\
  \frac{1}{T_\phi} &=&  \frac{1}{\hbar}|{\bf m}\cdot \Delta{\bf Q}|^2 \left(\frac{\epsilon}{\omega_{01}}\right)^2 \left.\frac{J(\omega)}{\hbar\omega}\right|_{\omega\rightarrow 0} \!\!\!\!\!\!\!\!\! 2k_B T. \quad\quad\label{Tphi-sc}
\end{eqnarray}
The leakage rates from the logical state $k=0,1$ to states $n=2,3,\ldots$ outside
the computational subspace can be estimated as
\begin{equation}
  \label{Tleak-charge}
  \frac{1}{T_L} = \frac{4}{\hbar}\sum_{n=2}^\infty|\langle k|{\bf m}\cdot{\bf Q}|n\rangle|^2 J(\omega_{nk}) \coth\frac{\hbar\omega_{nk}}{2k_B T}.
\end{equation}

\subsubsection{Single charge box}
\label{ssec-chbox}
\begin{figure}
\begin{minipage}{6.3cm}
\includegraphics[width=6cm]{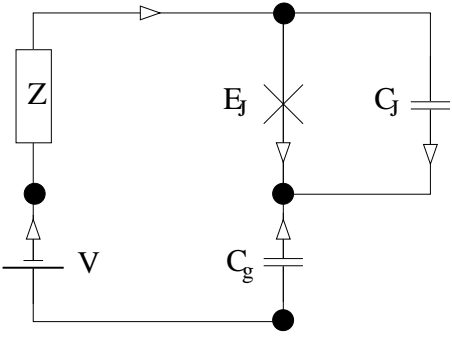}
\end{minipage}
\hfill
\begin{minipage}{9cm}
\caption{Circuit graph of a single voltage-biased charge box.
Branches represent a Josephson junction ($E_J$), capacitances
($C_J$ and $C_g$), a voltage source $V$, and the impedance $Z$.
The nodes are shown as black dots;  the node
connecting the junction ($E_J$) to the gate capacitance $C_g$
represents the SC island.\label{fig:chbox}}
\end{minipage}
\end{figure}
We now illustrate the circuit theory for charge qubits with
some examples.
The first example is the voltage-biased charge box, shown in Fig.~\ref{fig:chbox}.
The inductance of the leads has been neglected for simplicity (no $L$ and $K$ branches).
The tree of the graph is given by the Josephson, voltage source,
and impedance branches. For the loop matrices, we simply find
${\bf F}_{JC} = {\bf F}_{VC} = {\bf F}_{ZC} = 1$.
With the capacitances ${\cal C} \equiv C_{\rm tot} = C_J+C_g$ and $C_V = C_g$,
we arrive at the Hamiltonian,
\begin{equation}
  \label{eq:chbox-HS}
  {\cal H}_S = \frac{(Q_J+C_g V)^2}{2C_{\rm tot}}+E_J\cos\varphi .
\end{equation}
The coupling to the environment is characterized by
${\bf m}=( C_g/C_{\rm tot})$.
As an example, we give here the relaxation and dephasing times,
with $m=|{\bf m}|=C_g/C_{\rm tot}$,
\begin{eqnarray}
  \frac{1}{T_1} &=& 2\pi m^2 4 |\langle 0|n|1\rangle|^2 \frac{4{\rm Re}Z(\omega_{01})}{R_Q} \omega_{01} \coth\frac{\hbar\omega_{01}}{2k_B T}, \label{T1-chbox}\\
  \frac{1}{T_\phi} &=& 2\pi m^2 |\langle 0|n|0\rangle-\langle 1|n|1\rangle|^2 \frac{4{\rm Re}Z(0)}{R_Q}  \frac{2k_B T}{\hbar}, \quad\quad\label{Tphi-chbox}
\end{eqnarray}
where $n=Q/2e$ and $R_Q=h/e^2$.
In the semiclassical limit, $\langle 0|n|1\rangle\approx (1/2)(\Delta/\omega_{01})\Delta n$
and $\langle 0|n|0\rangle-\langle 1|n|1\rangle\approx (\epsilon/\omega_{01})\Delta n$.
With $\Delta n\approx 1$, we reproduce the results in \cite{makhlin01}.
Typical leakage rates are of the form of $1/T_1$, with the matrix element replaced by
$|\langle 0|n|k\rangle|$ and $|\langle 1|n|k\rangle|$, where $k\ge 2$ labels a state
other than the two qubit states, and with $\omega_{01}$ replaced by $\omega_{lk}$ ($l=0,1$).

\subsubsection{Flux-controlled Josephson junction}.
\begin{figure}
\begin{minipage}{6.3cm}
\includegraphics[width=6cm]{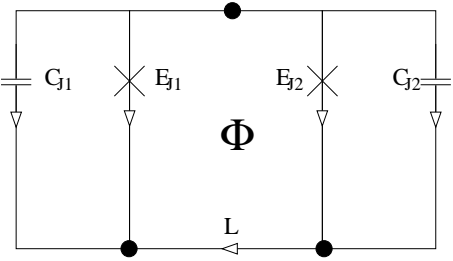}
\end{minipage}
\hfill
\begin{minipage}{9cm}
\caption{A flux-controlled Josephson junction.\label{fig:flux-controlled}}
\end{minipage}
\end{figure}
\begin{figure}
\begin{minipage}{6.3cm}
\includegraphics[width=7cm]{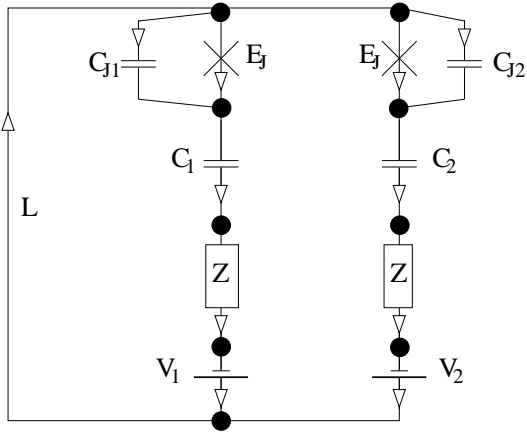}
\end{minipage}
\hfill
\begin{minipage}{9cm}
  \caption{Two inductively coupled charge boxes.\label{fig:ind-coupl}}
\end{minipage}
\end{figure}
A flux-controlled Josephson junction is a SC loop with two 
junctions  which acts as
an effective Josephson junction with a flux-dependent Josephson 
energy \cite{makhlin99}.  The circuit Fig.~\ref{fig:flux-controlled}
we use to describe the the flux-controlled junction comprises a
chord inductance ($K$) with inductance $L$.  
The tree consists of the two Josephson branches.
The only relevant loop matrix is
${\bf F}_{JK}=\left(\begin{array}{c c} 1 & -1 \end{array}\right)^T$.
In the limit $L\rightarrow 0$, and if $E_{J1}=E_{J2}$, we find 
${\bf F}_{JK}^T\bphi +\Phi_x = \varphi_1-\varphi_2 + \Phi \rightarrow 0$,
which leads us to the Hamiltonian
\begin{equation}
  \label{eq:fc-HS}
  {\cal H}_S = \frac{Q^2}{2 \bar{C}} - E_J(\Phi) \cos\varphi,
\end{equation}
where $\varphi = \varphi_1 + \pi \Phi/\Phi_0$, $\bar C = C_{J1} + C_{J2}$, and
$E_J(\Phi)  =  2 E_J \cos(2\pi\Phi/\Phi_0)$.

\subsubsection{Inductively coupled charge boxes}
We now turn to the case of two charge boxes of the type discussed in
Sec.~\ref{ssec-chbox}, coupled via an inductive loop \cite{makhlin01,makhlin99},
as shown in Fig.~\ref{fig:ind-coupl}.
Here, the tree consists of all Josephson, voltage source, and impedance
branches, plus the inductive branch $L$,
and the loop matrices are
\begin{equation}
\label{eq:icoupl-FJ}
  {\bf F}_{JC} = {\bf F}_{VC} = {\bf F}_{ZC} = 
\left(\begin{array}{c c}
      1 & 0\\
      0 & 1
\end{array}\right),
\:
  {\bf F}_{LC} = \left(\begin{array}{c c}
      1 & 1
\end{array}\right).
\end{equation}
With the two capacitance matrices
${\bf C} = {\rm diag}(C_1,C_2)$ and ${\bf C}_J = {\rm diag}(C_{J1},C_{J2})$,
we find
${\bf C}_{\rm tot} = {\bf C} + {\bf C}_J$,
${\bf C}_{JV} = {\bf C}$, 
${\bf C}_{JL} = {\bf C}_{LV}^T = (C_{1}, C_{2})^T$, and
${\bf C}_L = C_1+C_2$.
The vector $\bar{\bf m}$ consists of the two parts
${\bf m}_J = C$ and
${\bf m}_L = \left(\begin{array}{c c} C_1 & C_2\end{array}\right)$.
With Eq.~(\ref{eq:HS}) and the inverse of the total capacitance matrix,
\begin{widetext}
\begin{equation}
  \label{eq:3}
  {\cal C}^{-1}
= \frac{1}{\gamma}
\left(\begin{array}{c c c}
(C_1 +C_2) C_{J2}-C_2^2 & C_1 C_2 & -C_1 C_{J2}\\
C_1 C_2 & (C_1+C_2)C_{J1}-C_1^2 & -C_2 C_{J1}\\
-C_1 C_{J2} & -C_2 C_{J1} &  C_{J1} C_{J2}
\end{array}\right)
\equiv\left(\begin{array}{c c c}
C_{\rm eff,1}^{-1}    &   C_{\rm eff,12}^{-1}   &   C_{{\rm eff},L1}^{-1} \\
C_{\rm eff,12}^{-1}   &   C_{\rm eff,2}^{-1}    &   C_{{\rm eff},L2}^{-1} \\
C_{{\rm eff},L1}^{-1} &   C_{{\rm eff},L2}^{-1} &   C_{{\rm eff},L}^{-1}
\end{array}\right),
\end{equation}
\end{widetext}
where $\gamma=(C_1+C_2)C_{J1}C_{J2}-C_1^2 C_{J2}-C_2^2 C_{J1}$,
the Hamiltonian of the coupled system can be written as,
\begin{eqnarray}
  {\cal H}_S &=& \sum_{i=1,2}\left(\frac{(Q_{Ji}+C_i V_i)^2}{2 C_{{\rm eff},i}}
                           +E_{Ji}\cos\varphi_i \right) 
                           +\frac{(Q_L + C_1 V_1 + C_2 V_2)^2}{2 C_{{\rm eff},L}}
                           +\frac{\Phi_L^2}{2L}    \label{eq:icoupl-HS}\\
                & &        +\frac{(Q_{J1}+C_1 V_1)(Q_{J2}+C_2 V_2)}{C_{\rm eff,12}}
                           -\sum_{i=1,2}\frac{(Q_{Ji}+C_i V_i)(Q_L + C_1 V_1 + C_2 V_2)}{C_{{\rm eff},Li}}.\nonumber
\end{eqnarray}
While the last term in Eq.~(\ref{eq:icoupl-HS}) couples each qubit to the
$LC$ mode associated with the inductor $L$, and is thus responsible for the
inductive coupling of the qubits, the second last term provides a direct
capacitive coupling between the qubits.
In the limit $C_i \ll C_{Ji}$, we reproduce the results of \cite{makhlin01};
however, there are additional terms of order $C_i/C_{Ji}$, in particular
the new term $\propto 1/C_{\rm eff,12}$ in the Hamiltonian that capacitively couples
the qubits directly.
Since the coupled system involves at least four levels (more if excited states of the $LC$
coupling circuit or higher qubit levels are included), it can no longer be described
by a two-level Bloch equation with parameters $T_1$ and $T_2$.  We can however fix one of the
qubits to be in a particular state, say $|0\rangle$, and then look at the ``decoherence rates''
of the other qubit.
To lowest order in $C_i/C_{Ji}$, these rates due to the impedance $Z_i$ have the form
($q_i = C_i/(C_1+C_2)$)
\begin{eqnarray}
  \frac{1}{T_1} &=& 2\pi q_i^2 4 |\langle 00|n_L|10\rangle|^2 \frac{4{\rm Re}Z_i(\omega_{01})}{R_Q} \omega_{01} \coth\frac{\hbar\omega_{01}}{2k_B T}, \label{T1-icoupl}\\
  \frac{1}{T_\phi} &=& 2\pi q_i^2 |\langle 00|n_L|00\rangle-\langle 10|n_L|10\rangle|^2 \frac{4{\rm Re}Z_i(0)}{R_Q}  \frac{2k_B T}{\hbar}. \quad\quad\label{Tphi-icoupl}
\end{eqnarray}

\subsection{Multiple sources of decoherence}

In this Section, we show that the total decoherence and relaxation rates of a quantum system
in the presence of several decoherence sources are \textit{not} necessarily the 
sums of the rates due to each of the mechanisms separately,
and that the corrections to additivity (mixing terms) can have both signs
\cite{burkard04c}.
To this end, we investigate the decoherence due to several sources in 
superconducting (SC) flux qubits; the general idea of the present analysis
may however be applied to other systems as well.
\begin{figure}[b]
\begin{minipage}{7.3cm}
\includegraphics[width=7cm]{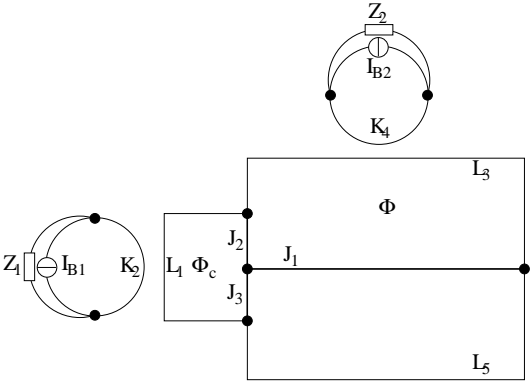}
\end{minipage}
\hfill
\begin{minipage}{9cm}
\caption{\label{circuit}
Circuit graph of the gradiometer qubit \cite{koch-u}, under the influence
of noise from two sources $Z_1$ and $Z_2$.
Branches of the graph denote Josephson junctions $J_i$,
inductances $L_i$ and $K_i$, current sources $I_{Bi}$,
and external impedances $Z_i$, and are connected by the nodes (black dots) of the
graph.
Inset:  A resistively-shunted Josephson junction (RSJ) $J_i$, represented by a thick 
line in the circuit graph, is modeled by an ideal junction (cross) with critical current 
$I_{ci}$, shunt resistance $R_i$, and junction capacitance $C_i$.}
\end{minipage}
\end{figure}

\subsubsection{Dissipative dynamics}
As an example, the gradiometer qubit drawn in Fig.~\ref{circuit}
with $n=3$ junctions will be discussed.
The gradiometer qubit is controlled by applying a magnetic flux $\Phi_c$ to the small loop
on the left by driving a current $I_{B1}$ in a coil next to it, and simultaneously 
by applying a magnetic flux $\Phi$ on one side of the gradiometer using $I_{B2}$.
The classical equations of motion of the SC circuit have the form
\begin{equation}
  \label{eq-mot}
  {\bf C} \ddot {\bf \bphi} = -\frac{\partial U}{\partial\bphi} - {\bf M}*\bphi ,
\end{equation}
where ${\bf C}$ is the capacitance matrix, $U$ the potential,
and ${\bf M}(t)$ the dissipation matrix.  
The convolution is defined as $(f*g)(t) = \int_{-\infty}^t f(t-\tau)g(\tau)d\tau$.
The dissipation matrix in the Fourier representation
can be found from circuit theory \cite{burkard04a} and has the form
\begin{equation}
  \label{Mcircuit}
  {\bf M}(\omega) = \bar{\bf m}\bar{\bf L}_Z(\omega)^{-1}\bar{\bf m}^T
                  = \bar{\bf m}\left({\bf L}_Z(\omega)+{\bf L}_{\rm c}\right)^{-1}\bar{\bf m}^T,
\end{equation}
where $\bar{\bf m}$ and ${\bf L}_{\rm c}$ denote 
real matrices that can be obtained from the circuit inductances.
One can assume the matrices ${\bf Z}$ and ${\bf L}_Z$ to be diagonal because
the impedances $Z_i$ are independent.

A Caldeira-Leggett Hamiltonian ${\cal H} = {\cal H}_S + {\cal H}_B + {\cal H}_{SB}$
can be constructed
that reproduces the classical dissipative equation of motion, Eq.~(\ref{eq-mot}),
and that is composed of parts for the system (S), given in Eq.~(\ref{Hamiltonian-S}),
for $m \ge 1$ harmonic oscillator baths (B),  and for the system-bath (SB) coupling,
\begin{eqnarray}
  {\cal H}_B &=& \sum_{j=1}^m \sum_\alpha \left(\frac{p_{\alpha j}^2}{2 m_{\alpha j}}+\frac{1}{2}m_{\alpha j} \omega_{\alpha j}^2 x_{\alpha j}^2\right),\label{HB}\\
  {\cal H}_{SB} &=&  \sum_\alpha \bphi ^T {\bf c}_\alpha {\bf x}_\alpha , \label{HSB}
\end{eqnarray}
where ${\bf x}_\alpha = (x_{\alpha 1}, \ldots, x_{\alpha m})$,
and ${\bf c}_\alpha$ is a real $n\times m$ matrix.
Defining the matrix spectral density ${\bf J}(\omega)$ of the environment,
where $\bdelta_{ij}({\bf X})\equiv \delta({\bf X}_{ij})$,  one obtains the relation
\begin{equation}
  {\bf J}(\omega) \equiv  \frac{\pi}{2}\sum_\alpha {\bf c}_\alpha{\bf m}_{\alpha}^{-1}{\bomega}_{\alpha}^{-1}
                                                \bdelta({\omega-\bomega_\alpha}){\bf c}_\alpha^T,
   = \left(\frac{\Phi_0}{2\pi}\right)^2\!\!{\rm Im}{\bf M}(\omega) 
                  = \sum_{j=1}^m J_j(\omega) {\bf m}_j(\omega){\bf m}_j(\omega)^T,  \label{J-mult}
\end{equation}
where the spectral decomposition (with the eigenvalues $J_j(\omega)>0$
and the real and normalized eigenvectors ${\bf m}_j(\omega)$)
of the real, positive, and symmetric matrix 
${\rm Im} {\bf M}(\omega)$ has been used.
The integer $m\leq n,n_Z$ denotes the 
maximal rank of ${\rm Im}{\bf M}(\omega)$, i.e., 
$m=\max_\omega \left({\rm rank} \left[ {\rm Im}{\bf M}(\omega)\right]\right)$.
Using Eq.~(\ref{J-mult}), 
and choosing $c_{\alpha ij} = \gamma_{\alpha j}{\bf m}_i(\omega_{\alpha j})$,
we find that $J_j(\omega)$ is the spectral density of the $j$-th
bath of harmonic oscillators in the environment,
$J_j(\omega) = (\pi/2)\sum_\alpha (\gamma_{\alpha j}^2/m_{\alpha j}\omega_{\alpha j})
\delta(\omega-\omega_{\alpha j})$.

From this model one obtains the Redfield equation, Eqs.~(\ref{Redfield-equation}), 
with a Redfield tensor, Eq.~(\ref{RGamma}), of the form
\begin{eqnarray}
{\rm Re}\Gamma_{lmnk}^{(+)} &=& \bphi_{lm}^T {\bf J}(|\omega_{nk}|) \bphi_{nk} \frac{e^{-\beta\omega_{nk}/2}}{\sinh(\beta |\omega_{nk}|/2)},\label{ReGamma}\\
{\rm Im}\Gamma_{lmnk}^{(+)} &=& -\frac{2}{\pi}P\!\!\int_0^\infty 
                                 \frac{\bphi_{lm}^T {\bf J}(\omega)\bphi_{nk}}{\omega^2-\omega_{nk}^2}
                                 \left(\omega-\omega_{nk}\coth\frac{\beta\omega}{2}\right),\nonumber\label{ImGamma}
\end{eqnarray}
where $\bphi_{nk} = \langle n|\bphi|k\rangle$.
For two levels $n=0,1$, and within the secular approximation, the relaxation
and decoherence rates $T_1^{-1}$ and $T_2^{-1}$ are found to be
\begin{eqnarray}
  T_1^{-1} &=& 4 \bphi_{01}^\dagger {\bf J}(\omega_{01})\bphi_{01} \coth\left(\frac{\beta\omega_{01}}{2}\right) 
            =  4 \sum_{j=1}^m |\bphi_{01}\!\cdot {\bf m}_j(\omega_{01})|^2 J_j(\omega_{01}) \coth\left(\frac{\beta\omega_{01}}{2}\right) ,\label{T1-mult}\\
  T_\phi^{-1} &=& \frac{2}{\beta}\lim_{\omega\rightarrow 0} (\bphi_{00}-\bphi_{11})^\dagger\frac{{\bf J}(\omega)}{\omega}(\bphi_{00}-\bphi_{11}) 
            =  \frac{2}{\beta} \sum_{j=1}^m  |{\bf m}_j(0)\cdot (\bphi_{00}-\bphi_{11})|^2  \left.\frac{J_j(\omega)}{\omega}\right|_{\omega\rightarrow 0},\label{Tphi-mult}
\end{eqnarray}
where we have used the spectral decomposition, Eq.~(\ref{J-mult}), to obtain the second 
equality in each line.

\subsubsection{Mixing Terms}

In the case where ${\bf L}_{\rm c}$ is diagonal, or if its off-diagonal elements
can be neglected because they are much smaller than ${\bf L}_Z(\omega)$ for
all frequencies $\omega$,
one finds, using Eq.~(\ref{Mcircuit}), that the contributions due to different 
impedances $Z_i$ are independent, thus $m=n_Z$ and
${\bf M}(\omega) = \bar{\bf m}\bar{\bf L}_Z(\omega)^{-1}\bar{\bf m}^T 
= \sum_j \bar{\bf m}_j \bar{\bf m}_j^T i\omega/(Z_j(\omega)+i\omega L_{jj})$,
where ${\bf m}_j = \bar{\bf m}_j$ is simply the $j$-th column of the matrix $\bar{\bf m}$
and $L_{jj}$ is the $j$-th diagonal entry of ${\bf L}_{\rm c}$.
As a consequence, the total rates $1/T_1$ and $1/T_\phi$ are the sums of the individual rates,
$1/T_1^{(j)}$ and $1/T_\phi^{(j)}$, where
\begin{eqnarray}
    \frac{1}{T_1^{(j)}} &=& 4 \left(\frac{\Phi_0}{2\pi}\right)^2 \!\! |\bphi_{01}\cdot \bar{\bf m}_j|^2 {\rm Re}\frac{\omega_{01}\coth\left(\beta\omega_{01}/2\right)}{Z_j(\omega_{01})+i\omega_{01} L_{jj}}, \quad\quad\label{T1-i}\\
    \frac{1}{T_\phi^{(j)}} &=& \frac{2}{\beta}\left(\frac{\Phi_0}{2\pi}\right)^2 |\bar{\bf m}_j\cdot (\bphi_{00}-\bphi_{11})|^2 {\rm Re}\frac{1}{Z_j(0)}. \label{Tphi-i}
\end{eqnarray}
In general, the situation is more complicated because current fluctuations
due to different impedances are mixed by the presence of the circuit.
In the regime ${\bf L}_{\rm c} \ll {\bf L}_Z(\omega)$, 
$\bar{\bf L}_Z^{-1}$ can be expanded as
\begin{equation}
  \label{series1}
  \bar{\bf L}_Z^{-1}  =  \left({\bf L}_Z(\omega)+{\bf L}_{\rm c}\right)^{-1}
          = {\bf L}_Z^{-1} - {\bf L}_Z^{-1}{\bf L}_{\rm c}{\bf L}_Z^{-1}
                           + {\bf L}_Z^{-1}{\bf L}_{\rm c}{\bf L}_Z^{-1}{\bf L}_{\rm c}{\bf L}_Z^{-1}
                           - \cdots \,.
\end{equation}
The series Eq.~(\ref{series1}) can be partially resummed,
\begin{equation}
  \label{partialresum}
  \bar{\bf L}_Z^{-1}(\omega) = {\rm diag}\left(\frac{i\omega}{Z_j(\omega) + i\omega L_{jj}} \right) 
                               + {\bf L}_{\rm mix}^{-1}(\omega),
\end{equation}
where the first term simply gives rise to the sum of the individual 
rates, as in Eqs.~(\ref{T1-i}) and (\ref{Tphi-i}), while the second term
gives rise to mixed terms in the total rates.   
The rates can therefore be decomposed as ($X=1,2,\phi$)
\begin{equation}
  \label{Tdec}
  \frac{1}{T_X}    = \sum_j \frac{1}{T_X^{(j)}} + \frac{1}{T_X^{\rm (mix)}}.
\end{equation}
One finds for the mixing term in the relaxation rate
\begin{equation}
  \frac{1}{T_1^{\rm (mix)}} = 4 \!\! \left(\frac{\Phi_0}{2\pi}\right)^2\!\!\!\bphi_{01}^\dagger \bar{\bf m}{\rm Im}{\bf L}_{\rm mix}^{-1}(\omega_{01})\bar{\bf m}^T\bphi_{01} \coth\!\!\left(\frac{\beta\omega_{01}}{2}\right).\label{T1-mix}
\end{equation}
One can show that there is no mixing term in the pure dephasing rate, i.e., $1/T_\phi^{\rm (mix)} = 0$,
and hence $T_2^{\rm (mix)} = 2T_1^{\rm (mix)}$.

In the case of two external impedances, $n_Z=2$,
Eq.~(\ref{series1}) can be completely resummed, with the result
\begin{widetext}
\begin{equation}
  \label{Lmix-grad}
  {\bf L}_{\rm mix}^{-1}(\omega) = \frac{L_{12}}{(Z_1(\omega)/i\omega+L_{11})(Z_2(\omega)/i\omega+L_{22})-L_{12}^2}
                           \left(\begin{array}{c c}
                               \frac{L_{12}}{Z_1(\omega)/i\omega+L_{11}}  &  -1\\
                               -1                           &  \frac{L_{12}}{Z_2(\omega)/i\omega+L_{22}}
                           \end{array}\right)
                         \approx  -\frac{\omega^2 L_{12}}{Z_1(\omega) Z_2(\omega)} \sigma_x ,
\end{equation}
\end{widetext}
where $L_{ij}$ are the matrix elements of ${\bf L}_{\rm c}$ and
where the approximation in Eq.~(\ref{Lmix-grad}) holds up to $O({\bf Z}^{-3})$.
In lowest order in $1/Z_i$, one finds, 
\begin{equation}
  \frac{1}{T_1^{\rm (mix)}} = -\!\left(\!\frac{\Phi_0}{2\pi}\!\right)^2 
 {\rm Im}\frac{8\varphi_{12}\omega_{01}^2 L_{12} }{Z_1(\omega_{01}) Z_2(\omega_{01})}  
\coth\left(\!\frac{\beta\omega_{01}}{2}\!\right).
\end{equation}
with $\varphi_{12}=(\bphi_{01}\cdot \bar{\bf m}_1)(\bphi_{01}\cdot \bar{\bf m}_2)$.

\begin{figure}[b]
\begin{minipage}{8.8cm}
\includegraphics[width=8.5cm]{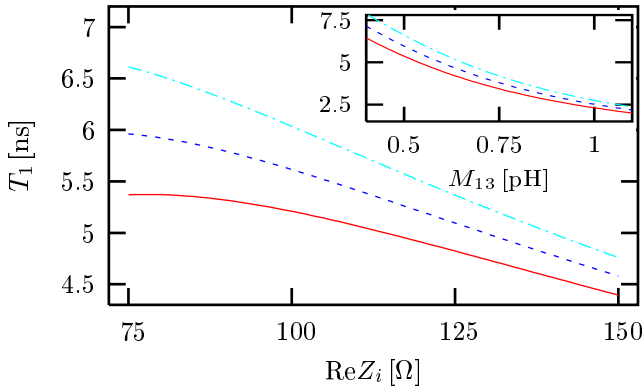}
\end{minipage}
\hfill
\begin{minipage}{8cm}
\caption{\label{fig2}
The relaxation rate $T_1$ without the mixing term (dashed blue line),
and including the mixing term for $R_{\rm im}=+10\,{\rm k}\Omega$ (solid red line)
and $R_{\rm im}=-10\,{\rm k}\Omega$ (dot-dashed light blue line), 
for $M_{13}=0.5\,{\rm pH}$ as a function of ${\rm Re}Z_i$.
Inset:  $T_1$ for $R={\rm Re Z}_i=75\,\Omega$
for a range of mutual inductances $M_{13}$.}
\end{minipage}
\end{figure}

If $R_i\equiv Z_i(\omega_{01})$ are real (pure resistances) then, as predicted above,
the imaginary part of the second-order term
in Eq.~(\ref{Lmix-grad}) vanishes, and we resort to third order,
\begin{equation}
  {\rm Im}{\bf L}_{\rm mix}^{-1} = \frac{\omega^3 L_{12}}{R_1 R_2}\left(\begin{array}{c c}
                               \frac{L_{12}}{R_1} & \frac{L_{11}}{R_1}+\frac{L_{22}}{R_2}\\
                               \frac{L_{11}}{R_1}+\frac{L_{22}}{R_2} & \frac{L_{12}}{R_2}
                           \end{array}\right),
\end{equation}
neglecting terms in $O(R_j^{-4})$.
If $L_{12}\ll L_{jj}$, we obtain
${\rm Im}{\bf L}_{\rm mix}^{-1} \approx 
(\omega^3 L_{12}/R_1 R_2) (L_{11}/R_1+L_{22}/R_2)\sigma_x$,
and
\begin{equation}
  \frac{1}{T_1^{\rm (mix)}} = \left(\frac{\Phi_0}{2\pi}\right)^2\frac{8\omega_{01}^3 L_{12}}{R_1 R_2} \left(\frac{L_{11}}{R_1}\!+\!\frac{L_{22}}{R_2}\right)\varphi_{12} \coth\left(\frac{\beta\omega_{01}}{2}\right).
\end{equation}

For the gradiometer qubit (Fig.~\ref{circuit}), we find
$L_{12}\approx M_{12}M_{13}M_{34}/L_1 L_3$, $L_{11}\approx L_2$, $L_{22}\approx L_4$,
where $L_k$ denotes the self-inductance of branch $X_k$ ($X$=$L$ or $K$) and $M_{kl}$
is the mutual inductance between branches $X_k$ and $X_l$,
and where we assume $M_{ij} \ll L_k$.
The ratio between the mixing the single-impedance contribution scales as
\begin{equation}
  \frac{1/T_1^{{({\rm mix})}}}{1/T_1^{(j)}} 
       \approx \frac{\omega_{01}^2 L_{12} L}{R^2},
\end{equation}
where we have assumed $R_1\approx R_2 \equiv R$, $L_{11}\approx L_{22} \equiv L$,
and $\bphi_{01}\cdot \bar{\bf m}_1 \approx \bphi_{01}\cdot \bar{\bf m}_2$.

The relaxation time $T_1$ was calculated at a temperature $T \ll \hbar\omega_{01}/k_B$ 
for the circuit Fig.~\ref{circuit},
for a critical current $I_c=0.3\,\mu {\rm A}$ for all junctions, and for the
inductances $L_1=30\,{\rm pH}$, $L_3=680\,{\rm pH}$, $L_2=L_4=12\,{\rm nH}$,
$M_{12}\simeq \sqrt{L_1 L_2}$, $M_{34}\simeq \sqrt{L_3 L_4}$ (strong inductive coupling),
$M_{35}=6\,{\rm pH}$, with $\omega_{01}=2\pi\cdot 30\,{\rm GHz}$, 
and with the impedances $Z_1=R$, $Z_2=R + i R_{\rm im}$, where $R$ and $R_{\rm im}=\pm 10\,{\rm k}\Omega$
are real ($R_{\rm im}>0$ corresponds to an inductive, $R_{\rm im}<0$ to a capacitive character of $Z_i$).  
In Fig.~\ref{fig2}, $T_1$ was plotted with and without mixing for a fixed value of $M_{13}=0.5\,{\rm pH}$
and a range of $R={\rm Re}Z_i$.
In the inset of Fig.~\ref{fig2}, $T_1$ (with mixing) and 
$((T_1^{(1)})^{-1}+(T_1^{(2)})^{-1})^{-1}$ (without mixing) are plotted for $R=75\,\Omega$ for a range
of mutual inductances $M_{13}$;  for this plot, the double minima of the 
potential $U$ and $\bphi_{01}$ were computed numerically for each value of $M_{13}$.
The plots (Fig.~\ref{fig2}) clearly show that summing the decoherence rates without taking into
account mixing term can both underestimate or overestimate the relaxation rate $1/T_1$, leading 
to either an over- or underestimate of the relaxation and decoherence times $T_1$ and $T_2$.



\section{Entanglement}
\label{entanglement}

A composite system is entangled if its wavefunction $\Psi$
 cannot be expressed as a tensor product $\Psi_A\otimes\Psi_B$
 of wavefunctions $\Psi_A$ and $\Psi_B$ for the parts A and B of the system
 (a more general, but similar, definition exists for mixed states).
A variety of quantum communication scenarios \cite{bennett00}, 
some of which have been
implemented successfully in quantum optics, require
  maximally entangled states of two qubits,
  also known as EPR pairs \cite{einstein35},
  such as the spin singlet,
  \begin{equation}
    \label{singlet}
    \ket{S} = \frac{1}{\sqrt{2}}\left(\spupdown - \spdownup \right).
  \end{equation}
The triplet state $\ket{T_0}= \spupdown + \spdownup$
  is another maximally entangled state, while the other two triplet states
  $\ket{T_+} = \spupup$ and $\ket{T_-} = \spdowndown$ are not entangled.
An important feature of these states is that they are non-local, in the 
sense that they violate Bell's inequalities \cite{bell66,mermin93}.
A universal quantum computer can, by definition, produce arbitrary quantum
 states, and, in particular, entangled ones such as the singlet $\ket{S}$.
E.g., the square root of swap gate $S$, see Eq.~(\ref{eq:SSWAP}), 
 has the ability to turn the
 unentangled state $\spupdown$ into a maximally entangled one,
 $S\spupdown = (\spupdown - i\spdownup)/(1-i)$, which is equivalent
 (up to a single-qubit operation) to the singlet.
There may be cases in which only certain entangled states are
 required for quantum communication, while 
 quantum computation itself does not need to be performed.  
In this case, a physical device
 dedicated to the task of producing entangled states of some sort may be sufficient.
We call devices of this sort {\it entanglers} \cite{burkard00b} and
 discuss a number of conceivable implementations of entanglers in solid state systems below.

Experiments with entangled 
photons have tested Bell's inequalities \cite{aspect82},
and various quantum communication protocols, such as 
dense coding \cite{mattle96} and quantum teleportation
\cite{bouwmeester97,boschi98}.  However, none of these protocols
been implemented so far with massive particles (such as electrons).

Unfactorizable states like Eq.~(\ref{singlet}) are very common in
solid-state systems.  Interacting many-particle systems possess very
complicated and entangled ground states.  Not all of these are
necessarily useful for quantum information processing, though,
because (i) it is essential that there is a physical mechanism to
extract and separate a pair of entangled particles from the many-body 
system in such a way that they can be used for quantum communication,
and (ii) for indistinguishable particles, not all states that ``look
entangled'' really are.  A measure of entanglement which excludes pure 
antisymmetrization was defined in \cite{schliemann01a,schliemann01b}.

\subsection{Production of entangled electrons}\label{production}

\subsubsection{Superconductor-normal junctions}\label{Andreev}
\begin{figure}[b]
\begin{minipage}{6.3cm}
\includegraphics[width=6cm]{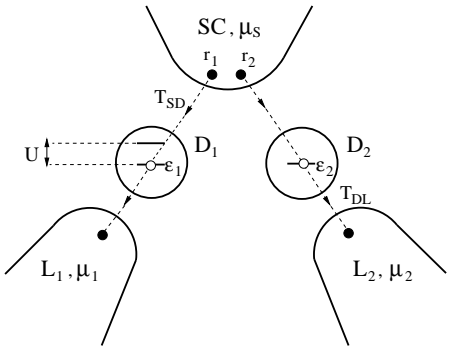}
\end{minipage}
\hfill
\begin{minipage}{9cm}
\caption{The Andreev entangler. 
A Cooper pair is split up into two entangled electrons 
which hop with amplitude $T_{SD}$ from two points ${\bf r}_{1}$,
${\bf r}_{2}$ of the superconductor, SC, 
(distance $\delta r=|{\bf r}_{1}-{\bf r}_{2}|$)
onto two dots $\rm D_{1,2}$ by means of Andreev tunneling.
The dots are coupled to normal leads ${\rm L_{1,2}}$
with tunneling amplitude $T_{DL}$. In order to maximize
the efficiency of the device, we require asymmetric barriers,
$|T_{SD}|/|T_{DL}|\ll 1$.
The chemical potentials of the SC and leads are $\mu_{l}$ and
$\mu_{S}$.
\label{entangler}}
\end{minipage}
\end{figure}

A superconductor (SC) with s-wave pairing symmetry contains an entire ``reservoir''
of spin singlet states as in Eq.~(\ref{singlet})
in the form of Cooper pairs that form the SC condensate \cite{schrieffer64}.
It is thus natural to think that such a system can act
as a source of spin-entangled electrons. 
A proposed setup \cite{recher01} is shown in Fig.~\ref{entangler}. 
The SC is held at the chemical potential $\mu_{S}$, and is weakly coupled by tunnel
barriers to two separate quantum dots $D_{1}$ and $D_{2}$ which
are in turn weakly coupled to Fermi liquid leads $L_{1}$ and
$L_{2}$, both held at the same chemical potential
$\mu_{1}=\mu_{2}$. The tunneling amplitudes between SC and dots,
and dots and leads, are denoted by $T_{SD}$ and $T_{DL}$.

Andreev tunneling is a process in which two electrons (one with spin
up and one with spin down) can tunnel coherently through a normal
barrier, while at the same time, single-particle tunneling as
suppressed \cite{glazman93}.
The setup Fig.~\ref{entangler} with the intermediate quantum dots
is designed to force two electrons
from a Cooper pair to tunnel coherently into {\em separate} leads
rather than both into the same lead.
The double occupation of a quantum dot is suppressed by the
Coulomb blockade mechanism \cite{kouwenhoven97c}.

The flow of entangled electrons from the SC via the dots to the leads
is controlled by applying a bias voltage $\Delta\mu=\mu_{S}-\mu_{l}>0$. 
The chemical potentials $\epsilon_{1}$ and $\epsilon_{2}$ of the two 
quantum dots can be
tuned by external gate voltages \cite{kouwenhoven97c} such that the
coherent tunneling of two electrons into different leads is at
resonance, while coherent tunneling of two electrons into the same lead is
suppressed by the on-site Coulomb repulsion $U$ of a quantum dot.

The relevant parameters describing the device and the desired regime 
of operation are discussed in \cite{recher01}.
It is required that the barriers of the dots are asymmetric, 
$|T_{SD}|\ll |T_{DL}|$, temperature is sufficiently small,
$\Delta\mu>k_{B}T$, and 
$\Delta, U, \delta\epsilon > \Delta\mu > \gamma_{l},  k_{B}T$,
and $\gamma_{l}>\gamma_{S}$,
where $\delta\epsilon$ is the single-level spacing of the dots,
$\Delta$ is the SC energy gap,
and $\gamma_{l}=2\pi\nu_{l}|T_{DL}|^2$ are the dot-lead tunnel rates.
The figure of merit for the device is the ratio
between the desired current $I_1$of pairwise entangled electrons
tunneling into {\em different} leads and 
the unwanted current $I_2$ of electron pairs that end up in 
the {\em same} lead \cite{recher01},
\begin{equation}
\label{current-ratio}
\frac{I_{1}}{I_{2}}=
\frac{2{\cal E}^2}{\gamma^2}
\left[\frac{\sin(k_{F}\delta r)}{k_{F}\delta r}\right]^2\,e^{-2\delta 
r/\pi\xi},
\quad\quad\quad
   \frac{1}{{\cal E}}=\frac{1}{\pi\Delta}+\frac{1}{U},
\end{equation}
where $k_F$ denotes the Fermi wavevector, 
$\gamma = \gamma_1 + \gamma_2$,
and $\xi$ the SC coherence length.

The desired current $I_1$ decreases exponentially with increasing
distance $\delta r=|{\bf r}_1-{\bf r}_2|$ between the tunneling points
on the SC, the scale given by the superconducting coherence length $\xi$.
With $\xi$ typically being on the order of $\mu {\rm m}$, 
this does not pose severe restrictions for a conventional s-wave SC.
In the important case $0\leq\delta r\sim\xi$ the suppression is only
polynomial $\propto 1/(k_{F}\delta r)^2$, with $k_{F}$ being the Fermi
wavevector in the SC.
One also observes that the effect of the quantum dots consists
in the suppression
factor $(\gamma/{\cal E})^2$ for tunneling into the same lead.
One therefore has to impose the additional condition
$k_F\delta r < {\cal E}/\gamma$,
which can be satisfied for small dots with ${\cal E}/\gamma\approx 100$ 
and $k_F^{-1}\approx 1\, {\rm \AA}$.

As an alternative to quantum dots as a means to separate the
entangled electrons from the SC, it hsa been proposed to use
a Luttinger liquid (see Sec.~\ref{LL} below) or a resistive 
lead where the dynamical Coulomb blockade effect helps to separate
the electron pair \cite{recher03}.

\subsubsection{Superconductor--Luttinger liquid junctions} \label{LL}

In the Andreev entangler (Sec.~\ref{Andreev}), entangled electron 
pairs are separated by the Coulomb repulsion in quantum dots that 
are attached to the SC which acts as a source of entangled spin singlets.
In related work \cite{recher02a,recher02b} and \cite{bena02}, 
it was suggested that
the strong Coulomb interactions in a one-dimensional conductor,
forming a Luttinger liquid \cite{tsvelik} can play the same role.
There is good experimental evidence for Luttinger liquid (LL) 
behavior in carbon nanotubes \cite{bockrath99}.

The setting discussed in \cite{recher02a,recher02b} consists of a
conventional s-wave SC tunnel-coupled to the center (bulk) of two 
spatially separated, for all practical purposes infinitely extended,
one-dimensional wires (e.g., carbon nanotubes) each forming a 
separate LL.  While the Coulomb interaction within each wire is
essential for the separation of entangled pairs into distinct
wires, it is assumed that the interaction between carriers in 
different wires is negligible. 
In the absence of backscattering, the low energy excitations of 
the LL are long-wavelength charge and spin density oscillations 
propagating with velocities $u_{\rho}=v_{F}/K_{\rho}$ for the charge 
and $u_{\sigma}=v_{F}$ for the spin \cite{schulz99}, 
where $v_{F}$ is the Fermi velocity and $K_{\rho}<1$ due to 
interaction. 
Transfer of electrons from the SC to the LL-leads is described by a
tunneling Hamiltonian,
\begin{equation}
H_{T} = t_{0}\,\sum_{ns}\,\psi_{ns}^{\dagger}\Psi_{s}({\bf r}_{n})
+ {\rm H.c.},
\end{equation}  
where $\Psi_{s}({\bf r}_{n})$ annihilates an
electron with spin $s$ at the point ${\bf r}_{n}$ on the SC nearest to 
the LL-lead $n=1,2$, and
$\psi_{ns}^{\dagger}$ creates it again with same spin and amplitude
$t_{0}$ at the point $x_{n}$ in LL $n$.  
By applying a bias $\mu=\mu_{S}-\mu_{l}$ between the SC,
with chemical potential $\mu_{S}$, and the leads, held at the same 
chemical
potential $\mu_{l}$, a stationary current of pairwise spin-entangled 
electrons
can flow from the SC to the leads. 

As in the case of the Andreev entangler with attached quantum dots,
the performance of this device can be quantified by the ratio between
the two competing currents $I_1$ and $I_2$ (see Sec.~\ref{Andreev}).
From a T-matrix calculation \cite{recher02a,recher02b},
one obtains
in leading order in $\mu/\Delta$ and at zero temperature,
\begin{equation}
I_{1}=\frac{I_{1}^{0}}{\Gamma(2\gamma_{\rho}+2)}\frac{v_{F}}{u_{\rho}}
\left[\frac{2\Lambda\mu}{u_{\rho}}\right]^{2\gamma_{\rho}}\!\!,
\quad
I_{1}^{0}
=4\pi e\gamma^{2}\mu\,\frac{\sin^2(k_{F}\delta r)}{(k_{F}\delta r)^2}
e^{-2\delta r/\pi\xi},
\label{i_{1}}
\end{equation}
where $\Gamma$ is the Gamma function, $\Lambda$ a short distance
cut-off on the order of the lattice spacing in the LL,
$\gamma=2\pi\nu_{S}\nu_{l}|t_{0}|^2$ the probability per spin to
tunnel from the SC to the LL-leads, $\nu_{S}$ and $\nu_{l}$ the
energy DOS per spin for the superconductor and the LL-leads at the
chemical potentials $\mu_{S}$ and $\mu_{l}$, resp., and $\delta r$
the separation between the tunneling points on the SC. 
The current $I_{1}$ has a characteristic non-linear dependence 
on the voltage (electro-chemical potential $\mu$), $I_{1}\propto
\mu^{2\gamma_{\rho}+1}$, with an interaction dependent exponent
$\gamma_{\rho}=(K_{\rho}+K_{\rho}^{-1})/4-1/2>0$, which is the
exponent for tunneling into the bulk of a single LL, i.e.\
$\rho(\varepsilon)\sim|\varepsilon|^{\gamma_{\rho}}$,  where
$\rho(\varepsilon)$ is the single-particle
density of states \cite{schulz99}. 
In the non-interacting limit $\gamma_{\rho}=0$ the current is
given by $I_{1}^{0}$.
As in Sec.~\ref{Andreev}, the coherence length $\xi$ of the Cooper 
pairs should exceed $\delta r$ in order to obtain a finite measurable 
current. Note that the suppression of the current by 
$1/(k_{F}\delta r)^{2}$ can be considerably reduced with the use 
of lower dimensional SCs \cite{recher02a,recher02b}. 
The \textit{desired} current $I_1$ now has to be compared with
\textit{unwanted} current consisting of electron pairs tunneling
into the same lead and having $\delta r=0$. 
It is found \cite{recher02a,recher02b}
that the current $I_{2}$ for tunneling into the same
lead (1 or 2) is suppressed if $\mu<\Delta$ with the result, again
in leading order in $\mu/\Delta$,
\begin{equation}
I_{2}=I_{1}\sum\limits_{b=\pm 1}\,A_{b}
\,\left(\frac{2\mu}{\Delta}\right)^{2\gamma_{\rho b}},
\label{currentI222}
\end{equation}
where $A_{b}$ is an interaction dependent constant of order
one, and where $\gamma_{\rho +}=\gamma_{\rho}$, and
$\gamma_{\rho -}=\gamma_{\rho +}+(1-K_{\rho})/2>\gamma_{\rho +}$.
Note that in Eq.~(\ref{currentI222}) the current $I_{1}$ needs
to be evaluated at $\delta r=0$. In the non-interacting limit,
$I_{2}=I_{1}=I_{1}^{0}$ is obtained by putting
$\gamma_{\rho}=\gamma_{\rho b}=0$, and $u_{\rho}=v_{F}$. The
expression Eq.~(\ref{currentI222}) 
for $I_{2}$ shows that the unwanted injection of two
electrons into the same lead is suppressed compared to $I_{1}$ by
a factor of $(2\mu/\Delta)^{2\gamma_{\rho +}}$, where
$\gamma_{\rho+}=\gamma_{\rho}$, if both electrons are injected
into the {\em same} direction (left or right movers), or by
$(2\mu/\Delta)^{2\gamma_{\rho -}}$ if the two electrons travel in
{\em different} directions.  It is more likely that the two 
electrons move in the same direction than in opposite directions,
because $\gamma_{\rho -}>\gamma_{\rho +}$.  The suppression of the
current $I_{2}$ by $1/\Delta$ is a manifestation the two-particle
correlations in the LL when the electrons tunnel into the
same lead, which is similar to the Coulomb blockade effect
in the case of tunneling into quantum dots in Sec.~\ref{Andreev}. 
As the SC gap $\Delta$ becomes larger, the delay time between the 
arrivals of the two partner electrons of a Cooper pair becomes shorter, 
and the effect the first electron in the LL has on the second
electron tunneling into the LL increases.
Increasing the bias $\mu$ opens more available channels into
which the electron can tunnel, and therefore the effect of the SC gap
$\Delta$ is less pronounced. This correlation
effect disappears when interactions in the LL are absent,
$\gamma_{\rho}=\gamma_{\rho b}=0$.  
Experimental systems with LL behavior are e.g.\ metallic
carbon nanotubes with similar exponents as derived here
\cite{egger97,kane_c97}.

\subsubsection{Transport through quantum dots}
Entanglers with a single quantum dot attached to leads
with a very narrow bandwidth \cite{oliver02}
or with three coupled quantum dots \cite{saraga03}
have been proposed.
The idea behind these proposals is the harness the singlet
ground state of a single two-electron quantum dot by extracting 
the two electrons into two separate leads.
In both proposals, the separation is enhanced due to
two-particle energy conservation.
A double-dot turnstile device with time-dependent barriers
was proposed in \cite{hu04}.

\subsubsection{Coulomb scattering in a 2D electron system}
Scanning probe techniques can be applied to a two-dimensional (2D)
electron system formed in a semiconductor heterostructure
in order to monitor and control the flow of electrons
\cite{topinka00,topinka01}.
It has been proposed to generate spin-entangled pairs of
electrons using this technique to control Coulomb scattering in a interacting
2D electron system \cite{saraga04}.
At a scattering angle of $\pi/2$, it is expected that constructive two-particle 
interference leads to a enhancement of the spin-singlet scattering probability,
while the triplet scattering is suppressed. 
The scattering amplitudes have been calculated within the Bethe-Salpeter 
equation for small $r_s$ and allow an estimate of the achievable current 
of spin-entangled electrons \cite{saraga04}.

\subsubsection{Entangled Electrons in a Fermi Sea}
\label{ssecFS}

A particularly appealing aspect of the electron spin as a
carrier of quantum information is that it is attached to a charge, and
thus it can---in principle---be transported in a conductor.
One can therefore envision 
solid-state structures (e.g., on a microchip) where 
entanglement is produced in one location 
by one of the previously discussed methods,
and subsequently conveyed through a wire to the
location where entanglement is ``used up'' in some 
quantum information protocol.  While transporting qubits
is quite unproblematic in the case of photons as the
quantum information carriers (photons have been used
in many experiments to carry quantum information, 
even over distances of kilometers), 
it is less trivial for electrons.  When an electron is injected
into a metallic wire, it is immersed into a sea of other
electrons that (i) are indistinguishable from the injected 
electron, and (ii) constantly interact with all other electrons
(including the injected one) via the Coulomb interaction.  
In this Section, the stability of spin entanglement 
in the Fermi sea will be
discussed \cite{divincenzo99a,burkard00b}.  
It turns out that (i) the indistinguishability
of particles in the Fermi sea is actually not a problem
for the transport of spin qubits and (ii) the Coulomb
interaction does have some effect, which is however mitigated
by the phenomenon of \textit{screening} which is well-known 
in interacting Fermi liquids \cite{mahan93}.
More precisely, when an electron in the orbital state $q$ is 
added to a Fermi sea (lead), the quasiparticle weight of that 
state will be renormalized by $0\leq z_q\leq 1$, i.e.\ some 
weight $1-z_q$ to find the electron in
the original state $q$ will be distributed among all the other
electrons.  Such a rearrangement of the Fermi
system due to the Coulomb interaction happens very quickly, on a
time scale given by the inverse plasma frequency.

In order to analyze this effect quantitatively, the
entangled two-electron state injected into two distinct leads
1 and 2 can be written in second quantized notation,
\begin{equation}
\ket{ \psi_{{\bf n}{\bf n'}}^{t/s} } =\frac{1}{\sqrt{2}}\,
(a_{{\bf n}\uparrow}^\dagger a_{{\bf n'}\downarrow}^\dagger \pm
a_{{\bf n}\downarrow}^\dagger\, a_{{\bf n'}\uparrow}^\dagger\,)\,
  \ket{\psi_0},
\label{state}
\end{equation}
  where $s$ and $t$ stand for the singlet and triplet,
  $\ket{\psi_0}$ for the filled Fermi sea,
  and ${\bf n}=({\bf q},l)$, where ${\bf q}$ denotes the momentum 
  and $l$ the lead quantum number of an electron.
As usual, the operator $a^\dagger_{{\bf n}\sigma}$
  creates an electron in state ${\bf n}$ with spin $\sigma$.
After their injection, the two
electrons of interest are no longer distinguishable from the
electrons of the leads, and consequently the two electrons taken out of
the leads will, in general, not be the same as the ones injected.

The time evolution of the triplet or singlet states,
  interacting with all other electrons in the Fermi sea,
  is described by the 2-particle Green's function
$G^{t/s}({\bf 1}{\bf 2},{\bf 3}{\bf 4};t)
  =\langle\psi^{t/s}_{{\bf 1}{\bf 2}}, t|\psi^{t/s}_{{\bf 3}{\bf
  4}}\rangle$, which is related to the standard
2-particle Green's function $G( 1 2, 3 4;t)$ by
\begin{equation}
G^{t/s}({\bf 1}{\bf 2},{\bf 3}{\bf 4};t)
=-{\frac{1}{2}}\sum_{\sigma} \!\left[ G( 1 \bar{2}, 3 \bar{4};t) 
\pm G( 1\bar{2}, \bar{3} 4;t)\right] ,
\label{generaloverlap}
\end{equation}
where $n=({\bf n},\sigma)$ and $\bar{n}=({\bf n},-\sigma)$.
Assuming that at time $t=0$, a triplet (singlet) is prepared, then
the {\em fidelity of transmission}
\begin{equation}
P(t)=|G^{t/s}({\bf 1}{\bf 2},{\bf 1}{\bf 2};t)|^2
\end{equation}
is defined
as the probability for finding a triplet (singlet) after time $t$.

With Eq.~(\ref{generaloverlap}), the problem reduces to that of
evaluating the standard two-particle Green's function
$G( 1 2, 3 4;t) = - \langle T a_1(t) a_2(t) a_3^\dagger a_4^\dagger\rangle$
for a time- and spin-in\-de\-pen\-dent
Hamiltonian, $H=H_0+\sum_{i<j} V_{ij}$, where $H_0$ describes the free
motion of the $N$ electrons, and $V_{ij}$ is the bare Coulomb
interaction between electrons $i$ and $j$,
$\langle ...\rangle$ denotes the zero-temperature expectation value,
and $T$ is the time ordering operator.
One can assume that the leads are sufficiently separated, so that
the mutual Coulomb interaction can be neglected,
and thus the problem of finding an explicit value for $G(12,34;t)$ is 
simplified since the 2-particle vertex part vanishes
(i.e. the Hartree-Fock approximation is exact),
\begin{equation}
G(12,34;t)=G(13,t)G(24,t)-G(14,t)G(23,t),
\end{equation}
with the interacting single-particle Green's functions 
\begin{equation}
G({\bf n},t) =- i\langle \psi_0| T a_{{\bf n}}(t)a_{{\bf n}}^\dagger|\psi_0\rangle
\equiv G_l({\bf q},t),
\end{equation}
for each lead $l=1,2$. 
Substituting this back into Eq.~(\ref{generaloverlap}), one obtains
\begin{equation}
G^{t/s}({\bf 1}{\bf 2},{\bf 3}{\bf 4};t)= -
G({\bf 1},t)\, G({\bf 2},t)[ \delta_{{\bf 13}}\delta_{{\bf 24}}
\mp \delta_{{\bf 14}} \delta_{{\bf 23}}],
\end{equation}
where the upper (lower) sign refers to the spin triplet (singlet).
For the initial state ($t=0$), or in the absence of interactions,
one finds $G({\bf n},t)=-i$, and thus $G^{t/s}$ reduces to 
$\delta_{{\bf 13}}\delta_{{\bf 24}} \mp \delta_{{\bf 14}} \delta_{{\bf 23}}$, 
and $P=1$.
The evaluation of the (time-ordered) single-particle
Green's functions $G_{1,2}$ close to the Fermi surface yields the
standard result \cite{mahan93}
$G_{1,2}({\bf q},t) \approx -i z_q
\theta(\epsilon_q -\epsilon_F) e^{-i \epsilon_q t- \Gamma_qt}$,
which holds for $0\leq t\lesssim 1/\Gamma_q$, where $1/\Gamma_q$ is the
quasiparticle lifetime, $\epsilon_q=q^2/2m$ the quasiparticle energy
(of the added electron), and $\epsilon_F$ the Fermi energy.
For a two-dimensional electron system (2DES),
as e.g.\ in GaAs heterostructures,
$\Gamma_q \propto (\epsilon_q -\epsilon_F)^2
\log(\epsilon_q -\epsilon_F)$ \cite{quinn82} within the random phase
approximation (RPA), which accounts for screening and which is
obtained by summing all polarization diagrams \cite{mahan93}.  
The quantity of interest here is the quasiparticle weight,
$z_F=(1-\frac{\partial}{\partial \omega}
{\rm Re}\,\Sigma_{\rm ret}( k_F,\omega))^{-1}\big|_{\omega=0}$,
evaluated at the Fermi surface,
where $\Sigma_{\rm ret}(q,\omega)$ is the retarded irreducible self-energy.
For momenta ${\bf q}$ close to the Fermi surface and for identical leads 
($G_1=G_2$) we find $|G^{t/s}({\bf 1}{\bf 2},{\bf 3}{\bf 4};t)|^2
=\, z_F^4 \, |\, \delta_{{\bf 13}}\delta_{{\bf 24}}
\mp \delta_{{\bf 14}} \delta_{{\bf 23}}|^2$,
for times satisfying $0< t\lesssim 1/\Gamma_q$.
The fidelity therefore turns out to be
\begin{equation}
P=z_F^4
\end{equation}.

The irreducible self-energy $\Sigma_{\rm ret}$ and from it the quasiparticle
weight factor in two dimensions were evaluated explicitly \cite{burkard00b},
with the final result
\begin{equation}
z_F=\frac{1}{1+r_s \left(1/2 + 1/\pi \right)},
\label{qpweight}
\end{equation}
in leading order of the interaction parameter $r_s=1/k_F a_B$, where
$a_B=\epsilon_0\hbar^2/me^2$ is the Bohr radius.  In particular, in a
GaAs 2DES we have $a_B=10.3$ nm, and $r_s=0.614$, and thus we obtain
$z_F=0.665$.   The expansion in powers of $r_s$ for the exact
RPA self-energy can be summed up and evaluated numerically, with the
(more accurate) result $z_F=0.691155$ for GaAs.
The fidelity of transmission of the injected singlet in this case
is around $P\approx 0.2$. However, for large electron
density (small $r_{s}$), $P$ is closer to unity. Note the fidelity
of the (``postselected'') singlet pairs which can successfully be
removed from the Fermi sea, is equal to $1$, provided that (as
assumed here) the spin-scattering effects are negligible. That
this is indeed the case in GaAs 2DEGs is supported by experiments
where the electron spin has been transported phase-coherently over
distances of up to 
$100\,\mu {\rm m}$ \cite{kikkawa97,kikkawa98,awschalom99}.

\subsection{Detection of spin entanglement}
\label{detection}

Efforts are being made to produce spin entanglement in solid-state 
structures;  therefore, it is only natural to ask how one can \textit{test}
for the presence of entanglement in such a setting.
Here, a variety of tests for spin entanglement are described.
This investigation touches on fundamental
issues such as the non-locality of quantum mechanics, especially
for massive particles,
and genuine two-particle Aharonov-Bohm effects which are
fascinating topics in their own right.
The main idea in all of the following detection schemes
is to exploit the unique relation between
the symmetry of the orbital state and the two-electron 
spin state which makes it possible to detect an electron spin state
via the orbital (charge) degrees of freedom.

\subsubsection{Coupled quantum dots}

The first setup to be considered can be used to probe
  the entanglement of two electrons localized in a double-dot
  by measuring a transport current and its fluctuations, or current noise \cite{loss00}.
It is assumed that the double-dot is weakly coupled to in- and outgoing leads
  (at chemical potentials $\mu_{1,\,2}$) with tunneling amplitude ${\cal T}$,
  where the dots are shunted in parallel.
The regime of interest is (i) the Coulomb blockade regime \cite{kouwenhoven97c} 
where the charge on the dots is quantized and (ii) the cotunneling regime
\cite{averin92,konig97}, where single-electron tunneling is forbidden by
energy conservation.  The latter regime is defined by 
$U>|\mu_1\pm\mu_2|>J>k_BT, 2\pi \nu{\cal T}^2$
  where $U$ is the single-dot charging energy,
  $\nu$ the lead density of states, and $J$ the exchange coupling between the dots.
The current in the cotunneling regime is generated by a coherent virtual process 
  where one electron tunnels from a dot to, say, lead 2
  and then a second electron tunnels from lead 1 to this dot.
If the bias voltage is larger than the exchange coupling, $|\mu_1-\mu_2|>J$, 
elastic as well as inelastic cotunneling will occur. 
It will be assumed that ${\cal T}$ is small enough for the 
  double-dot to return to its equilibrium state after each tunneling event.
An electron can either pass through the upper or lower dot,
therefore a closed loop is formed by these two paths.
A magnetic flux then gives rise to an
  Aharonov-Bohm phase $\phi=ABe/\hbar$ ($A$ being the loop area)
  between the upper and the lower paths leading
  to quantum interference effects.
This transport setting is sensitive to the spin symmetry of 
the two-electron state on the double dot;
if the two electrons on the double-dot are in the {\it singlet state},
  then the tunneling current
  acquires an additional phase of $\pi$
  leading to a sign reversal
  of the coherent contribution
  compared to that for triplets.
In cotunneling current, this additional phase manifests
itself in the sign of an interference term \cite{loss00}
\begin{equation}
\label{eqnIDD}
I=e\pi\nu^2{\cal T}^4 \,
\frac{\mu_1-\mu_2} {\mu_1\mu_2}
\,\left(2 \pm\cos\phi\right),
\end{equation}
where the upper sign refers to the triplet states in the double-dot
  and the lower sign to the singlet state.
The shot noise is Poissonian with power $S(0) = -e|I|$, hence it 
has the same dependence on the state on the double dot.

The shot noise has also been calculated for finite frequencies
in \cite{loss00}, and it was found that again $S(\omega)\propto (2\pm\cos\phi)$, 
  and that the odd part of $S(\omega)$
  leads to slowly decaying oscillations of the noise
  in real time,
  $S(t) \propto \sin(\mu t)/\mu t$, $\mu=(\mu_1 +\mu_2)/2$,
  which can be ascribed to a charge imbalance on the double dot
  during an uncertainty time $\mu^{-1}$.

Note that while the scheme described above is able to distinguish
the states $\ket{S}$ and $\ket{T_0}$ on the dots, the three triplets 
$\ket{T_0}$, $\ket{T_+}$, and $\ket{T_-}$, can be further distinguished
by an orientationally inhomogeneous magnetic field which results in a
spin-Berry phase \cite{loss92,loss00} that leads to left, right or no
phase-shift in the Aharonov-Bohm oscillations of the current (noise).

\subsubsection{Coupled dots with SC leads}
\label{ssecSCLead}

A related scenario of double-dots (DD) has been considered in
 \cite{choi00}, where two quantum dots are again shunted
 in parallel between the leads, but without any direct coupling between them.
The two dots are assumed to be coupled via tunneling (with amplitude ${\cal T}$)
  to two superconducting leads.
It turns out that the s-wave SC energetically favors an entangled singlet-state 
on the dots.
In addition to this, the coupling to the SC provides a mechanism for detecting 
the spin state via the Josephson current through the double dot system.
In leading order $\propto {\cal T}^4$, the spin coupling is
described by a Heisenberg Hamiltonian \cite{choi00}
\begin{equation}
H_{\rm eff}
\approx J\,(1+\cos\varphi)\,
  \left({\bf S}_a\cdot{\bf S}_b-\frac{1}{4}\right) ,
\end{equation}
where $J\approx 2{\cal T}^2/\epsilon$,
  $\epsilon$ is the energy difference from the dot state to the Fermi level,
and $\varphi$ is the average phase difference across the SC-DD-SC junction.
The exchange coupling between the spins can be controlled by tuning the external
parameters ${\cal T}$ and $\varphi$, thus providing
another implementation of a two-qubit quantum gate (Sec.~\ref{universal-set}) 
or entangler (Sec.~\ref{production}).
The spin state (singlet or triplet) on the dot can be probed if the 
SC leads are joined with one additional (ordinary) Josephson junction with
  coupling $J'$ and phase difference $\theta$ into a SQUID.
The SC current $I_S$ flowing in this ring is given by \cite{choi00}
\begin{equation}
\label{eqnIsIjDD}
I_S/I_J
= \left\{\begin{array}{ll}
  \sin(\theta-2\pi{f}) + (J'/J)\sin\theta\, ,
  &\mbox{singlet}, \\
  (J'/J)\sin\theta\, ,
  &\mbox{triplets},
  \end{array}\right.
\end{equation}
where $I_J = 2eJ/\hbar$.
The spin state of the DD is now probed by
measuring the spin- and flux-dependent critical current
  $I_c=\max_{\theta} \{|I_S|\}$  by
biasing the system with a dc current $I$
until a finite voltage $V$ appears for $|I|>I_c$ (the SC goes into
the normal state).
Another interesting effect is long-distance coupling between
spins residing in dots separated by $\delta r$ which is induced by the 
presence of the SC.  The resulting exchange coupling is
\begin{equation}
J(\delta r) = J(0) \left[ \frac{\sin(k_F \delta r)}{k_F \delta r}  
\right] ^2
e^{-2\delta r/\xi}.
\end{equation}

\subsubsection{Beam splitter shot noise}\label{bsnoise}

Pairwise spin entanglement between electrons in two mesoscopic wires
can be detected from a charge current measurement after
transmission through an electronic beam splitter \cite{burkard00b}.
In this scheme, the singlet EPR pair Eq.~(\ref{singlet}) gives rise to 
an enhancement of the shot noise power (``bunching'' behavior), whereas 
the triplet EPR pair $\ket{T_0}$ leads to a suppression of noise 
(``antibunching''). 
This behavior can be anticipated from a textbook example:
the scattering theory of two identical particles in 
vacuum \cite{feynman65,ballentine90}.
There, the differential scattering cross-section $\sigma$ in the
center-of-mass frame can be expressed in terms of the scattering amplitude
$f(\theta)$  and scattering angle $\theta$ as
\begin{equation}
\sigma(\theta)=|f(\theta) \pm f(\pi-\theta)|^2
=|f(\theta)|^2+|f(\pi-\theta)|^2 \pm 2 Re f^*(\theta)f(\pi-\theta),
\end{equation}
the first two terms on the right being the
contributions which would be obtained if the particles were
distinguishable, and the third (exchange) term the contribution
due to the particles' indistinguishability. 
This last term gives rise to genuine two-particle interference effects.
Here, the plus (minus) sign applies to spin-1/2 particles in the
singlet (triplet) state, described by a (anti)symmetric orbital wave
function. The very same two-particle interference mechanism which
is responsible for the enhancement (reduction) of the scattering cross
section $\sigma(\theta)$ near $\theta=\pi/2$ also leads to an increase
(decrease) of the correlations of the particle number in the output
arms of a beam splitter \cite{loudon98}.
\begin{figure}
\centerline{
\includegraphics[width=11cm]{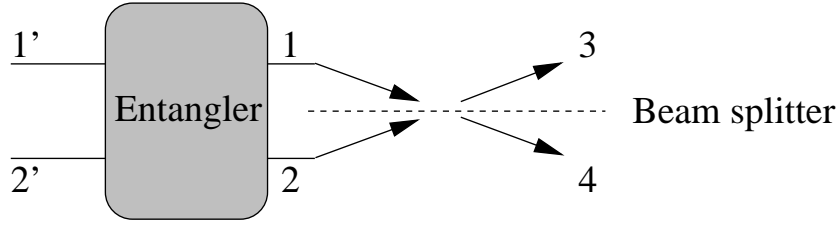}
}
\caption{\label{noise-fig1}\small
Setup for measuring the noise of entangled states.
Uncorrelated electrons are fed into the entangler (see text)
through the Fermi leads $1'$, $2'$ and are transformed into
pairs of electrons in the entangled singlet (triplet) state
$|\mp\rangle$, which are injected into leads $1$, $2$
(one electron of undetermined spin state into each lead).
The entanglement of the, say, spin singlet can then be detected
in an interference experiment using a beam splitter (with
no backscattering): Since the orbital wave function of
the singlet is symmetric, the electrons leave the scattering
region preferably in the same lead ($3$ or $4$). This correlation
(``bunching'') is revealed by an enhancement of the noise by a factor
of $2$ in the outgoing leads.}
\end{figure}

For the detection of spin entanglement of electrons carried
by two mesoscopic wires, we propose a
non-equilibrium transport measurement based on the set-up shown in
Fig.~\ref{noise-fig1}.
The beam splitter ensures that the electrons leaving
the entangler (see Sec.~\ref{production}) 
have an amplitude $t$ to be interchanged (without mutual interaction)
such that $0<|t|^2<1$. 
In the absence of spin scattering the noise measured in the
outgoing leads 3 and 4 exhibits bunching behavior for pairs of electrons
with a symmetric orbital wave function \cite{hanburybrown56}, i.e., for spin
singlets, while spin triplets induce antibunching behavior, 
due to their antisymmetric orbital wave function.
The latter situation has been considered for electrons in the
normal state both in theory \cite{buttiker90,buttiker92,martin92} and in
recent experiments \cite{liu98,henny99,oliver99,torres99}. 
The experiments 
have been performed in semiconductor nanostructures with geometries that are
closely related to the set-up proposed in Fig.~\ref{noise-fig1}
but without entangler.
It should be stressed here that if bunching (enhancement
of shot noise) is detected, it can be interpreted as a unique signature
for entanglement of the injected electrons,
since the maximally entangled singlet is the
only state leading to this effect.
The effect of interactions in the leads have already been assessed in
Sec.~\ref{ssecFS}.  In order to determine the shot noise of spin-entangled
electrons, the standard scattering matrix approach \cite{buttiker90,buttiker92},
extended to a situation with entanglement, is applied. 

We start by writing down the entangled incident state
$|\pm\rangle \equiv |\psi_{{\bf 1}{\bf 2}}^{t/s}\rangle$,
where we set ${\bf n}=(\varepsilon_n,n)$,
now using the electron energies $\varepsilon_n$ instead of the
momentum as the orbital quantum number in Eq.~(\ref{state}) and
where the operator
$a^\dagger_{\alpha \sigma}(\varepsilon)$ creates an incoming
electron in lead $\alpha$ with spin $\sigma$ and energy
$\varepsilon$.
The theory for the current correlations in a
multiterminal conductor \cite{buttiker90,buttiker92}
can easily be generalized
to the case of entangled scattering states, with the important consequence
that Wick's theorem does not apply.
The operator for the current carried by electrons
in lead $\alpha$ of a multiterminal conductor can be written as
\begin{equation}
I_{\alpha}(t) =
\frac{e}{h\nu} \sum_{\varepsilon \varepsilon ' \sigma} \left[
a_{\alpha \sigma}^\dagger(\varepsilon)a_{\alpha \sigma}(\varepsilon ')
- b_{\alpha \sigma}^\dagger(\varepsilon)
b_{\alpha \sigma} (\varepsilon ')\right]
\exp\left[i(\varepsilon-\varepsilon ')t/\hbar\right]\!,
\label{current_def}
\end{equation}
where the operators
$b_{\alpha \sigma}(\varepsilon)$ for the outgoing electrons are related to the operators
$a_{\alpha \sigma}(\varepsilon)$ for the incident electrons via
$b_{\alpha\sigma}(\varepsilon)=\sum_{\beta} s_{\alpha\beta}a_{\beta \sigma}(\varepsilon)$, 
where
$s_{\alpha\beta}$ denotes the scattering matrix.
The scattering matrix is assumed to be spin- and energy-independent.
Note that for a discrete energy spectrum in the leads, one can
normalize the operators $a_{\alpha \sigma}(\varepsilon)$ such that
$\{a_{\alpha \sigma}(\varepsilon),a_{\beta \sigma'}^\dagger(\varepsilon ')\}=
\delta_{\sigma\sigma'}\delta_{\alpha\beta}\delta_{\varepsilon\varepsilon '}/\nu$,
where the Kronecker symbol $\delta_{\varepsilon\varepsilon '}$ equals $1$ if 
$\varepsilon=\varepsilon '$ and $0$ otherwise.
Here, $\nu$ stands for the density of states in the leads.
Each lead is assumed to consist of only a single quantum channel;
the generalization to leads with several channels is straightforward but
not required here.
The current Eq.~(\ref{current_def}) can be expressed in terms of the
scattering matrix as
\begin{eqnarray}
  I_{\alpha}(t) &=& \frac{e}{h\nu}\sum_{\varepsilon \varepsilon ' \sigma} 
  \sum_{\beta\gamma}
 a_{\beta \sigma}^\dagger (\varepsilon) A_{\beta\gamma}^\alpha
a_{\gamma \sigma}(\varepsilon ') e^{i(\varepsilon-\varepsilon ')t/\hbar} , 
\label{current}\\
A_{\beta\gamma}^{\alpha} &=&
\delta_{\alpha\beta}\delta_{\alpha\gamma}
-s_{\alpha\beta}^{*} s_{\alpha\gamma}. 
\label{A2}
\end{eqnarray}
The correlation function between the currents $I_\alpha (t)$ 
and $I_\beta (t)$ in two leads $\alpha,\beta = 1,..,4$ of the BS
\begin{equation}
  \label{cross1}
  S_{\alpha\beta}^\chi(\omega) = \lim_{\tau\rightarrow\infty}
  \frac{h\nu}{\tau}\int_0^\tau\!\!\!dt\,\,e^{i\omega t}
  \,{\rm Re}\,
  \Tr \left[ \delta I_\alpha(t)\delta I_\beta(0) \chi \right],
\end{equation}
where $\delta I_\alpha = I_\alpha - \langle I_\alpha\rangle$,
$\langle I_\alpha\rangle=\Tr ( I_\alpha \chi )$,
$\nu$ is the density of states in the leads, and $\chi$
is the density matrix of the injected electron pair.
Here, only the zero-frequency correlator $S_{\alpha\beta}\equiv S^{\chi}_{\alpha\beta}(0)$
will be of interest (the dependence on $\chi$ was omitted).
Further evaluation with $\chi=\ket{\pm}\bra{\pm}$ yields
\begin{equation}
  S_{\alpha\beta}
   = \!\frac{e^2}{h\nu}\Big[\sum_{\gamma\delta}\!{}^{'}
    A_{\gamma\delta}^{\alpha}A_{\delta\gamma}^{\beta}
   \mp \delta_{\varepsilon_1,\varepsilon_2}
    \big(A_{12}^{\alpha}A_{21}^{\beta}\! +\!A_{21}^{\alpha}A_{12}^{\beta}
\big)\Big],\label{cross5}
\end{equation}
where $\sum_{\gamma\delta}^{\prime}$ denotes the sum over $\gamma=1,2$ and
all $\delta\neq\gamma$, and where again the upper (lower) sign refers to 
triplets (singlets).
Note that the autocorrelator $S_{\alpha\alpha}$ is the shot noise in lead $\alpha$.

The result Eq.~(\ref{cross5}) can now be applied to the set-up shown in
Fig.~\ref{noise-fig1} involving four leads, described by the single-particle
scattering matrix elements, $s_{31}=s_{42}=r$, and $s_{41}=s_{32}=t$,
where $r$ and $t$ denote the reflection and transmission amplitudes
at the beam splitter.
In the absence of backscattering, $s_{12}=s_{34}=s_{\alpha\alpha}=0$,
the noise correlations for the incident state $|\pm\rangle$ are
\begin{equation}
  \label{noise}
S_{33}=S_{44}=-S_{34}=2\frac{e^2}{h\nu}T\left(1-T\right)
  \left(1\mp \delta_{\varepsilon_1\varepsilon_2}\right),
\end{equation}
where $T=|t|^2$ denotes the transmittivity of the beam splitter.
For the remaining two triplet states $|\!\uparrow\uparrow\rangle$ 
and $|\!\downarrow\downarrow\rangle$ one also obtains Eq.~(\ref{noise})
with the upper sign.
The mean current in lead $\alpha$ is, both for singlets and triplets,
$\left|\langle I_\alpha\rangle\right| = e/h\nu$.
The noise-to-current ratio, or Fano factor, 
$F = S_{\alpha\alpha} /\left|\langle I_\alpha\rangle\right|$
is thus found to be
\begin{equation}
  \label{fano}
  F =  2eT(1-T)\left(1\mp \delta_{\varepsilon_1\varepsilon_2}\right),
\end{equation}
and correspondingly for the cross correlations.
Eq.~(\ref{fano}) implies that if electrons in the singlet
state $|-\rangle$ with equal energies,
$\varepsilon_1=\varepsilon_2$, are injected pairwise into the leads $1$ and $2$,
then the zero frequency noise is {\it enhanced} by a factor of two,
$F=4eT(1-T)$, compared to the shot noise of uncorrelated particles
\cite{buttiker90,buttiker92,martin92,khlus87,landauer89,lesovik89},
$F=2eT(1-T)$. 
This noise enhancement is due to {\it bunching} of electrons in the 
outgoing leads, caused by the symmetric orbital wavefunction of the spin 
singlet $|-\rangle$.
The triplet states $|+\rangle$, $|\!\uparrow\uparrow\rangle$, and
$|\!\downarrow\downarrow\rangle$ exhibit {\it antibunching}, i.e. a complete 
suppression of the noise, $S_{\alpha\alpha}=0$.
As already mentioned above, the noise enhancement for the singlet $|-\rangle$ is a
unique signature for entanglement (no unentangled state exists which can
lead to this phenomenon), therefore entanglement can be observed by
measuring the noise power of a mesoscopic conductor
as shown in Fig.~\ref{noise-fig1}.
The various triplet states $|+\rangle$, $|\!\uparrow\uparrow\rangle$, and
$|\!\downarrow\downarrow\rangle$ cannot be distinguished by the noise
measurement alone;  this distinction requires a measurement of the spins
of the outgoing 
electrons, e.g. by inserting spin-selective tunneling devices \cite{prinz98}
into leads $3$ and $4$.  
The same signature of entanglement as for the shot noise
can also be seen in the full counting statistics of the charge transport
\cite{taddei02}.

\subsubsection{Lower bounds for entanglement}\label{bounds}
\begin{figure}
  \begin{minipage}{6.3cm}
\includegraphics[width=6cm]{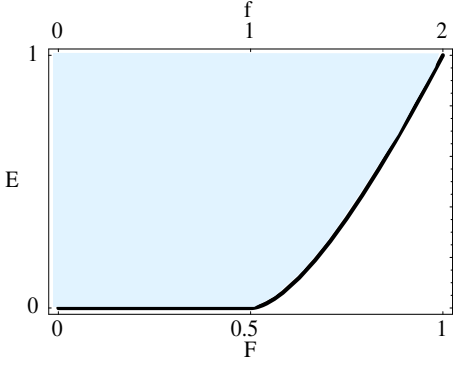}
  \end{minipage}
\hfill
  \begin{minipage}{8cm}
\caption{\label{fig:ent-bound}
Entanglement of formation $E$ of the electron spins
versus singlet fidelity $F$ and the reduced correlator
$f=S_{33}/2eI T(1-T)$.
The curve illustrates the exact relation for Werner states.  
For general states, the curve is a lower bound for $E$;
allowed values for $E$ and $f$ (or $F$) 
are in the shaded region.}
\end{minipage}
\end{figure}

Here, the result of the previous Section is extended
by providing a quantitative \textit{lower bound} for 
the \textit{amount} $E$ of spin entanglement
carried by individual pairs of electrons, related to the zero-frequency current correlators
when measured in a beam splitter setup (Fig.~\ref{fig:ent-bound}).
This result \cite{burkard03} therefore relates experimentally accessible quantities with
a measure for entanglement, the entanglement of formation $E$ \cite{bennett96a}.
Having information about $E$ is important since it quantifies the usefulness of
a bipartite state for quantum communication.

Starting form a general state, expressed in the singlet-triplet basis as
$\chi = F \ket{-}\bra{-} + G_0 \ket{+}\bra{+}
+ \sum_{i=\uparrow,\downarrow} G_i \ket{ii}\bra{ii} +  \Delta\chi$,
where $\Delta\chi$ are off-diagonal terms, one can decompose the
current correlators Eq.~(\ref{cross1}) as
\begin{equation}
S_{\alpha\beta}  \equiv  S^\chi_{\alpha\beta}  
                    =    F S_{\alpha\beta}^{\ket{\Psi_-}} + G_0 S_{\alpha\beta}^{\ket{\Psi_+}} 
                       + \sum_{i=\uparrow,\downarrow} G_i S_{\alpha\beta}^{\ket{ii}},
\end{equation}
where $S_{\alpha\beta}^{\ket{\Psi}}  \equiv  S_{\alpha\beta}^{\ket{\Psi}\bra{\Psi}}$.
The off-diagonal terms in $\Delta\chi$ do not enter $S_{\alpha\beta}$
because the operators $\delta I_\alpha(t)$ conserve total spin.
The coefficients $F$, $G_0$, $G_\uparrow$, and $G_\downarrow$ depend on the state
preparation and therefore on the entangler.
The more information about these coefficients can be gained,
the better the chance to measure the entanglement of $\chi$.
In Sec.~\ref{bsnoise} it was shown that the singlet state $\ket{-}$ 
gives rise to enhanced
shot noise (and cross-correlators) at zero temperature,
$S^{\ket{-}}_{33} = -S^{\ket{-}}_{34}= 2eI T(1-T) f$,
with the \textit{reduced correlator} $f=2$, as compared to the
``classical'' Poissonian value $f=1$.
All triplet states are noiseless,
$S^{\ket{+}}_{\alpha\beta} = S^{\ket{\uparrow\uparrow}}_{\alpha\beta} 
= S^{\ket{\downarrow\downarrow}}_{\alpha\beta} = 0$ ($\alpha,\beta=3,4$).
Therefore, both the auto- and cross-correlations are only due to the 
singlet component of the incident two-particle state,
\begin{equation}
  \label{eq:Sf}
  S_{33}= -S_{34} = F S^{\ket{\Psi_-}} = 2eI T(1-T) f,
  \quad\quad\quad f=2F.
\end{equation}

The entanglement of a bipartite {\em pure} state 
$\ket{\psi}\in{\cal H}_A\otimes{\cal H}_B$ can be measured by 
the von Neumann entropy $S_{\rm N}(\ket{\psi}) = -\Tr_B \rho_B \log\rho_B$
($\log$ in base 2) of the reduced density matrix 
$\rho_B=\Tr_A\ket{\psi}\bra{\psi}$, where $0\le S_{\rm N}\le 1$, 
$S_{\rm N}(\ket{\Psi_\pm})=S_{\rm N}(\ket{\Phi_\pm})=1$,
and $S_{\rm N}(\ket{\psi})=0 \Leftrightarrow \ket{\psi}=\ket{\psi}_A\otimes\ket{\psi}_B$.
Physically, if $S_{\rm N}(\ket{\psi}) \simeq N/M$ then $M\ge N$ 
copies of $\ket{\psi}$
are sufficient to perform, e.g., quantum teleportation of $N$ qubits
for $N,M\gg 1$ (similarly for other quantum communication protocols).
Generally, for a bipartite {\em mixed} state $\chi$
the entanglement of formation \cite{bennett96a} is
$E(\chi)=\min_{\{(\ket{\chi_i},p_i)\}\in \cal{E}(\chi)} 
\sum_i p_i S_{\rm N}(\ket{\chi_i})$,
where ${\cal E}(\chi)=\{(\ket{\chi _i} ,p_i) 
| \sum_i p_i \ket{\chi_i}\bra{\chi_i} =\chi \}$,
i.e., the least expected entanglement of any ensemble of pure states
realizing $\chi$.
A state with $E>0$ ($E=1$) is (maximally) entangled, and
neither local operations nor classical communication (LOCC) 
between A and B can increase $E$.

For an arbitrary mixed state of two qubits $\chi$, 
$E(\chi)$ is not a function
of only the singlet fidelity $F=\bra{-}\chi\ket{-}$.
However, $E(\chi)=E(F)$ for the Werner states \cite{werner89}
\begin{equation}
  \label{werner}
  \rho_F  =  F \ket{-}\bra{-}  +  \frac{1 - F}{3}
  \left( \ket{+}\bra{+} + \sum_{i=\pm}\ket{\Phi_i}\bra{\Phi_i}\right),
\end{equation}
the unique rotationally invariant states with singlet fidelity $F$,
where $\ket{\Phi_\pm} =  (\spupup \pm \spdowndown)/\sqrt{2}$.
The entanglement of formation of the Werner states is 
known \cite{bennett96a} as
$E(F)\equiv E(\rho_F)=H_2(1/2 +\sqrt{F(1-F)}) $ 
if $1/2<F\le 1$ and $E(F)\equiv E(\rho_F)=0$ if $0 \le F < 1/2$,
with the dyadic Shannon entropy $H_2(x)=-x\log x -(1-x)\log(1-x)$.
With Eq.~(\ref{eq:Sf}), one can now express $E(\rho_F)$ in terms
of the reduced correlator $f$ (Fig.~\ref{fig:ent-bound}).

This result can be generalized to arbitrary mixed states $\chi$ of two spins (qubits)
using the following trick.
Any state $\chi$ can be transformed into $\rho_F$ with $F=\langle \Psi_-|\chi|\Psi_-\rangle$ 
by a random bipartite rotation (``twirl'') \cite{bennett96a,bennett96b},
i.e.\ by applying $U\otimes U$ with a random $U\in SU(2)$.
Entanglement can only decrease (or remain constant) under the twirl
operation because it involves only LOCC, 
\begin{equation}
  \label{eq:bound}
  E(F)\le E(\chi).
\end{equation}
Thus, the entanglement of formation $E(F)$ of the 
corresponding Werner state represents
a \textit{lower bound} on $E(\chi)$ (Fig.~\ref{fig:ent-bound}).
A noise power exceeding $f=2F>1$ in the BS setup Fig.~\ref{noise-fig1} 
can therefore be interpreted as a sign of entanglement, $E(F)>0$, 
between the electron spins injected into leads 1 and 2.

The lower bounds that have been discussed so far are only useful
if one is assessing a source that aims at producing spin spinglet
entanglement.  However, it is possible in principle to extend
this result to arbitrary entangled states if a means of rotating
the spin of the carriers in one of the ingoing arms of the BS
is available;  such spin rotators could, e.g., be implemented
by making use of the Rashba spin orbit effect \cite{egues02}.

The relation between the shot noise $f$ and the entanglement $E$
has also been explored in a number of non-ideal situations,
in the presence of decoherence, backscattering, and thermally 
mixed input states \cite{burkard03}.

\subsubsection{Proposed tests of Bell's inequalities}
There have been a number of proposals to test Bell's inequalities \cite{bell66,mermin93}
with \textit{spin}-entangled electrons directly with a spin-sensitive detection scheme 
\cite{maitre00,kawabata01,lesovik01}
in contrast to the 
detection scheme in Secs.~\ref{bsnoise} and \ref{bounds} which only involves
the measurement of a charge current.
Combinations of the Andreev entangler setup with a Bell test were 
studied in \cite{chtchelkatche02,samuelsson04b}.

Bell tests for \textit{orbital} entanglement
with electron charge qubits in ballistic conductors
have also been proposed \cite{ionicioiu01}.  A scheme to generate 
two-particle orbital entanglement in a mesoscopic 
normal-superconductor system and to detect it via a violation
of a Bell inequality was analyzed in \cite{samuelsson03}.
A violation of a Bell inequality due to orbital entanglement
in a Hanbury Brown--Twiss setting was also proposed \cite{samuelsson04a}.

\subsection{Production of entangled photons}

The use of entangled photons from parametric down-conversion
in non-linear crystals \cite{kwiat95,kwiat99} has become a routine process
\cite{mattle96,bouwmeester97,boschi98}.
However, two disadvantages of these sources are (i) that they are 
quite inefficient (between $10^6$ and $10^{10}$ pump photons per
yield photon) and (ii) that they are stochastic, i.e., although the
rate (mean number of entangled pairs per second) is known, the 
precise instant of the emission cannot be controlled.
It would be desirable to have a source of entangled photons
that is both \textit{deterministic} and \textit{efficient}.
The use of electron-hole recombination in a single
QD was suggested in \cite{benson00,moreau01}.
Non-resonant excitation of a QD is expected to produce pairs of entangled 
photons with an efficiency (production rate/pump rate) that 
is about four orders of magnitude bigger 
than for parametric down-conversion \cite{moreau01}.
Ultraviolet entangled photons have recently been generated in the
semiconductor CuCl in a process called resonant hyper-parametric
scattering (RHPS) which involves the creation of a virtual 
biexciton state in the crystal \cite{edamatsu04}.  Although this is still
a stochastic source, RHPS is a very efficient process.

The production of polarization-entangled photons using the biexcitonic 
ground state has been investigated in \textit{two tunnel-coupled} QDs \cite{gywat02}.
Biexcitons are bound states of two excitons in a semiconductor, where each
exciton is the bound state of a negatively charged conduction-band electron 
and a positively charged valence-band hole.
Excitonic absorption in single QDs has been studied theoretically \cite{efros82},
and biexcitonic states in single QDs have been investigated
\cite{banyai88,takagahara89,bryant90,hu90,nair96,hawrylak99,kiraz02,santori02}.
Single excitons in coupled QDs have been observed in experiment \cite{bayer01,schedelbeck97}
and spin spectroscopy of excitons in QDs was performed using
polarization-resolved magnetophotoluminenscence \cite{johnston01}.
When discussing confined excitons, one needs to distinguish two regimes \cite{efros82}:
(i) The \textit{weak confinement} limit $a_X\ll a_e, a_h$, where $a_X$ is the radius of the 
free exciton and $a_e, a_h$ the electron and hole effective Bohr radii in the QD. 
In this regime, an exciton can be considered (as in the bulk material) as a boson
in an external confinement potential.
(ii) The \textit{strong confinement} limit $a_X\gg a_e, a_h$, where
electrons and holes are separately confined in the QD.
In this regime, electron-hole pairs cannot be considered as bosons anymore.
In bulk GaAs $a_X\approx 10$ nm, one is often in an intermediate regime 
$a_X\approx a_e, a_h$ for typical QD radii.  

\begin{figure}
  \begin{minipage}{11.3cm}
    \includegraphics[width=11cm]{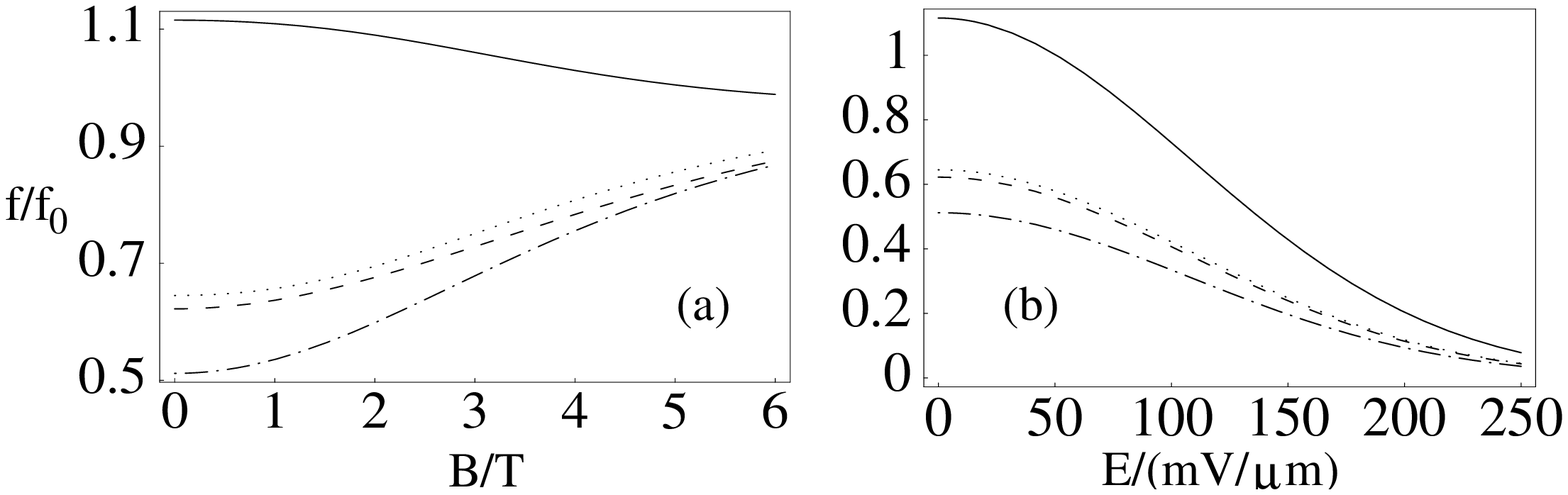}
  \end{minipage}
 \hfill
  \begin{minipage}{6.3cm}
    \includegraphics[width=6cm]{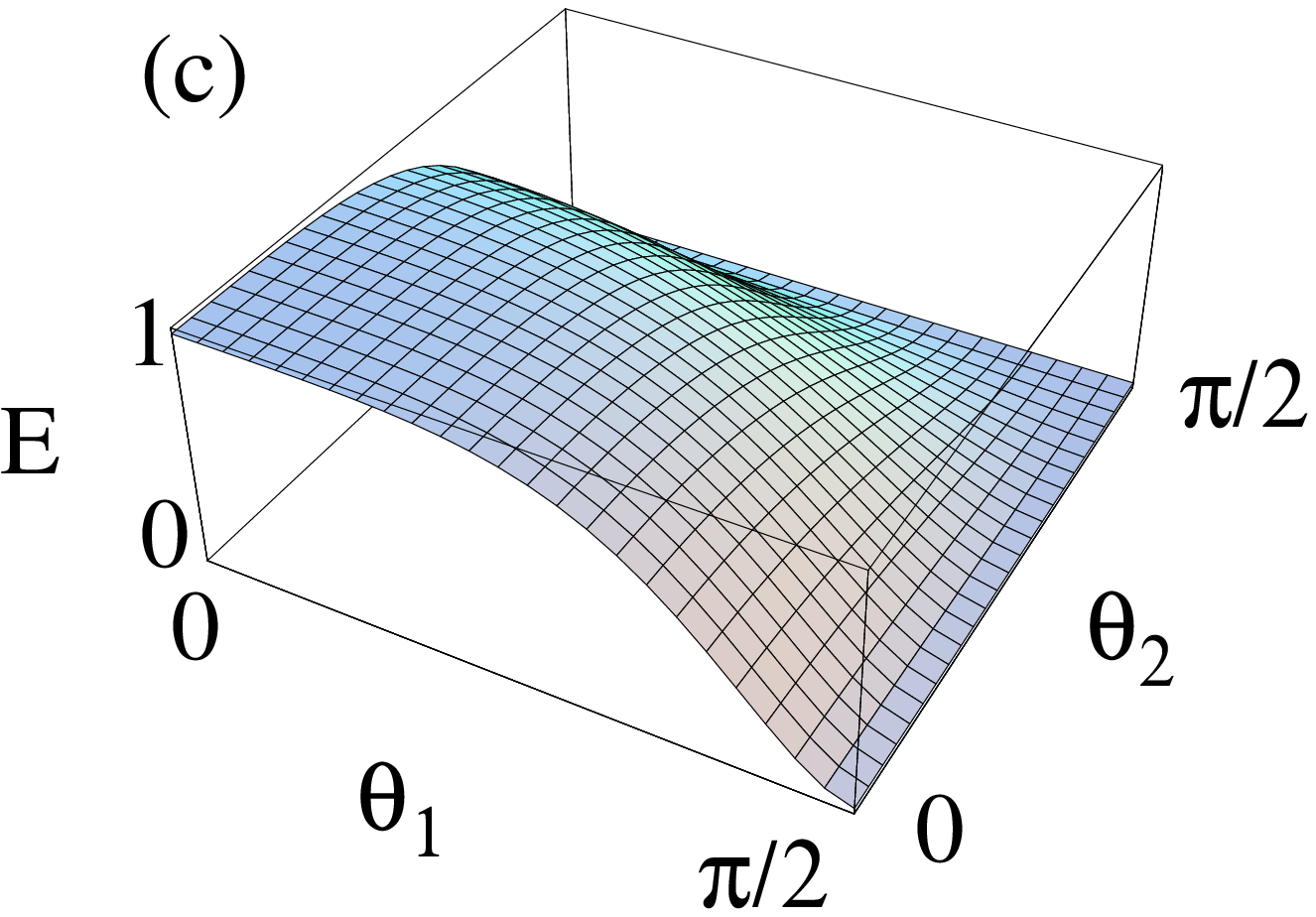}
  \end{minipage}
\caption{
(a) and (b): Oscillator strengths $f_{XX,X}$
for transitions between the biexciton states $|XX\rangle =|IJ\rangle$ 
and a single remaining
exciton on one QD in units of $f_0$  as a function of
(a) the magnetic field $B$ (in Tesla) at $E=0$ and 
(b) the electric field $E$ (in ${\rm mV/}\mu {\rm m}$) at $B=0$.
The plotted HL energies $E_{IJ}$ are $E_{ss}$ (solid line), 
$E_{st}$ (short-dashed), $E_{ts}$ (dot-dashed), 
and $E_{tt}$ (dotted), neglecting the
Zeeman energy.
The parameters were chosen for GaAs with $\eta=\omega_{e}/\omega_{h}=1/2$.
(c) Entanglement of formation $E$ as a function of the 
emission angles $\theta_1$ and $\theta_2$ of the two photons.
Only if both photons are emitted perpendicular to the plane 
($\theta_{1}=\theta_{2}=0$), the entanglement is maximal ($E=1$).
If at least one of them is emitted in-plane ($\theta_{i}=\pi/2$),
then the two photons are not entangled ($E=0$).}
\label{os1}
\end{figure}

Starting from a strong confinement ansatz and using the
Heitler-London (HL) approximation, the low-energy (spin-resolved)
biexciton spectrum  (in which the electrons
and holes each form either a spin singlet or triplet)
and the oscillator strengths Fig.~\ref{os1},
being a measure for the optical transition rates,
have been calculated \cite{gywat02}.
The HL ansatz is similar to the one used for electrons
in Sec.~\ref{HL} with the Hamiltonian
\begin{equation}
H  =  \sum_{\alpha=e,h}\sum_{i=1}^{2} h_{\alpha i} 
+H_{C}+H_{Z}+H_{E}, \label{h}
\end{equation}
where $h_{\alpha i} = (\mathbf{p}_{\alpha i}+q_{\alpha} \mathbf{A}(\mathbf{r}_{\alpha 
i})/c)^{2}/2m_{\alpha}+V_{\alpha}(\mathbf{r}_{\alpha i})$ is the 
single-particle Hamiltonian for the $i$-th electron
($\alpha=e$, $q_e=-e$) or  hole ($\alpha=h$, $q_h=+e$).
The Coulomb interaction is included by 
$H_{C}=(1/2)\sum_{(\alpha,i)\neq (\beta, j)} 
q_{\alpha}q_{\beta}/\kappa |\mathbf{r}_{\alpha i}-\mathbf{r}_{\beta j}|$, 
with a dielectric constant $\kappa$ (for bulk GaAs, $\kappa =13.18$).  
A magnetic field $\mathbf{B}$ in $z$ direction leads to 
orbital effects via the vector potential (in the symmetric gauge)
$\mathbf{A}=B(-y,x,0)/2$ and to the Zeeman term 
$H_{Z}= \sum_{\alpha,i}g_{\alpha}\mu_{B} B S_{z}^{\alpha i}$, 
where $g_{\alpha}$ is the effective g-factor of the electron (hole) 
and $\mu_{B}$ is the Bohr magneton. 

Two-particle wave functions for electrons and for holes separately 
are constructed according to the HL method, 
i.e. a symmetric ($|s \rangle^{\alpha}\equiv|I=0\rangle^{\alpha}$, 
spin singlet) and an 
antisymmetric ($|t \rangle^{\alpha}\equiv|I=1\rangle^{\alpha}$, spin triplet)
linear combination of two-particle states 
$|DD' \rangle_{\alpha}=|D \rangle_{\alpha}\otimes|D' \rangle_{\alpha}$,
\begin{equation}
|I\rangle^{\alpha} = 
N_{\alpha I}(|12\rangle_{\alpha} +(-1)^{I} |21 \rangle_{\alpha}),
\end{equation}
where $N_{\alpha I}=1/\sqrt{2(1 + (-1)^{I}|S_{\alpha}|^{2})}$ and $S_{\alpha}=\, _{\alpha}\langle 1 | 2 \rangle_{\alpha}$ denotes the overlap between the two orbital wave functions $|1 \rangle_{\alpha}$ and $|2 \rangle_{\alpha}$.  
From the electron and hole singlet and triplet, 
the four biexciton states $|IJ \rangle  =  |I\rangle ^{e} \otimes |J\rangle ^{h}$ 
can be formed, 
where $I=0$ ($1$) for the electron singlet (triplet) and $J=0$ ($1$) for the hole singlet (triplet).
The biexciton energies
\begin{equation}
E_{IJ}=\langle IJ|H|IJ\rangle = E^{0}+E^{Z}+E_{IJ}^{W}+E_{IJ}^{C},
\end{equation}
with $E_{IJ}^A\equiv \langle IJ|H_A|IJ\rangle$, can be calculated analytically within HL \cite{gywat02}.

The exciton and biexciton recombination rates are determined by the
oscillator strength $f$ which is a measure for the coupling between the dipole moment of
the exciton states and the electromagnetic field.
For a transition between the $N+1$ and $N$ exciton states $|N+1\rangle$ 
and $|N\rangle$, the oscillator strength is defined as 
\begin{equation}
f_{N+1,N}=\frac{2|p_{N {\bf k} \lambda}|^{2}}{m_{0}\hbar\omega_{N+1,N}}, 
\end{equation}
where $m_{0}$ is the bare electron mass, 
$\hbar\omega_{N+1,N}=E_{N+1}-E_{N}$, and
$p_{N {\bf k} \lambda} = \langle N+1|{\bf e}_{{\bf k} \lambda}\!\cdot {\bf p}|N\rangle$,
where ${\bf e}_{{\bf k} \lambda}$ is the unit polarization vector for a photon 
with momentum ${\bf k}$ and helicity $\lambda =\pm 1$, and ${\bf p}$ is the electron
momentum operator.
The inter-band momentum matrix element for a cubic crystal symmetry is given by
$M_{\sigma\lambda}(\theta)={\bf e}_{{\bf k} \lambda} \cdot {\bf p}_{cv}(\sigma) = 
p_{cv}(\cos(\theta)-\sigma\lambda)/2
\equiv p_{cv}m_{\sigma\lambda}(\theta)$,
where $\theta$ is the angle between $\mathbf{k}$ and the
normal to the plane of the 2D electron system
(assuming that the latter coincides with
one of the main axes of the cubic crystal), 
and $E_p=2p_{cv}^2/m_0$ ($=25.7\,{\rm eV}$ for GaAs).

The orbital momentum matrix element for the recombination
of an exciton to the vacuum and for transitions from a biexciton
state $|XX\rangle$ to a single exciton state $|X\rangle$ are
given in \cite{gywat02};
we give here our result for a
transition between the HL biexciton states $|XX\rangle = |IJ\rangle$ with one 
exciton on each QD and a single exciton in the final state 
$|X\rangle$,
\begin{equation}
|\langle IJ|{\bf e}_{{\bf k}\lambda}\!\cdot{\bf p}|X\rangle | 
  =  2 M_{\sigma\lambda}(\theta) \sqrt{N_{IJ}}
\left(C_{eh}\left[(\!-1\!)^{I\!+\!J}\!\!\!+\!S_{e}S_{h}\right]
+S_{eh}\left[(\!-1\!)^{J}\!S_{e}\!+\!(\!-1\!)^{I}\!S_{h}\right]\right).
\label{DD}
\end{equation}
The corresponding oscillator strength $f$ is plotted in Figs.~\ref{os1}a 
and~\ref{os1}b, in the approximation  $\hbar\omega_{XX,X}\approx E_{g}$.
The effect of an electric field 
is to spatially separate the electrons from the holes,
which leads to a reduction of the oscillator strengths.
Hence, the optical transition rate can be efficiently switched off and on,
thus allowing the deterministic emission of one photon pair.

The HL biexciton state $|IJ\rangle$ can be written as
a superposition of dark ($S_{z}=\pm 2$) and bright ($S_{z}=\pm 1$) exciton states.
Upon recombination of the biexciton,
the emitted photon states are (up to normalization)
\begin{equation}
|\chi_{IJ}\rangle \propto
|\! +\!1,\theta_{1}\rangle|\! -\! 1,\theta_{2}\rangle + 
(-1)^{I\! +\! J}|\! -\! 1,\theta_{1}\rangle|\! +\!1,\theta_{2}\rangle,
\label{photons}
\end{equation}
where $|\sigma,\theta\rangle
=N(\theta)(m_{\sigma,+1}(\theta)|\sigma_{+}\rangle 
+m_{\sigma,-1}(\theta)|\sigma_{-}\rangle)$
is the state of a photon emitted from the recombination of
an electron  with spin $S_{z}=\sigma/2=\pm 1/2$ and a heavy hole with
spin $S_{z}=3\sigma/2$ in a direction which
encloses the angle $\theta$ with the normal to the plane of the
2D electron and hole motion.
The states of right and left circular
polarization are denoted $|\sigma_{\pm}\rangle$.
The entanglement between the two photon polarizations in the state Eq.~(\ref{photons}) can be
quantified by the von Neumann entropy $E$. For $|ss\rangle$ or $|tt\rangle$ 
and emission of the two photons enclosing an azimuthal angle $\phi=0$ or $\pi$,
we obtain
\begin{equation}
E=\log_{2}(1+x_{1}x_{2})-\frac{x_{1}x_{2}\log_{2}(x_{1}x_{2})}{1+x_{1}x_{2}},
\end{equation}
where $x_{i}=\cos^2(\theta_{i})$.
We plot $E(\theta_1,\theta_2)$ in Fig.~\ref{os1}c.
Only the emission of both photons perpendicular to the plane 
($\theta_{1}=\theta_{2}=0$)
results in maximal entanglement ($E=1$).
The two photons are not entangled ($E=0$) if at least
one of them is emitted in-plane ($\theta_{i}=\pi/2$).
In order to observe the proposed effect, the relaxation rate to the biexciton 
ground state needs to be larger the biexciton recombination rate.
The existence of such a regime is suggested by experiments with low 
excitation densities, see e.g.~\cite{ohnesorge96,dekel00}.
An upper limit for the pair production rate is then given by 
$(\tau_{X}+\tau_{XX})^{-1}$, where $\tau_{X}$ is the exciton 
and $\tau_{XX}$ the biexciton lifetime.


\bibliographystyle{apsrmp}
\bibliography{ssqc}

\end{document}